\documentclass[sn-mathphys,Numbered]{sn-jnl}

\usepackage{graphicx}%
\usepackage{graphics}
\usepackage{multirow}%
\usepackage{amsmath,amssymb,amsfonts}%
\usepackage{amsthm}%
\usepackage{mathrsfs}%
\usepackage[title]{appendix}%
\usepackage{xcolor}%
\usepackage{textcomp}%
\usepackage{manyfoot}%
\usepackage{booktabs}%
\usepackage{algorithm}%
\usepackage{algorithmicx}%
\usepackage{algpseudocode}%
\usepackage{listings}%
\usepackage{hyperref}       
\usepackage{url}            
\usepackage{nicefrac}       
\usepackage{microtype}      
\usepackage{subcaption}
\usepackage{array}
\usepackage{colortbl}
\usepackage{rotating}
\usepackage{wrapfig}
\usepackage{chngcntr}
\usepackage{natbib}
\usepackage{listings}
\usepackage{xcolor} 
\usepackage{caption}
\usepackage{rotating}
\usepackage{array}
\usepackage{tikz}
\usepackage{caption}
\usepackage{comment}
\usepackage[normalem]{ulem}
\definecolor{mycyan}{rgb}{0.125,0.345,1}
\definecolor{myyellow}{rgb}{1,0.85,0.196}
\newcommand{\alphalph}[1]{%
  \ifcase#1\or a\or b\or c\or d\or e\or f\or g\or h\or i\or j\or k\or l\or m\or n\or o\or p\or q\or r\or s\or t\or u\or v\or w\or x\or y\or z\or
  aa\or ab\or ac\or ad\or ae\or af\or ag\or ah\or ai\or aj\or ak\or al\or am\or an\or ao\or ap\or aq\or ar\or as\or at\or au\or av\or aw\or ax\or ay\or az\or
  ba\or bb\or bc\or bd\or be\or bf\or bg\or bh\or bi\or bj\or bk\or bl\or bm\or bn\or bo\or bp\or bq\or br\or bs\or bt\or bu\or bv\or bw\or bx\or by\or bz\or
  \else\@ctrerr\fi}



\definecolor{codegray}{gray}{0.95}

\theoremstyle{thmstyleone}%
\theoremstyle{thmstyletwo}%

\theoremstyle{thmstylethree}%

\begin{document}

\title[\textsc{Lingo}]{Efficient and Scalable Fine-Tune of Language Models for Genome Understanding}

\author[1]{\fnm{Huixin} \sur{Zhan}}\email{huixin.zhan@cshs.org}
\author[2]{\fnm{Ying Nian} \sur{Wu}}\email{ywu@stat.ucla.edu}

\author*[1,3]{\fnm{Zijun} \sur{Zhang}}\email{zijun.zhang@cshs.org}

\affil[1]{\orgdiv{Division of Artificial Intelligence in Medicine}, \orgname{Cedars-Sinai Medical Center}, \orgaddress{\city{Los Angeles}, \postcode{90048}, \state{CA}, \country{USA}}}

\affil[2]{\orgdiv{Department of Statistics}, \orgname{University of California, Los Angeles}, \orgaddress{\city{Los Angeles}, \postcode{90095}, \state{CA}, \country{USA}}}

\affil[3]{\orgdiv{Department of Computational Biomedicine}, \orgname{Cedars-Sinai Medical Center}, \orgaddress{\city{Los Angeles}, \postcode{90048}, \state{CA}, \country{USA}}}


\abstract{

Although DNA foundation models have advanced the understanding of genomes,
they still face significant challenges in the limited scale and diversity of genomic data. This limitation starkly contrasts with the success of natural language foundation models, which thrive on substantially larger scales. 
Furthermore, genome understanding involves numerous downstream genome annotation tasks with inherent data heterogeneity, thereby necessitating more efficient and robust fine-tuning methods tailored for genomics. Here, we present \textsc{Lingo}: \textsc{L}anguage prefix f\textsc{In}e-tuning for \textsc{G}en\textsc{O}mes. 
Unlike DNA foundation models, \textsc{Lingo} strategically leverages natural language foundation models' contextual cues, recalibrating their linguistic knowledge to genomic sequences. \textsc{Lingo} further accommodates numerous, heterogeneous downstream fine-tune tasks by an adaptive rank sampling method that prunes and stochastically reintroduces pruned singular vectors within small computational budgets. Adaptive rank sampling outperformed existing fine-tuning methods on all benchmarked 14 genome understanding tasks, while requiring fewer than 2\% of trainable parameters as genomic-specific adapters.
Impressively, applying these adapters on natural language foundation models matched or even exceeded the performance of DNA foundation models.
\textsc{Lingo} presents a new paradigm of efficient and scalable genome understanding via genomic-specific adapters on language models.
}




\maketitle

\section{Main}
DNA foundation models, such as DNABERT~\citep{ji2021dnabert}, DNABERT-$2$~\citep{zhou2023dnabert}, and Nucleotide Transformer (NT)~\citep{dalla2023nucleotide}, have made significant progress in decoding the linguistic intricacies of the genome. An important paradigm of utilizing such DNA foundation models is ``pre-training+finetuning'', i.e., pre-training on unlabeled genomic sequences, and then adaptation to a particular genome understanding task.  A critical aspect of genome annotation and downstream tasks is their considerable number and diversity. For example, state-of-the-art deep learning models in epigenetics alone can encompass nearly 22,000 individual tasks~\citep{chen2022sequence}. This multiplicity of tasks, considered alongside the large parameter size of these models, poses a significant challenge. As models grow in size and complexity, the practice of full-model fine-tuning (FMFT) — entailing the retraining of every model parameter for each task — becomes increasingly impractical for genomic studies~\cite{ding2023parameter}. Using one of the largest NT models with $2.5$B parameters as an example — deploying independent instances of fine-tuned models, each with $2.5$B parameters, is prohibitively resource intensive. Moreover, as the proportion of model parameters relative to training data grows, the risk of overfitting during fine-tuning rises~\citep{valipour2022dylora}. 

\begin{figure}[!htbp]
    \centering
    \begin{minipage}[t]{0.45\textwidth}
            \begin{subfigure}[t]{\linewidth}
            \centering     \captionsetup{justification=raggedright,singlelinecheck=false}
            \caption{}
            \label{fig:frame-1}
            \includegraphics[height=2.4cm,width=6.5cm]{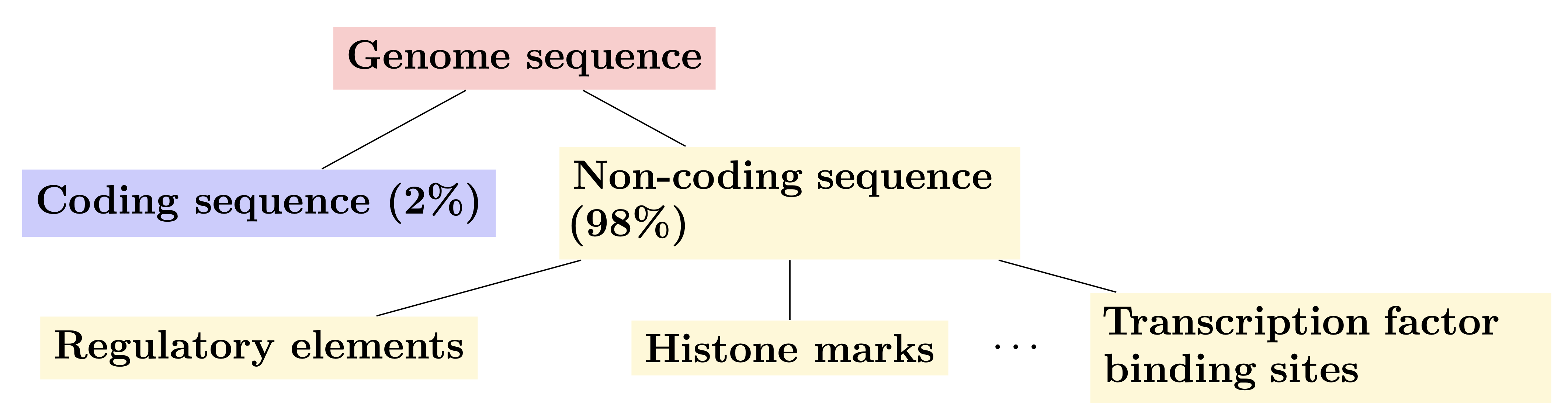}
        \end{subfigure}

            \begin{subfigure}[t]{\linewidth}
        \centering
        \captionsetup{justification=raggedright,singlelinecheck=false}
                \caption{}
        \label{fig:frame-2}
        \includegraphics[height=7.5cm,width=6.2cm]{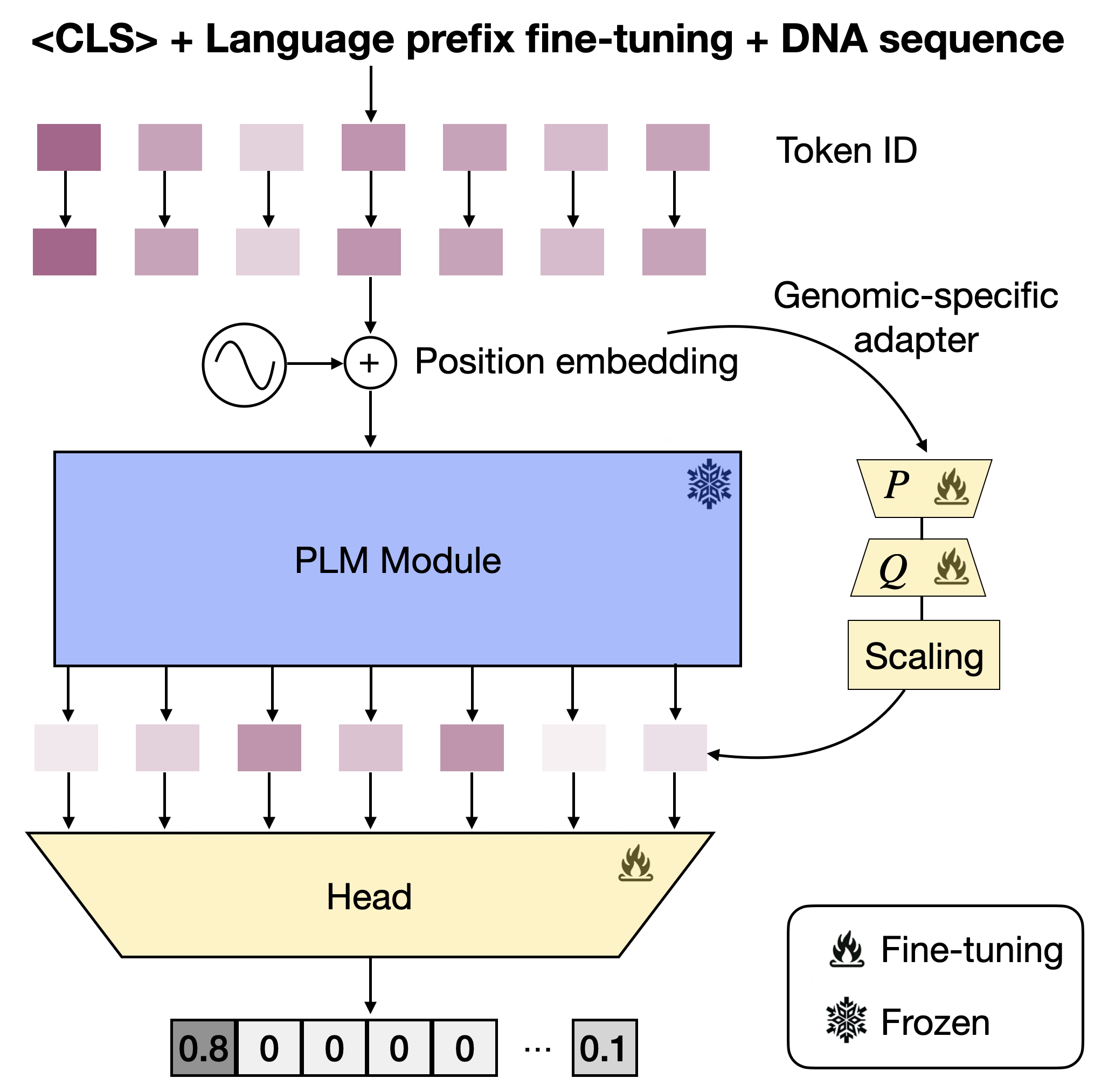}
        \end{subfigure}       
    \end{minipage}
    \hfill
    \begin{minipage}[t]{0.45\textwidth}

                \begin{subfigure}[t]{\linewidth}
            \centering
            \captionsetup{justification=raggedright,singlelinecheck=false}
            \caption{}
            \label{fig:frame-3}
            \includegraphics[height=2.5cm,width=6cm]{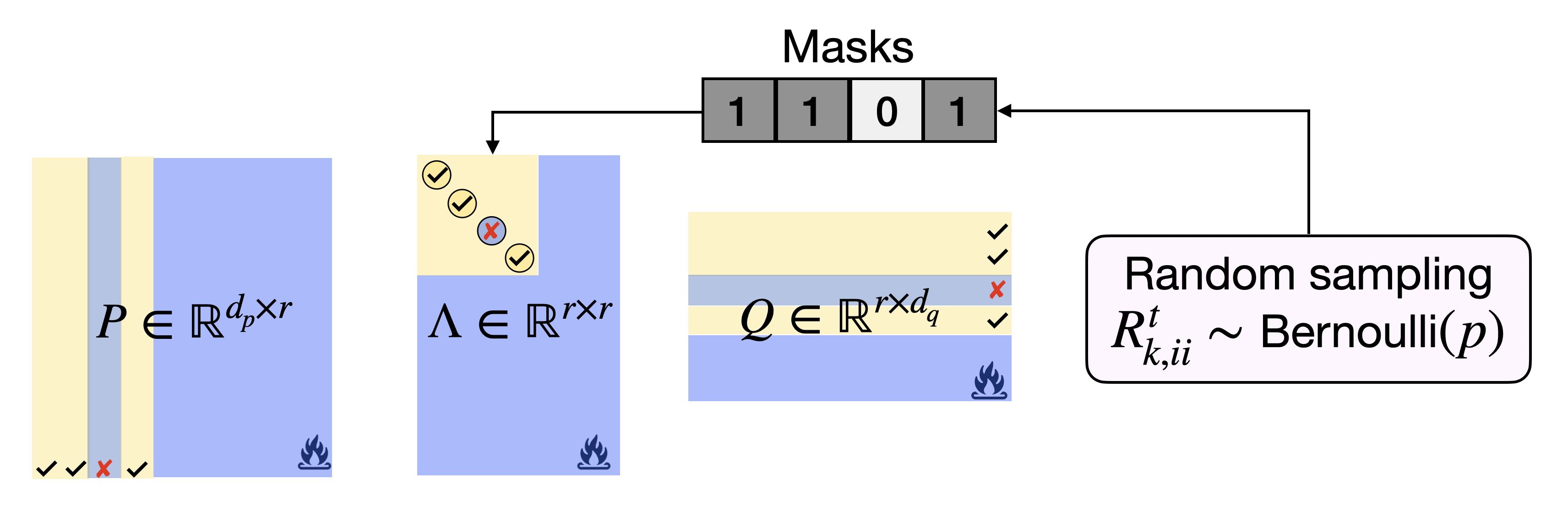} 

        \end{subfigure}

                \begin{subfigure}[t]{\linewidth}
            \centering
            \captionsetup{justification=raggedright,singlelinecheck=false}
                        \caption{}
            \label{fig:frame-4}
            \includegraphics[height=1.8cm,width=5cm]{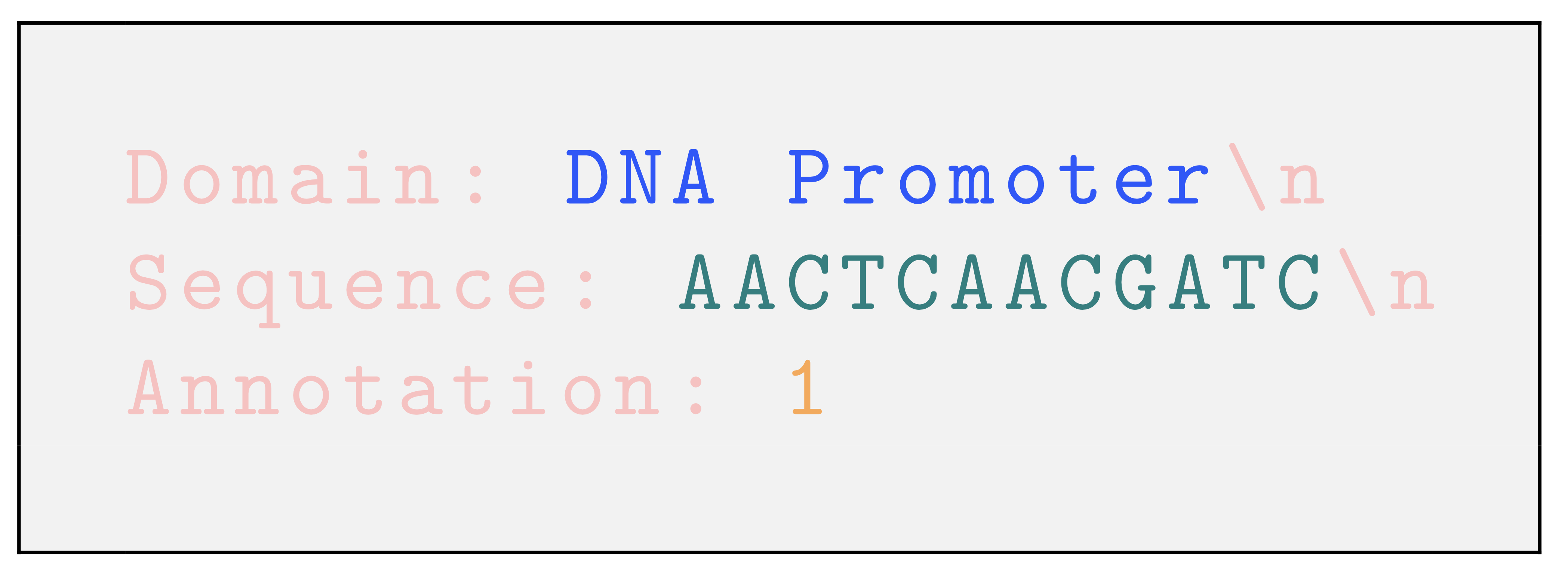}
        \end{subfigure}
                \begin{subfigure}[t]{\linewidth}
            \centering
            \captionsetup{justification=raggedright,singlelinecheck=false}
                        \caption{}
            \label{fig:frame-5}
            \includegraphics[height=4.9cm,width=6cm]{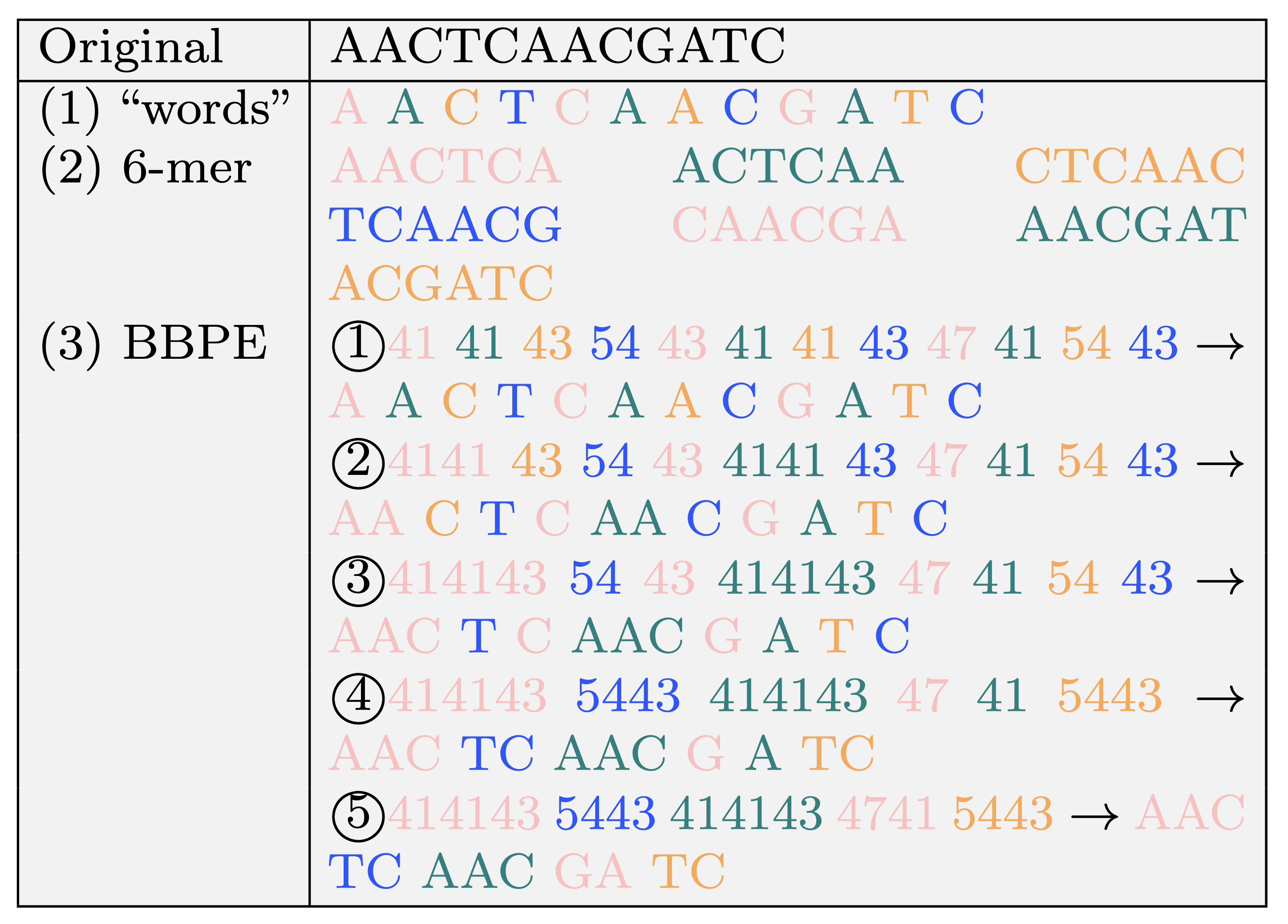}
        \end{subfigure}
     \end{minipage}
    \caption{\textbf{\large{a}} Heterogeneity inherent in genomic sequence data. Genome sequences consist of 2\% coding sequences and 98\% non-coding sequences, including regulatory elements, histone marks, and transcription factors, illustrating the complexity of genomic structure. \textbf{\large{b}} \textsc{Lingo} framework for domain-shift genome understanding. Instead of applying a full-model fine-tuning to the Pretrained Language Model (PLM) modules (illustrated in blue), our method uses low-rank approximation with adaptive rank sampling on Low-rank Adapters (LoRA) blocks (depicted in yellow). These LoRA blocks are integrated with the classification head, regulated by a scaling hyperparameter. \textbf{\large{c}} The adaptive rank sampling method. In this technique, we generate masks for every singular value based on a Bernoulli distribution. These masks are then applied to the singular values, enabling the pruning and stochastic reintroduction of pruned singular vectors. \textbf{\large{d}} Illustration of natural language prompting. These promptings encompass a prefix and a suffix, strategically positioned around the genomic sequence. The prefix is designed to transmit domain-specific information, while the suffix provides annotation details. \textbf{\large{e}} Different methodologies for tokenizing genomic sequences. Displayed for each row are the original genomic sequence and its respective transformations into ``words'' tokenization, 6-mer tokenization, and byte-level byte-pair encoding (BBPE) tokenization.}
\end{figure}

Computationally, there are two lines of solutions to efficiently fine-tune DNA foundation models on a large scale: first, model compression to reduce model size; second, parameter-efficient fine-tuning (PEFT) to add task-specific adapters in the small-parameter regime. While model compression approaches have been well-established in recent years, implementing them on large language models can be very expensive, as these techniques typically necessitate FMFT~\citep{ma2023llm}. As a countermeasure, PEFTs focus on fine-tuning the model on only a small number of additional parameters, significantly decreasing the computational costs. In PEFTs, low-rank adaptation, e.g., low-rank adapters (LoRA)~\citep{hu2021lora} and adaptive low-rank adaptation (AdaLoRA)~\citep{zhang2023adaptive} are increasingly prominent. While LoRA introduces fine-tuning through fixed-rank LoRA blocks, AdaLoRA adaptively decreases the total rank of all LoRA blocks, maintaining salient singular values based on their importance scores. Building on the foundations of low-rank adaptation methods, KronA~\citep{edalati2022krona} and FedPara~\citep{hyeon2021fedpara} introduce advanced techniques: KronA replaces LoRA projections with Kronecker factors to enhance representability, while FedPara leverages a novel re-parameterization with low-rank weights and a Hadamard product, achieving full-rank matrix and tensor expressivity with reduced parameters. However, all aforementioned methods operate deterministically based on pre-pruning states, leading to potential sub-optimal outcomes in genome understanding tasks, particularly when these pre-pruning states are not ideal. This deterministic nature of PEFTs becomes a significant limitation due to the considerable heterogeneity of genomic sequences, culminating in challenges posed by inherent data heterogeneity. For example, as illustrated in Figure~\ref{fig:frame-1}, the genetic elements encompass both coding sequences and non-coding regulatory sequences, such as histone modifications and transcription factor binding sites. This diverse array of elements highlights the complexity and multifaceted nature of genomic composition and regulation. Therefore, it is essential to introduce randomness to the inherently unstable pre-pruning states. 

Echoing the substantial parameters in DNA foundation models that hamper their downstream fine-tuning capability, their development has been limited by data scalability issues~\cite{dalla2023nucleotide,floridi2020gpt,byrska2022high}. GPT-3~\cite{floridi2020gpt}, trained on approximately $45$ terabytes of text data, vastly surpasses the NT with $173.9$ billion nucleotides in its multispecies dataset from NCBI~\footnote{\url{https://www.ncbi.nlm.nih.gov/}}, thereby demonstrating a significant difference in the scale of training data between a broad-scope pre-trained natural language foundation model (PLM) and a specialized DNA foundation model. Moreover, the relative constraint in genomic data availability and diversity hinders the ability of these models to reach the levels of efficacy and robustness that are evident in their natural language processing (NLP) and computer vision (CV) counterparts, underscoring a critical area for future development of DNA foundation models. For instance, whole-genome sequencing data from 20,314 individuals in the gnomAD consortium revealed 261.9 million distinct genetic variants, indicating a variant-to-base pair ratio of approximately 4.4e-6 in this cohort~\citep{karczewski2020mutational}.  As a consequence, the generalizability and therefore reasoning abilities of DNA foundation models has been questioned~\cite{tangbuilding}. In contrast, PLMs have achieved remarkable progress in NLP~\cite{hupkes2023taxonomy,touvron2023llama} and CV~\cite{wang2023internimage}. More importantly, recent advances have shown that PLMs possess surprising abilities of in-context reasoning and cross-modality learning~\citep{coskun2023can,hu2023zero,dinh2022lift}. PLMs that are only trained on natural languages exhibit compression ability of images that surpass established compression algorithms~\citep{deletang2023language}. The idea of large language models as universal compute machines has been demonstrated effective for NLPs and images, but not yet for DNAs. Therefore, empirically, our objective is to explore and substantially expand the domain-shift ability of PLMs on genome understanding. Nevertheless, the venture of PLMs into the field of genomic sequences introduces unique challenges. PLMs are initially trained on vast amounts of general language data and thus develop a strong understanding of natural language. However, genomic sequences, despite being sequential and preserving complex patterns, do not conform to the rules of human language. Moreover, the effective tokenization and context length in genomics are still debatable~\cite{dotan2023effect}.

Here, we introduce a new paradigm of efficient and scalable genome understanding via genomic-specific adapters on PLMs. To adapt the PLMs to genome domain, we develop a novel \textsc{Lingo}: \textsc{L}anguage prefix f\textsc{In}e-tuning for \textsc{G}en\textsc{O}mes approach to prime PLMs, i.e., open pre-trained transformers (OPTs), for genome understanding tasks. Unlike the direct input of DNA sequences, \textsc{Lingo} leverages the inherent contextual learning capabilities of PLMs to guide their transition from processing natural language to interpreting genomic sequences, thereby recalibrates their extensive linguistic knowledge to the intricacies of genomic sequences. Methodologically, \textsc{Lingo}'s adaptive rank sampling prunes and stochastically reintroduces pruned singular vectors, adhering to a cubic budget schedule. This technique is widely applicable across various foundation models, particularly useful in addressing the unstable pre-pruning frequently observed in genomics. In addition, we have selected byte-level byte-pair encoding (BBPE) tokenization for genomic sequences, incorporating the use of frequently occurring token IDs to effectively train on DNA sequences. 

We applied \textsc{Lingo} on a comprehensive set of genome understanding tasks. Across all PLMs we tested, \textsc{Lingo}'s adaptive rank sampling achieves superior performance among PEFT methods. On \textsc{Lingo}-trained OPT with 350M parameters, i.e. OPT-350M, its performances matched or surpassed FMFT on all 14 genomic sequence datasets we benchmarked, while utilizing under $2\%$ of the trainable parameters as genomic-specific adapters. More impressively, compared to state-of-the-art DNA foundation models trained on DNA sequences alone, \textsc{Lingo} also demonstrated superior performance by applying these genomic-specific adapters to PLMs. In the specific context of the ten datasets in histone marker prediction task in yeast, \textsc{Lingo}-trained OPT-350M is among the Top-2 performed models in 9/10 tasks, consistently surpassing either or both of DNABERT-2 and Nucleotide Transformer. Our \textsc{Lingo} framework provides a powerful new strategy for genome understanding, and a significant step for extending artificial general intelligence in genomics.

\section{Results}
\subsection{Language prefix fine-tuning framework for genome understanding}

We developed a language prefix fine-tuning framework, \textsc{Lingo}, for scalable and effective genome understanding tasks. 
Unlike conventional DNA foundation models that are only pre-trained on DNA sequence, \textsc{Lingo} leverages the natural language PLMs, incorporating a genomic-specific adapter through a novel and robust approach designed to accommodate the heterogeneity of genomic data.
As shown in Figure~\ref{fig:frame-2} and~\ref{fig:frame-3}, we introduce our adaptive rank sampling in \textsc{Lingo}. Adaptive rank sampling, rather than fine-tuning the entire PLM modules $W_0$ (in blue), effectively reduces the rank count (Figure~\ref{fig:frame-2}). This approach projects high-dimensional weight matrices into smaller subspaces, utilizing singular value decomposition (SVD)~\cite{stewart1993early} for a low-rank approximation of gradient updates, represented as $W_0 + \Delta W = W_0 + P \Lambda Q$. Here, $P \in \mathbb{R}^{d_p \times r}$ and $Q \in \mathbb{R}^{r \times d_q}$, where the rank $r \ll \min(d_p,d_q)$. Fine-tuning is then exclusively conducted on the genomic-specific adapter, specifically the matrices $P$ and $Q$, along with the classification heads (illustrated in olive). The genomic-specific adapter optimizes computational efficiency and model performance in high-dimensional data contexts. Adaptive rank sampling, as depicted in Figure~\ref{fig:frame-3}, involves the definition of masks $R_{k,ii}^t$, which are employed both for pruning each singular value and subsequently reintroducing it. These masks are conceptualized as random variables, each following a Bernoulli distribution characterized by the parameter $p$, denoted as $R_{k,ii}^t \sim \text{Bernoulli}(p)$. This stochastic approach introduces an element of randomness essential for our analysis and model robustness. Adaptive rank sampling adeptly retains crucial singular values, taking into account both their importance and sensitivity during current batch training. 

\textsc{Lingo} further leverages text prototypes, comprising a prefix and a suffix flanking the genomic sequence (Figure~\ref{fig:frame-4}). 
The prefix conveys domain information, while the suffix conveys annotation details. For the processing of these prefixes and suffixes, we employ the BBPE tokenizer of GPT-2~\cite{floridi2020gpt}. Furthermore, as illustrated in Figure~\ref{fig:frame-5}, although various typical methods exist for tokenizing genomic sequences, such as one-hot or k-mer tokenization, we opt for BBPE tokenization for genomic sequences as well. This tokenizer is well-suited for DNA sequences, adeptly capturing the frequent patterns of nucleotides. To achieve this, the BBPE tokenizer begins with an initialization of a dictionary containing all individual bytes in UTF-8 encoding. It then iteratively merges the most frequently occurring token pairs. Each newly formed pair is subsequently incorporated into the dictionary as a novel token, a process illustrated in steps $\textcircled{1}$ through $\textcircled{5}$.

We evaluated \textsc{Lingo} against FMFT, LoRA, AdaLoRA, adaptive rank sampling, on foundation models using three genome understanding tasks, i.e., histone marker prediction in yeast, i.e., Histone (Yeast), promoter detection in human, i.e., Promoter (Human), and histone mark prediction in human, i.e., Histone (Human). The statistics for all three datasets and the setup are shown in Supplementary Table~\ref{tab:dataset} and Supplementary Table~\ref{tab:hyperparameters}. All \textsc{Lingo} in the results section reflects language prefix fine-tuning with adaptive rank sampling.

\subsection{Full model fine-tune of OPTs matches DNA foundation models}

\begin{figure}[!htbp]
    \centering
            \begin{subfigure}[b]{0.49\linewidth}
        \captionsetup{justification=raggedright,singlelinecheck=false}
        \caption{}\label{fig:fmft_h3}
        \includegraphics[height=5cm]{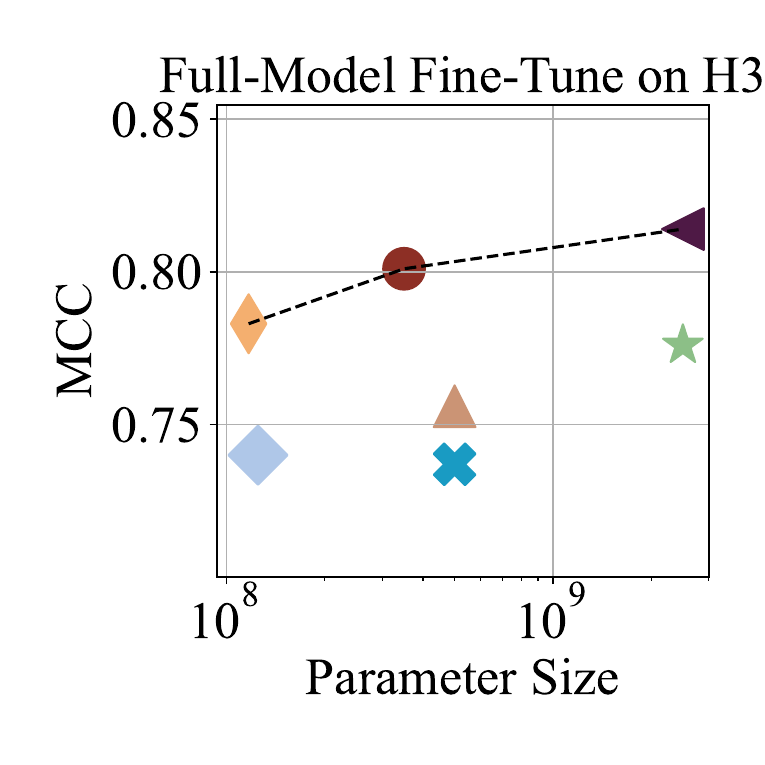}
    \end{subfigure}
    \begin{subfigure}[b]{0.49\linewidth}
       \captionsetup{justification=raggedright,singlelinecheck=false}
       \caption{}\label{fig:fmft_tata}
        \includegraphics[height=5cm]{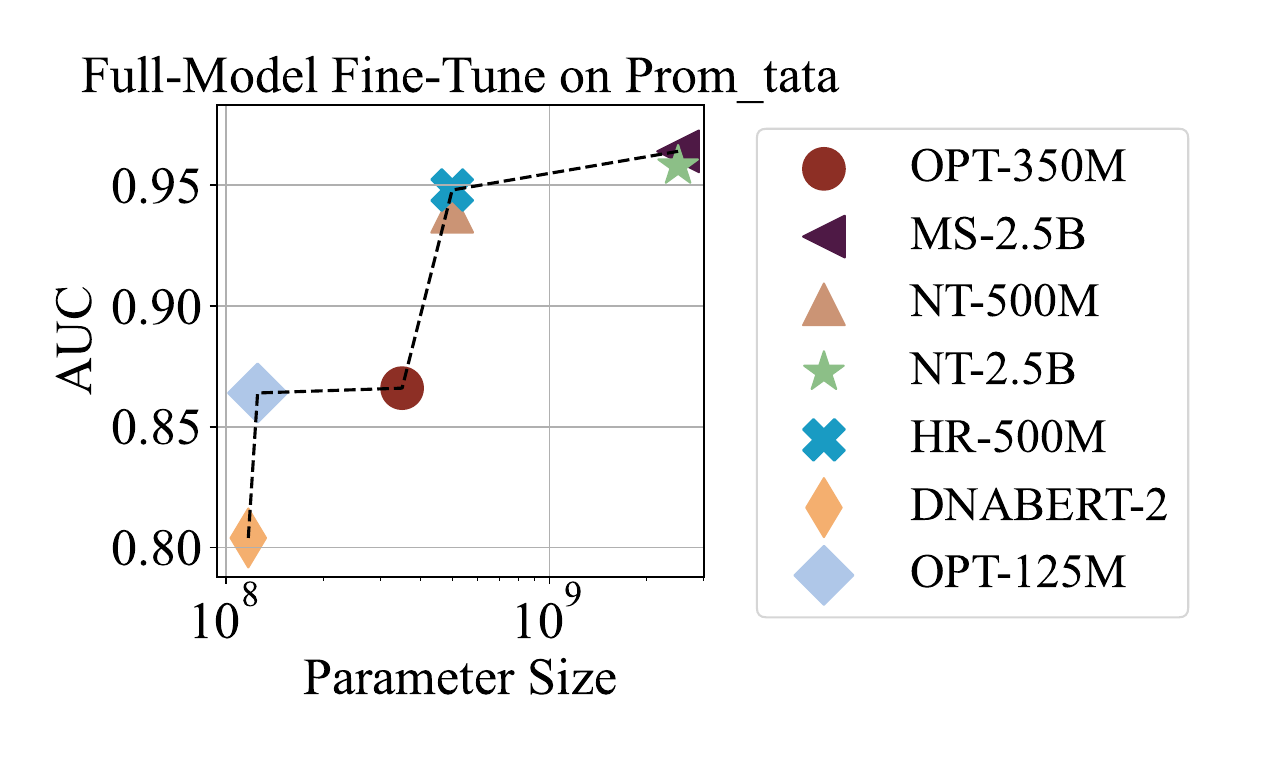}
    \end{subfigure}
        \begin{subfigure}[b]{0.49\linewidth}
        \captionsetup{justification=raggedright,singlelinecheck=false}
         \caption{}\label{fig:fmft_emp}
        \includegraphics[height=7cm]{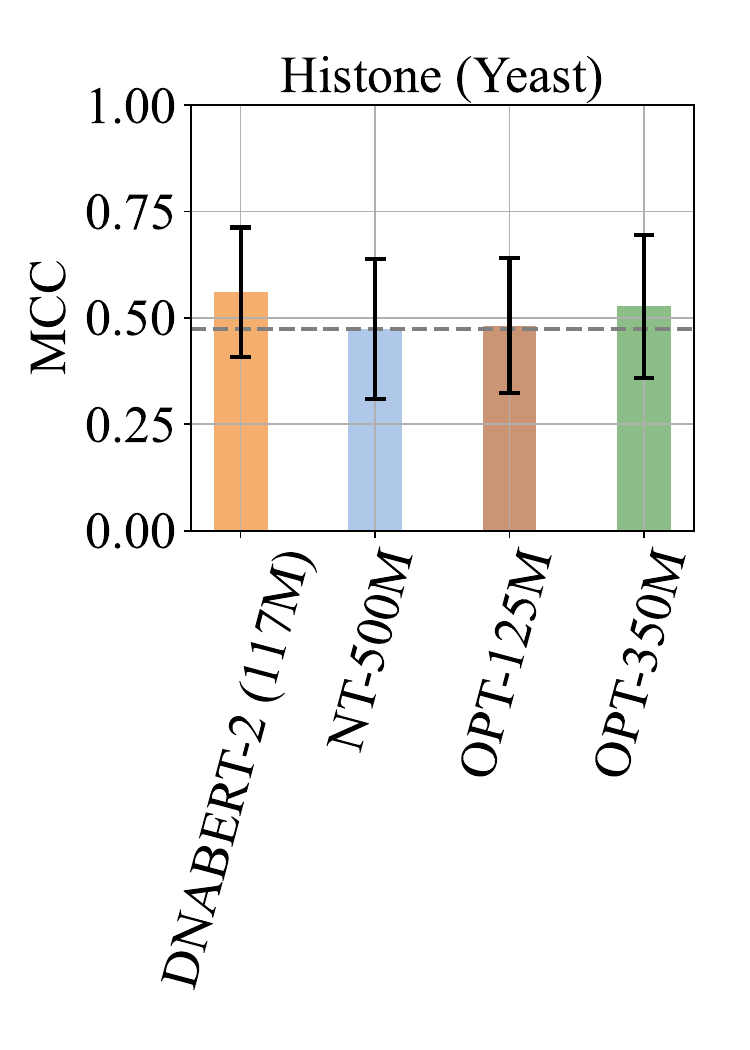}
    \end{subfigure}
    \begin{subfigure}[b]{0.49\linewidth}
    \captionsetup{justification=raggedright,singlelinecheck=false}
    \caption{}\label{fig:fmft_pd}
        \includegraphics[height=7cm]{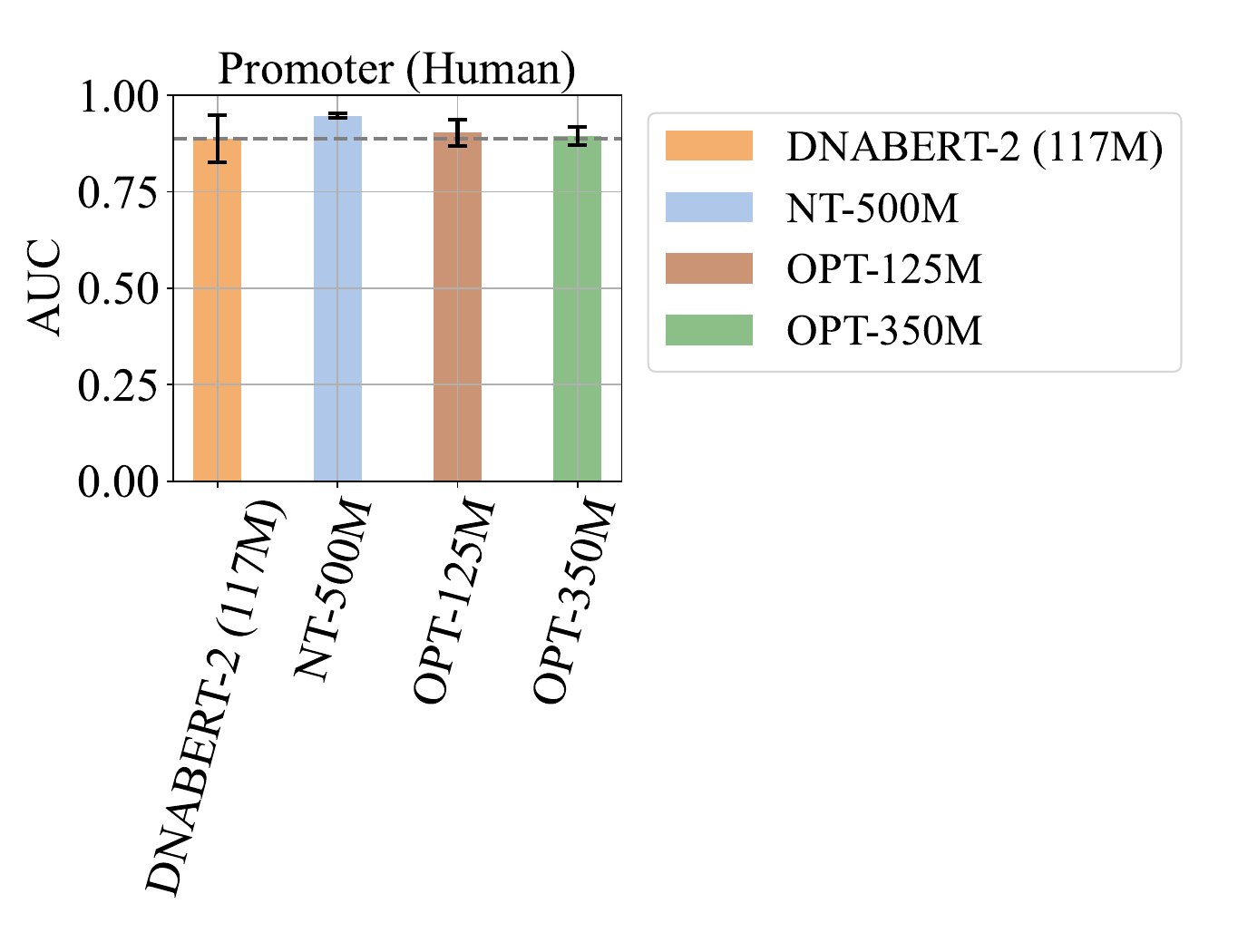}
    \end{subfigure}
    \caption{\textbf{\large{a}} Full-model fine-tuning performance of various models on the H3 dataset. The x-axis represents the parameter size, with a lower value being preferable, while the y-axis indicates the MCC, where higher values are desirable. The Pareto front is marked by a dotted line, with the upper left corner signifying models that achieve a balance of lower parameter size and higher MCC. Notably, OPT-350M aligns with the Pareto front alongside two DNA foundation models, DNABERT-2 and MS-2.5B. \textbf{\large{b}} Full-model fine-tuning performance on the Prom\_tata dataset. Here, the x-axis measures parameter size, and the y-axis shows the AUC, with the Pareto front again indicated by a dotted line. The upper left corner, indicating optimal performance, features both OPT-125M and OPT-350M models. \textbf{\large{c}} Illustration of the average MCC on the Histone (Yeast) dataset. OPT-125M and OPT-350M outperform the NT-500M model. \textbf{\large{d}} Average AUC on the Promoter (Human) dataset. OPT-125M and OPT-350M surpassing DNABERT-2 in performance.}
\end{figure}

To demonstrate the potential of applying natural language PLMs towards genome understanding tasks, we first performed FMFT on OPTs on previously established benchmark genome sequence datasets~\citep{zhou2023dnabert}.
The results for FMFT on different models are shown in Figure~\ref{fig:fmft_h3} and~\ref{fig:fmft_tata}. As baseline comparisons, we report the previously published evaluation metrics of DNA foundation models, including DNABERT-2, and four DNA foundation models from NT, with sizes spanning from $500$M to $2.5$B parameters. Consistent with previous evaluation metrics, the Matthews Correlation Coefficient (MCC)~\citep{chicco2020advantages} and Area Under the Curve (AUC) were computed for the OPT-125M and OPT-350M models, which were fully fine-tuned on Histone (Yeast) and Promoter (Human) tasks. The NT models have been pre-trained on three distinct datasets: the human reference genome (HR), the $1000$G dataset~\citep{byrska2022high}, and genomes from multiple species (MS); therefore they are referred to as HR-$500$M, NT-$500$M (with $1000$G data), NT-$2.5$B (with $1000$G data), and MS-$2.5$B. 

Full-model fine-tuned OPTs achieved on par performance with DNA foundation models.
In Figure~\ref{fig:fmft_h3}, on FMFT of H3 Histone (Yeast) task, we find that the OPT with $350$M parameters (OPT-$350$M) lies on the Pareto front (shown in black dotted line) with two DNA foundation models, i.e., DNABERT-$2$ and MS-$2.5$B. 
Specifically, the OPT-$350$M achieves a $2.5\%$ decrease in MCC while utilizing a parameter size of $14\%$ compared to MS-$2.5$B. The OPT model with $125$M parameters (OPT-$125$M) outperforms the DNA foundation model HR-$500$M with $25\%$ of HR-$500$M parameters. Similarly, in Figure~\ref{fig:fmft_tata}, in the context of FMFT for the Prom\_tata Promoter (Human) task, both OPT-$125$M and OPT-$350$M models are positioned on the Pareto front, as indicated by the black dotted line. This performance is comparable with three additional DNA foundation models: DNABERT-$2$, HR-$500$M, and MS-$2.5$B. 

In Figures~\ref{fig:fmft_emp} and~\ref{fig:fmft_pd}, we present the average MCC and AUC for the ten Histone (Yeast) datasets and three Promoter (Human) datasets, respectively. Specifically, Figure~\ref{fig:fmft_emp} reveals that the average MCC for OPT-125M and OPT-350M surpasses that of NT with 500M parameters, i.e., NT-500M, by 0.85\% and 5.39\%, respectively. Similarly, Figure~\ref{fig:fmft_pd} shows that the average AUC for OPT-125M and OPT-350M exceeds DNABERT-2 by 1.57\% and 0.7\%, respectively. Despite the pre-training of OPT models on natural language datasets, their FMFT performance proves highly competitive against DNA foundation models. This competitive edge is likely attributed to their robust reasoning abilities, as discussed in \citep{coskun2023can,hu2023zero,dinh2022lift}. In light of these findings, it is evident that the transferability of skills learned from NLP to genomic sequence understanding holds substantial promise.

\subsection{Adaptive rank sampling enhances genomic adapters across foundation models}

\begin{figure}[!htbp]
    \centering
    \begin{subfigure}[b]{0.22\linewidth}
    \captionsetup{justification=raggedright,singlelinecheck=false}
    \caption{}\label{fig:peft-1}
        \includegraphics[width=3cm,height=3cm]{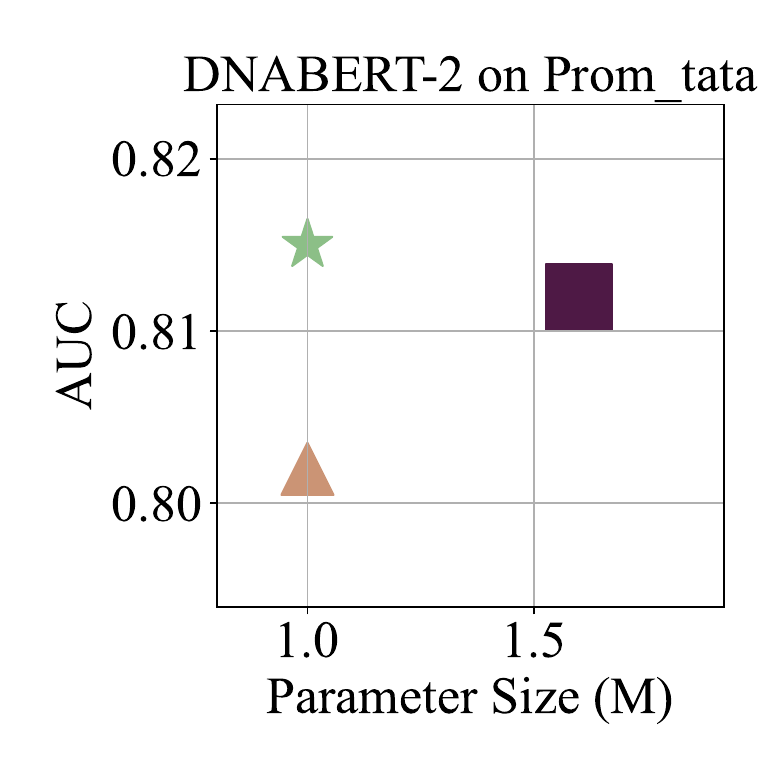}
    \end{subfigure}
    \begin{subfigure}[b]{0.22\linewidth}
    \captionsetup{justification=raggedright,singlelinecheck=false}
      \caption{}\label{fig:peft-3}
        \includegraphics[width=3cm,height=3cm]{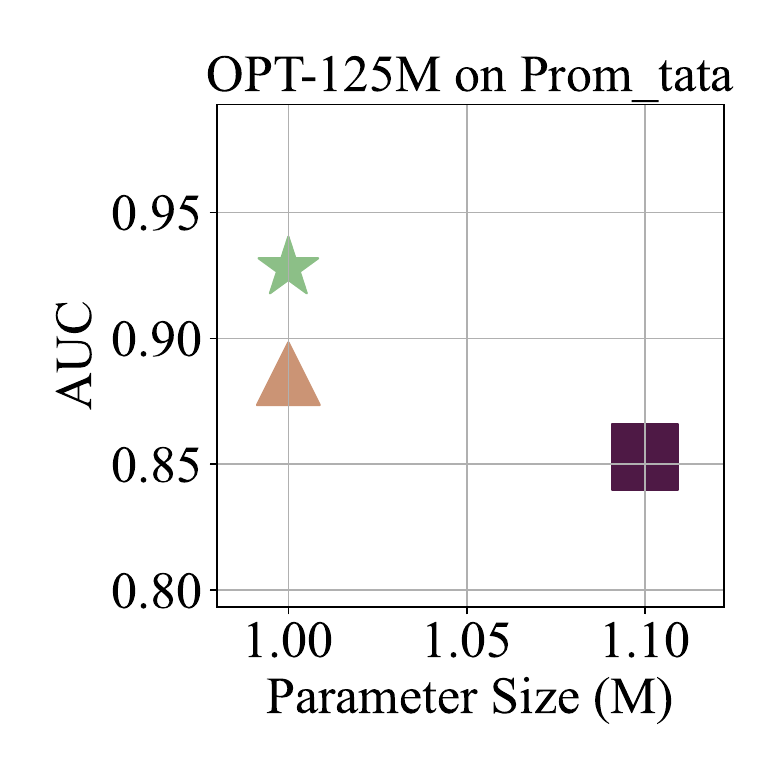}
    \end{subfigure}
    \begin{subfigure}[b]{0.22\linewidth}
    \captionsetup{justification=raggedright,singlelinecheck=false}
            \caption{}\label{fig:peft-4}
        \includegraphics[width=3cm,height=3cm]{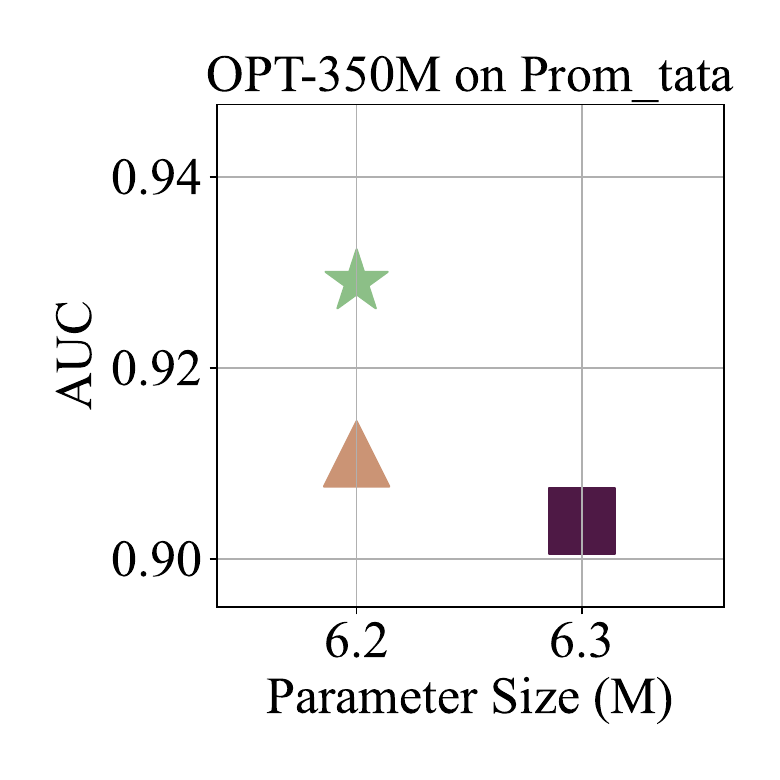}
    \end{subfigure}
     \begin{subfigure}[b]{0.22\linewidth}
    \captionsetup{justification=raggedright,singlelinecheck=false}
    \caption{}\label{fig:peft-2}
        \includegraphics[width=3cm,height=3cm]{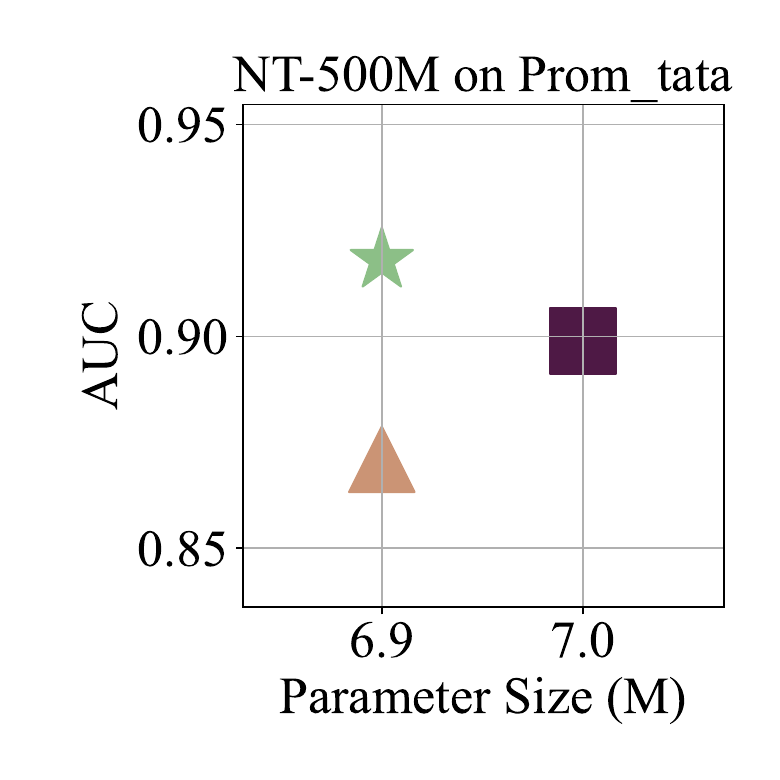}
    \end{subfigure}

        \begin{subfigure}[b]{0.22\linewidth}
        \captionsetup{justification=raggedright,singlelinecheck=false}
                \caption{}\label{fig:peft-5}
        \includegraphics[width=3cm,height=3cm]{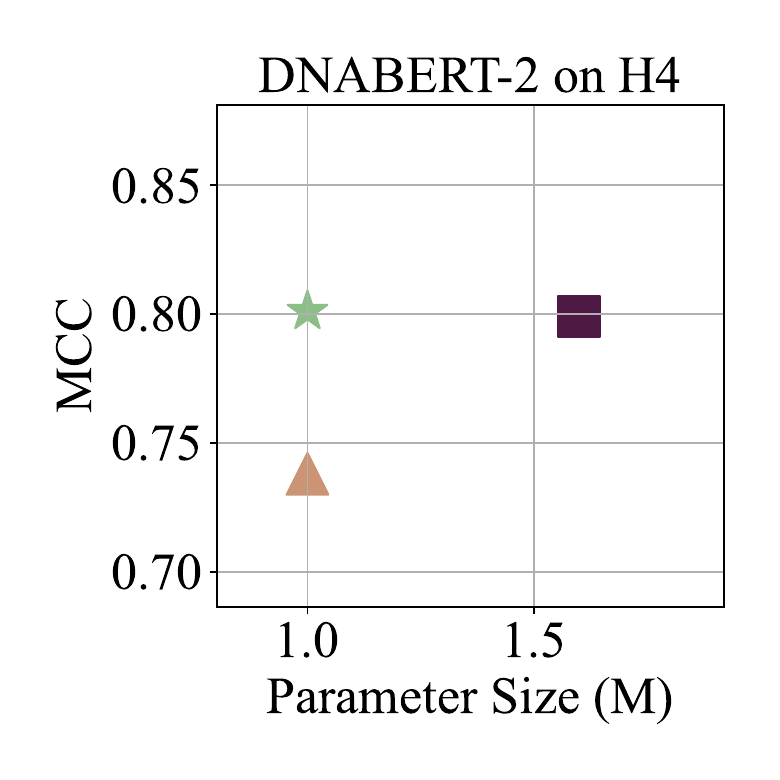}
    \end{subfigure}
    \begin{subfigure}[b]{0.22\linewidth}
    \captionsetup{justification=raggedright,singlelinecheck=false}
            \caption{}\label{fig:peft-7}
        \includegraphics[width=3cm,height=3cm]{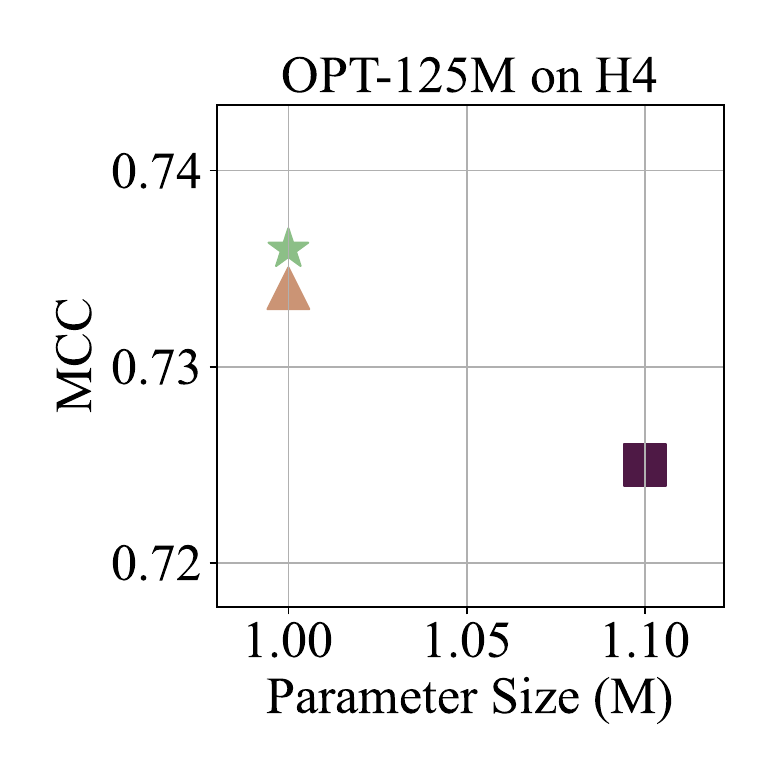}
    \end{subfigure}
    \begin{subfigure}[b]{0.22\linewidth}
    \captionsetup{justification=raggedright,singlelinecheck=false}
            \caption{}\label{fig:peft-8}
        \includegraphics[width=3cm,height=3cm]{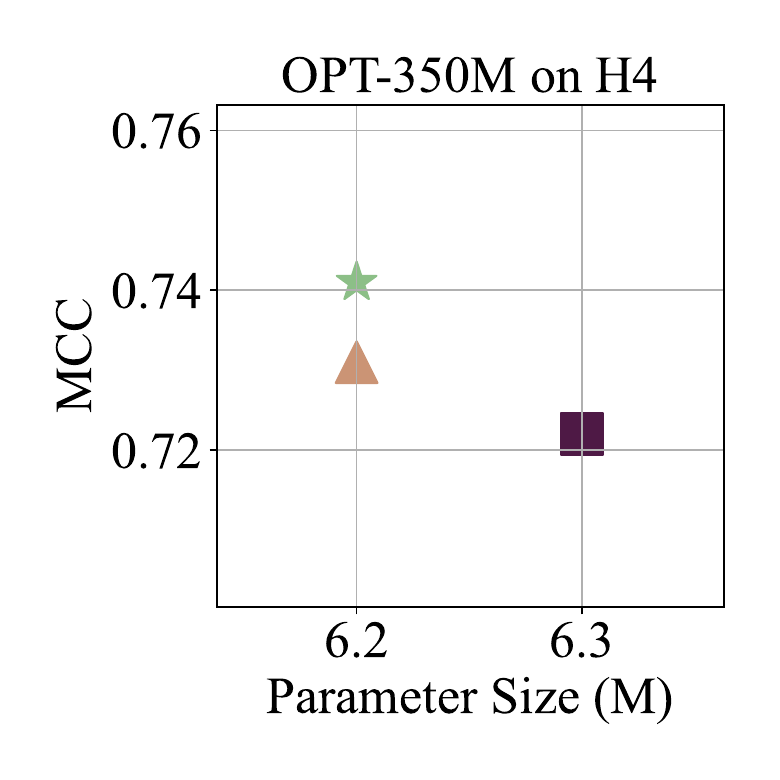}
    \end{subfigure}
        \begin{subfigure}[b]{0.22\linewidth}
    \captionsetup{justification=raggedright,singlelinecheck=false}
            \caption{}\label{fig:peft-6}
        \includegraphics[height=3cm]{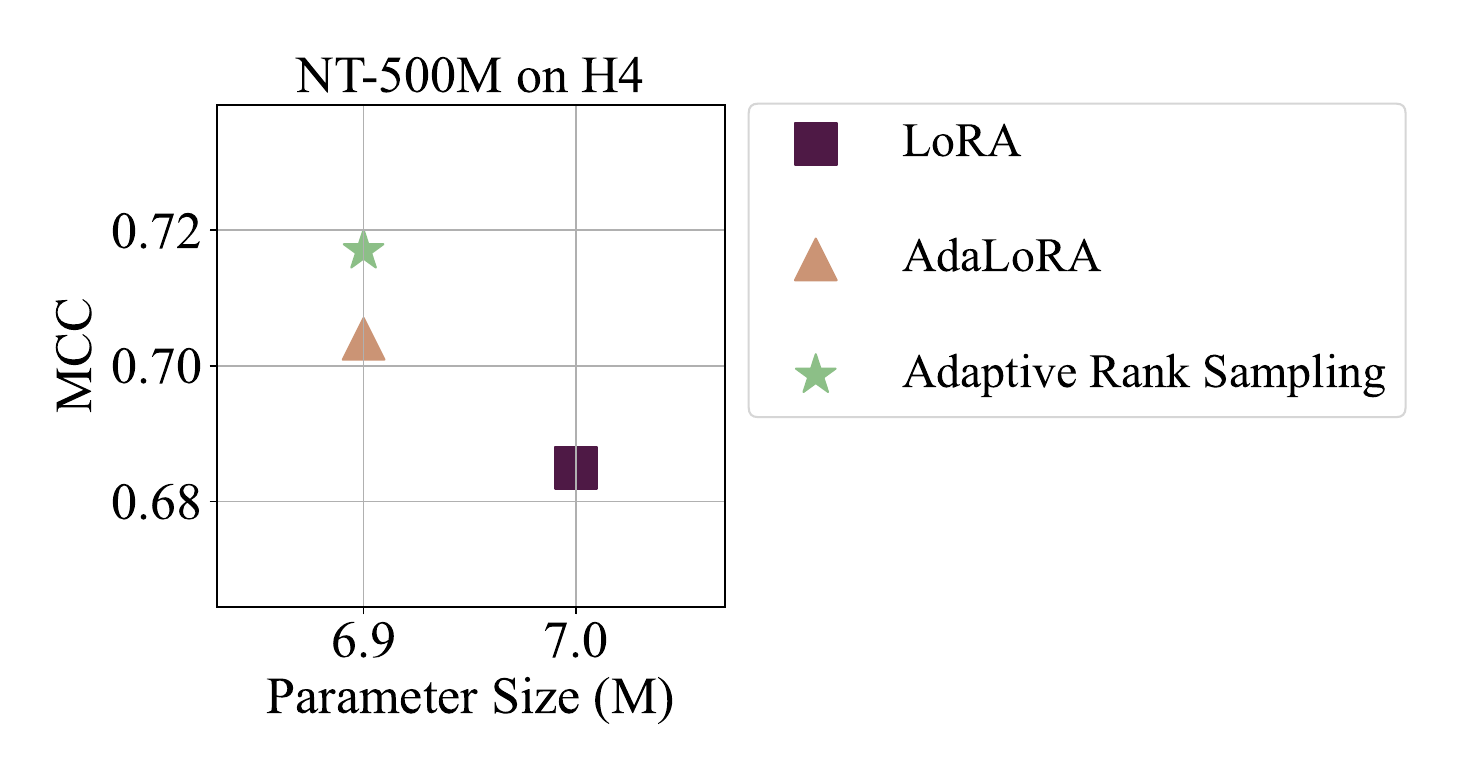}
    \end{subfigure}

    \caption{\textbf{\large{a}}, \textbf{\large{b}}, \textbf{\large{c}}, and \textbf{\large{d}}: Relationship between the AUC and parameter size for the Prom\_tata dataset within the Promoter (Human) task, focusing on DNABERT-2, OPT-125M, OPT-350M, and NT-500M models. Panels \textbf{\large{a}}, \textbf{\large{b}}, \textbf{\large{c}}, and \textbf{\large{d}} are arranged in ascending order of parameter size, ranging from 117M to 500M, to facilitate a comparative analysis based on model complexity. In each panel, models positioned in the upper-left indicate superior performance, characterized by higher AUC and lower parameter size. \textbf{\large{e}}, \textbf{\large{f}}, \textbf{\large{g}}, and \textbf{\large{h}}: Illustration of the MCC versus parameter size for the H4 dataset in the Histone (Yeast) task, again examining the same set of models: DNABERT-2, OPT-125M, OPT-350M, and NT-500M. }
\end{figure}

We then asked if OPTs could remain competitive to DNA foundation models in the small-parameter adapter regime using PEFT methods. Towards this end, we benchmarked the performance of three PEFT methods across both DNA foundation models and OPTs, while also evaluating the benefit of adaptive rank sampling across various types of DNA foundation models and PLMs (Figure ~\ref{fig:peft-1} - ~\ref{fig:peft-8}). The Pareto front is depicted with a dotted line. Across all tests for OPTs, adaptive rank sampling consistently outperforms the other two PEFT methods. For instance, Figure~\ref{fig:peft-4} shows OPT-350M surpassing LoRA with a 2.5\% increase in AUC and 1e5 fewer trainable parameters for the Prom\_tata Promoter (Human) task. Similarly, Figure~\ref{fig:peft-8} illustrates OPT-350M outperforming LoRA by 1.7\% in AUC, again with 1e5 fewer trainable parameters on the same task. 
The superior performance of adaptive rank sampling on OPTs is consistent across ten Histone (Yeast) datasets and three Promoter (Human) datasets (Supplementary Figure~\ref{fig:peft-more}aa - Figure~\ref{fig:peft-more}az). 

Across all DNA foundation models and PLMs we tested, our adaptive rank sampling consistently outperforms other PEFT methods for both Histone (Yeast) and  Promoter (Human) tasks. To provide a clearer illustration of the AUC and MCC in relation to the parameter size for LoRA, AdaLoRA and adaptive rank sampling, we have detailed these metrics of Histone (Yeast) in Supplementary Table~\ref{tab:MCC_EMP} and Promoter (Human) in Supplementary Table~\ref{tab:PD_AUC}. 

When applied to DNA foundation models, the performance of adaptive rank sampling aligns with the Pareto front observed in FMFT outcomes and is comparable to that of FMFT. As an example, in Supplementary Table~\ref{tab:PD_AUC}, adaptive rank sampling posts a 0.926 AUC with merely 6.9M trainable parameters (only 1.3\% of FMFT's parameters), whereas FMFT achieves a 0.95 AUC using 500M trainable parameters for NT-500M model. Supplementary Table~\ref{tab:MCC_EMP} shows the MCCs for various models and methods on the Histone (Yeast) task, where adaptive rank sampling's performance lies on a Pareto front with that of FMFT. 
This demonstrates that adaptive rank sampling, by accommodating the unstable pre-pruning states in genomic data, learns effective genomic-specific adapters across all foundation models.

\begin{figure}[!htbp]
    \centering
      \begin{subfigure}[b]{0.34\linewidth}
          \captionsetup{justification=raggedright,singlelinecheck=false}
          \caption{}\label{fig:peft-9}
        \includegraphics[height=3cm,width=4.7cm]{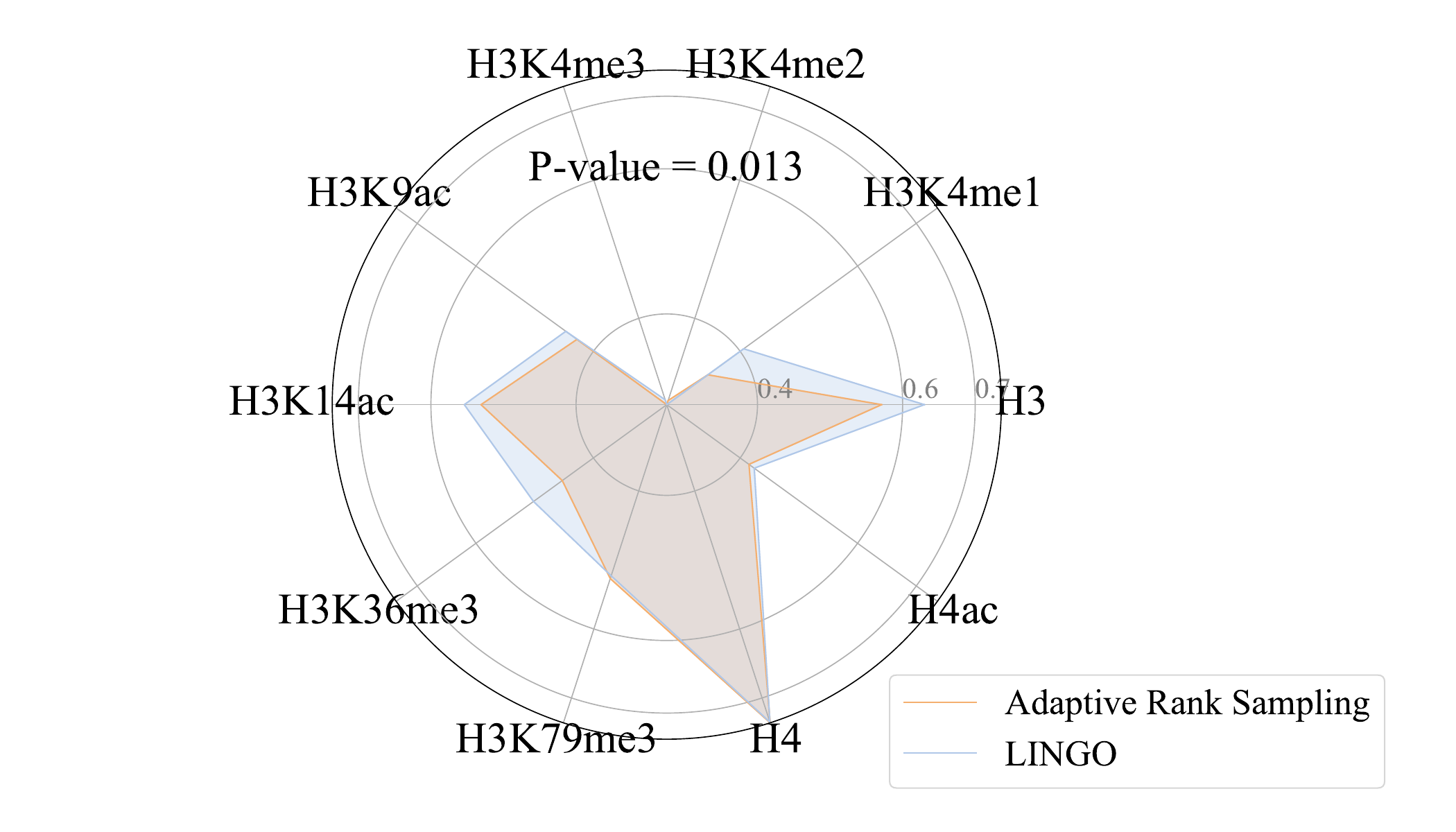}
    \end{subfigure}
     \begin{subfigure}[b]{0.34\linewidth}
         \captionsetup{justification=raggedright,singlelinecheck=false}
                 \caption{}\label{fig:peft-10}
        \includegraphics[height=3cm,width=4.7cm]{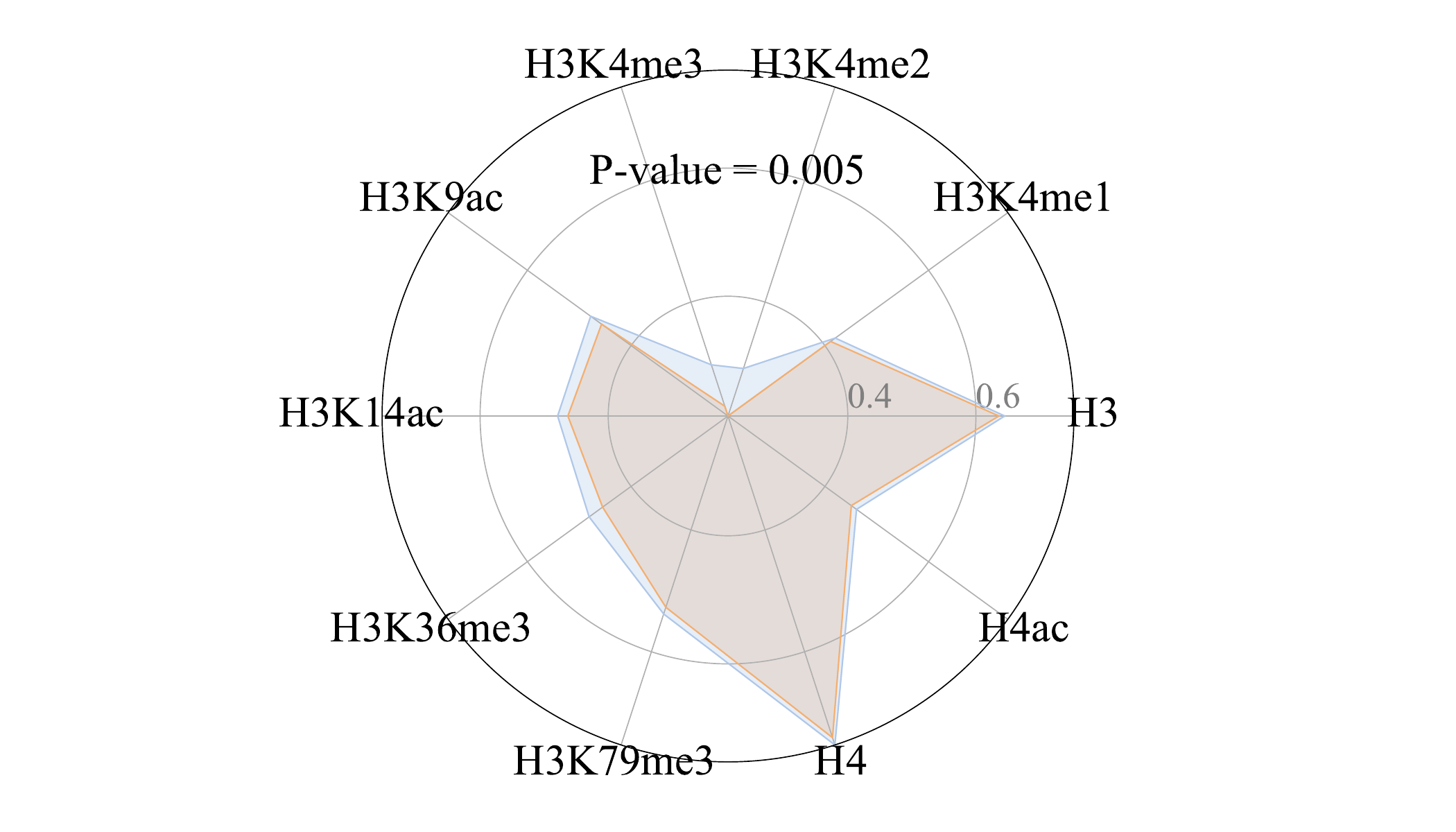}
    \end{subfigure}
                \begin{subfigure}[b]{0.3\linewidth}
                    \captionsetup{justification=raggedright,singlelinecheck=false}
                            \caption{}\label{fig:avg-auc}
        \includegraphics[height=3cm,width=4.5cm]{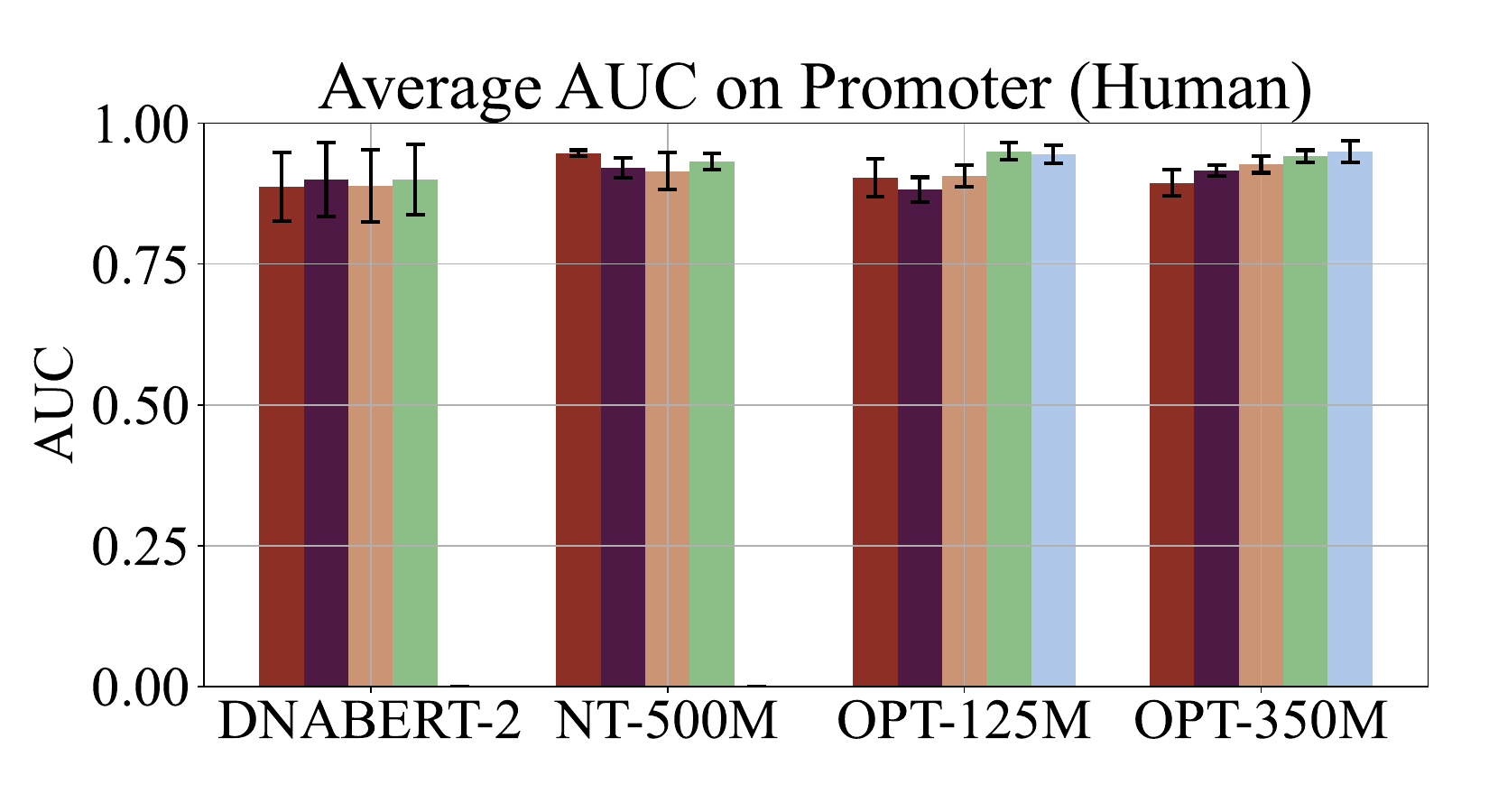}
    \end{subfigure}
    \\
     \begin{subfigure}[b]{0.34\linewidth}
         \captionsetup{justification=raggedright,singlelinecheck=false}
                 \caption{}\label{fig:avg-mcc}
        \includegraphics[height=3cm,width=4.5cm]{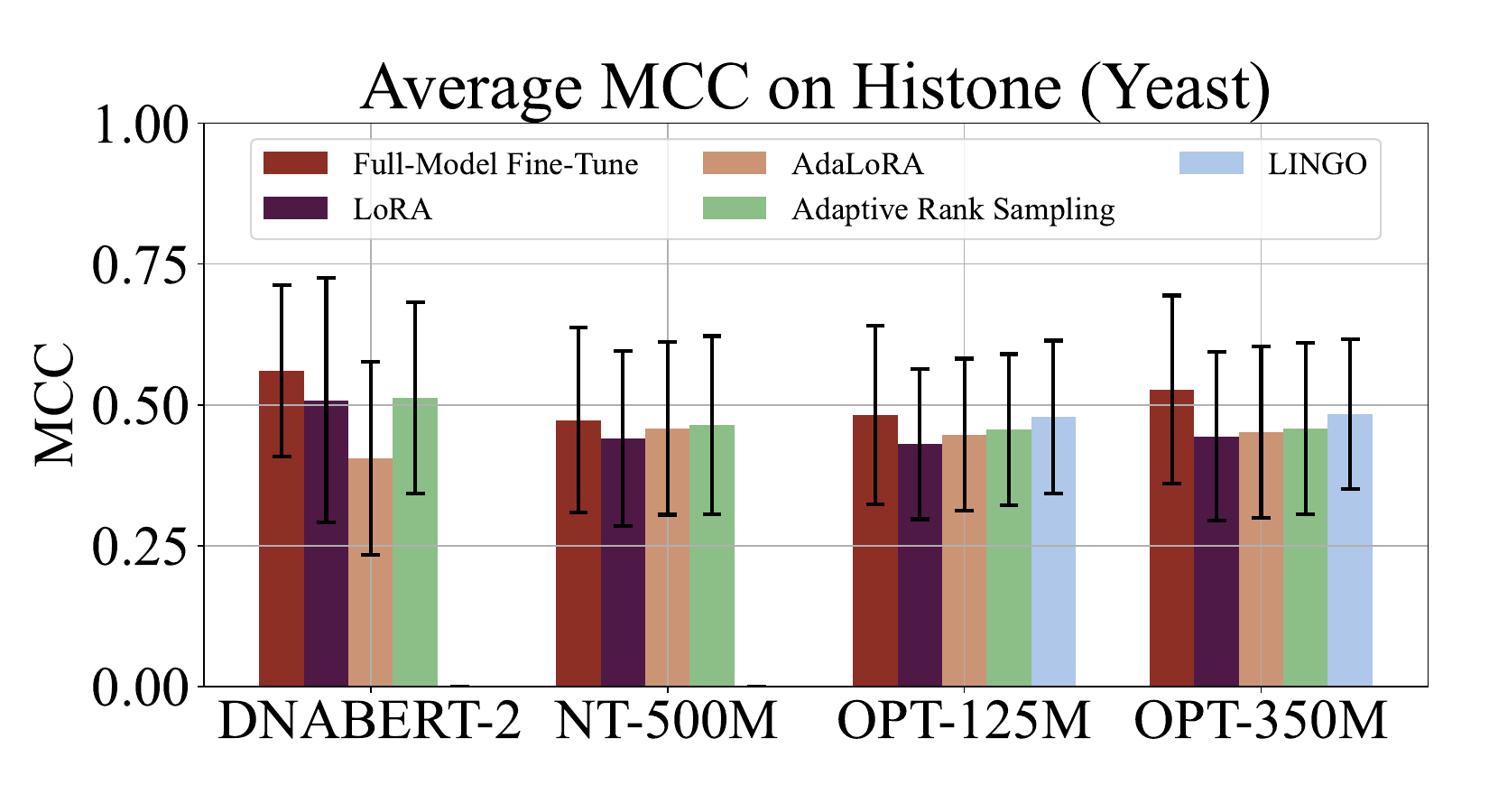}
    \end{subfigure}
                    \begin{subfigure}[b]{0.34\linewidth}
                \captionsetup{justification=raggedright,singlelinecheck=false}
                \caption{}\label{fig:top-2-125m}
        \includegraphics[height=3cm,width=4.4cm]{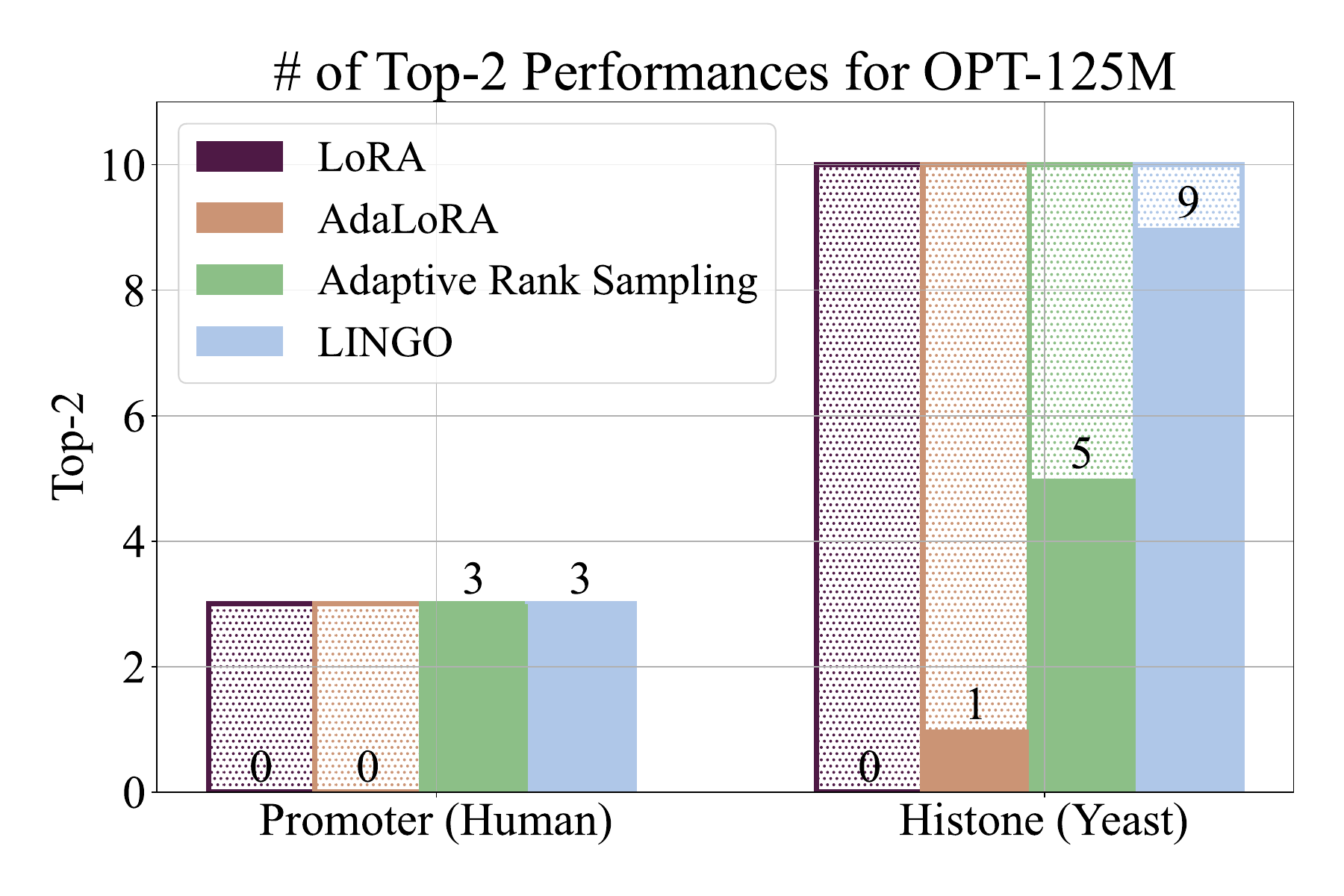}
    \end{subfigure}
     \begin{subfigure}[b]{0.3\linewidth}
     \captionsetup{justification=raggedright,singlelinecheck=false}
         \caption{}\label{fig:top-2-350m}
        \includegraphics[height=3cm,width=4.4cm]{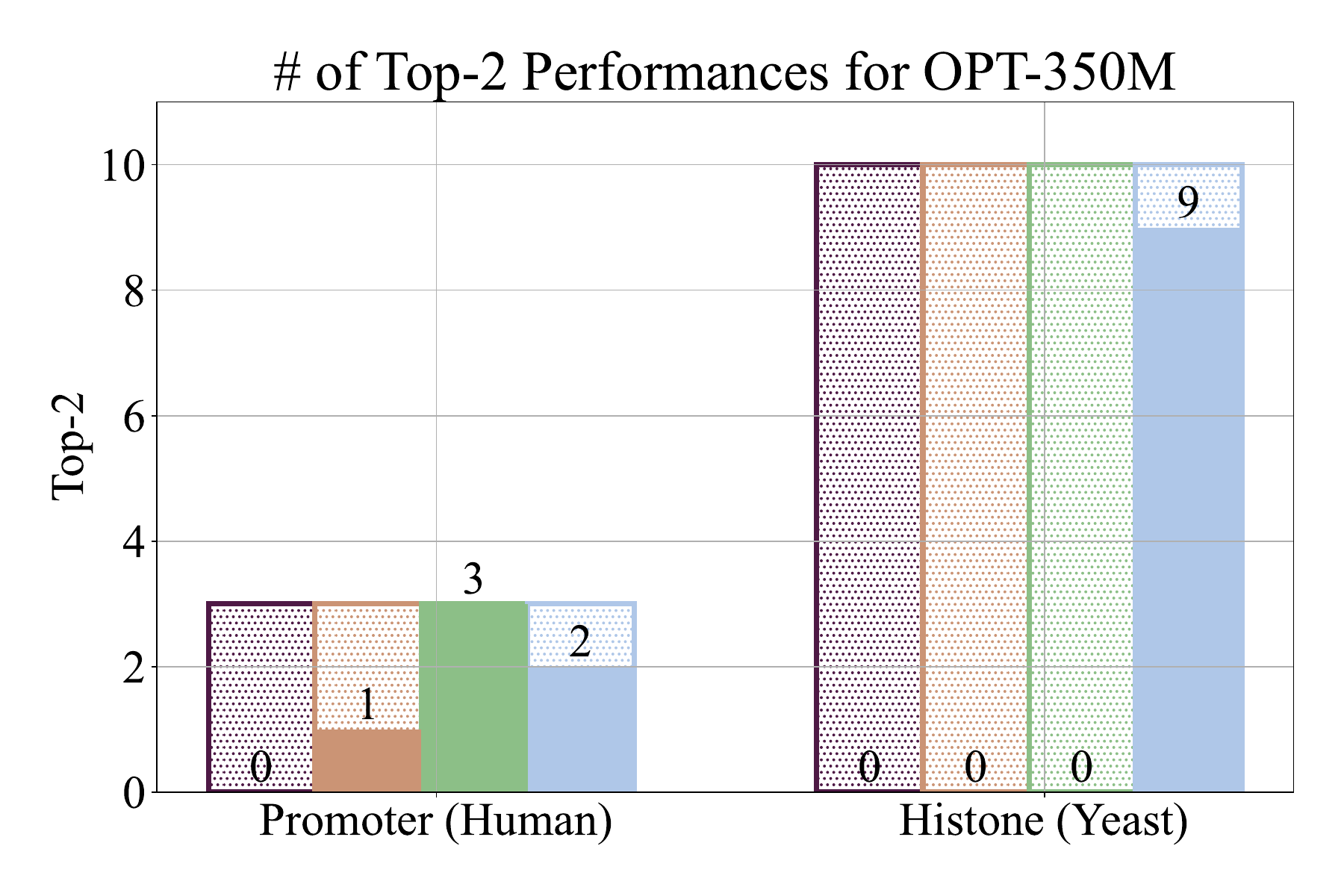}
          \end{subfigure}
    
    \caption{\textbf{\large{a}} MCC comparisons between adaptive rank sampling and \textsc{Lingo} across ten datasets on Histone (Yeast) for the OPT-125M model. A one-sided paired t-test was performed between \textsc{Lingo} and adaptive rank sampling. \textbf{\large{b}} MCC between adaptive rank sampling and \textsc{Lingo} is showed for the OPT-350M model on Histone (Yeast). A one-sided paired t-test was performed. \textbf{\large{c}} Average AUC on the Promoter (Human) task for five different methods – full-model fine-tune, LoRA, AdaLoRA, adaptive rank sampling, and \textsc{Lingo} – applied to a variety of models including DNABERT-2, NT-500M, OPT-125M, and OPT-350M. \textbf{\large{d}} Comparison of the average MCC on Histone (Yeast) for the same five methods across various models.  \textbf{\large{e}} Frequency of OPT-125M model registers Top-2 performance across four PEFT methods: LoRA, AdaLoRA, adaptive rank sampling, and \textsc{Lingo}. The solid bars in the graph represent the count of datasets where a method achieved top-2 performance, while the hatched bars indicate the total number of datasets considered. \textbf{\large{f}} Performance of the OPT-350M model with the same four PEFT methods. }\label{fig:avg-auc-mcc}
\end{figure}    

\subsection{Leveraging prompts for in-context learning in pre-trained natural language models}

To further leverage the in-context learning ability and address the NLP pretraining of OPTs, we applied and evaluated \textsc{Lingo} on OPTs. \textsc{Lingo} adeptly utilizes contextual signals inherent in PLMs, effectively reorienting their capabilities from standard natural language processing tasks to the intricate and specialized field of genomic sequence analysis. This redirection is achieved by fine-tuning PLMs to recognize and interpret the unique patterns and structures present in genomic data, thereby extending their applicative scope beyond conventional linguistic tasks to the nuanced domain of genomics.
We conducted further tests to determine whether \textsc{Lingo} demonstrates superior performance over DNA foundation models and other PEFT methods, including adaptive rank sampling alone. 
Notably, we define \textsc{Lingo} as applying adaptive rank sampling with language prefix exclusively on natural language PLMs, while adaptive rank sampling alone can be applied to both PLMs and DNA foundation models.

On both OPT-125M and OPT-350M, \textsc{Lingo} consistently outperforms adaptive rank sampling alone. In Figures~\ref{fig:peft-9} and~\ref{fig:peft-10}, both radar figures present the MCC across ten datasets for the Histone (Yeast) task. The performances of adaptive rank sampling alone and \textsc{Lingo} are represented by orange and blue areas, respectively. A one-sided paired t-test was performed between these two methods, revealing that \textsc{Lingo} significantly outperforms adaptive rank sampling, with P-values of 0.013 and 0.005 for both tests, respectively. Additionally, these tests were extended to three datasets in Promoter (Human), with corresponding results presented in Supplementary Figures~\ref{fig:peft-more}ba - Supplementary Figures~\ref{fig:peft-more}bb.
Details of these metrics can be found in Supplementary Table~\ref{tab:MCC_EMP} and Supplementary Table~\ref{tab:PD_AUC}. 
For example, in Supplementary Table~\ref{tab:MCC_EMP}, \textsc{Lingo}-trained OPT-125M  achieves an MCC of 0.735 for the H4 dataset. In comparison, AdaLoRA and LoRA attain AUCs of 0.734 and 0.725, respectively, on the Histone (Yeast) task. In Supplementary Table~\ref{tab:PD_AUC}, \textsc{Lingo}-trained OPT-125M  achieves an AUC of 0.98 for the prom\_notata dataset. In comparison, FMFT, AdaLoRA, and LoRA attain AUCs of 0.947, 0.931, and 0.907, respectively, on the Promoter (Human) task.

\textsc{Lingo} also outperforms other PEFT methods in terms of average performances for both tasks. Specifically, Figure~\ref{fig:avg-auc} shows that in the OPT-350M model, \textsc{Lingo} surpasses adaptive rank sampling by 0.9\%. Moreover, as illustrated in Figure~\ref{fig:avg-mcc}, \textsc{Lingo} demonstrates superior performance over adaptive rank sampling in both OPT-125M and OPT-350M models, with improvements of 2.2\% and 2.6\%, respectively.

We highlight the \textsc{Lingo} performances of OPTs in comparison to DNA foundation models (Figure~\ref{fig:top-2-125m} and Figure~\ref{fig:top-2-350m}). 
Compared to the two DNA foundation models, i.e. DNABERT-2 and NT-500M, \textsc{Lingo}-trained OPTs are consistently among the Top-2 for Promoter (Human) and Histone (Yeast) tasks.
This is significant, because PEFT is computationally efficient than training DNA foundation models from scratch, while a small-parameter genomic adapter on OPTs consistently outperforms at least one of such DNA foundation models. For the three datasets in Promoter (Human) task, \textsc{Lingo}-trained OPT-125M  is among the Top-2 performed models in 3/3 tasks in Figure~\ref{fig:top-2-125m}. Similarly, in Figure~\ref{fig:top-2-350m}, for the ten datasets in Histone (Yeast) task, \textsc{Lingo}-trained OPT-350M is among the Top-2 performed models in 9/10 tasks.

\subsection{One-hot encoding addresses the challenge of semantic disambiguation}

We sought to investigate the effect of semantic ambiguity in our framework, and whether a DNA-specific tokenization method could improve our framework's performance via mitigating semantic ambiguity.
So far for the previous PLMs, BBPE tokenization is utilized for encoding genomic sequences. Following the aggregation of the most frequent pairs, a notable observation is the potential overlap of token identifiers between genomic sub-sequences and conventional English lexicon. However, it is important to acknowledge that the semantic nature of these genomic sub-sequences, compared to their linguistic lexicon, is fundamentally different. Thus, we hypothesized that the coexistence of multiple semantic interpretations within the same token set could adversely impact downstream task efficacy. To further investigate this hypothesis, we apply an additional experimental approach that utilizes simple one-hot encoding for genomic sequence representations. 

Our experiments demonstrate that semantic disambiguation does not decrease \textsc{Lingo}'s performance; in contrast, one-hot encoding achieved subpar performance compared to BBPE, likely due to its lack of ability to capture contextual information. In Table~\ref{fig:one-hot}, we show the performances for \textsc{Lingo} for BBPE tokenization, i.e., \textsc{Lingo}+BBPE, and one-hot encoding,i.e. \textsc{Lingo}+One-hot, respectively, for Promoter (Human) task. For example, in the prom\_notata dataset, the OPT-125M model with \textsc{Lingo}+BBPE attains an AUC of 0.98. In comparison, the same model utilizing \textsc{Lingo}+One-hot achieves a commendable AUC of 0.971. The findings indicate that while one-hot encoding does not surpass BBPE tokenization in AUC, it nonetheless demonstrates sufficient efficacy. This can be attributed to its proficiency in mitigating the ambiguities arising from semantic multiplicity in PLMs. Therefore, our study concludes that despite the relative simplicity of one-hot encoding, which may not comprehensively capture contextual nuances and dependencies inherent in genomic sequences, it effectively addresses the challenge of semantic disambiguation, thereby necessitating future investigations of effective tokenization methods in our framework.

\begin{table}[!htbp]
       \caption{\small AUC comparisons between two tokenization methods used with \textsc{Lingo} on the OPT-125M model: one-hot tokenization (\textsc{Lingo} + One-hot) and BBPE tokenization (\textsc{Lingo} + BBPE). }
    \label{fig:one-hot}
    \centering
    \begin{tabular}{>{\centering\arraybackslash}p{1.7cm} >{\centering\arraybackslash}p{2.5cm}|  >{\centering\arraybackslash}p{1.7cm} >{\centering\arraybackslash}p{1.7cm} >{\centering\arraybackslash}p{1.7cm}}
    \hline
      \multirow{1}{*}{\textbf{Model}}   & \multirow{1}{*}{\textbf{Method}} & \multirow{1}{*}{\textbf{Prom\_all}} & \multirow{1}{*}{\textbf{Prom\_notata}} & \multirow{1}{*}{\textbf{Prom\_tata}}\\
      \hline
    \multirow{2}{*}{OPT-$125$M}  
         & \textsc{Lingo} + BBPE & $0.954$ & $0.980$ & $0.895$ \\
          & \textsc{Lingo} + One-hot & $0.921$ & $0.971$ & $0.850$ \\
          \hline
    \multirow{2}{*}{OPT-$350$M}  
         & \textsc{Lingo} + BBPE & $0.957$ & $0.983$ & $0.890$ \\
         & \textsc{Lingo} + One-hot & $0.947$ & $0.960$ & $0.807$ \\
         \hline
    \end{tabular}
\end{table}

\subsection{Efficient and accurate genome-scale prediction by language prefix fine-tuning }

\begin{figure}[!htbp]
    \centering
        \begin{subfigure}[b]{0.5\textwidth}
        \centering\captionsetup{justification=raggedright,singlelinecheck=false}
         \caption{}
        \label{fig:complexity-hmp}
        \includegraphics[height=3.5cm]{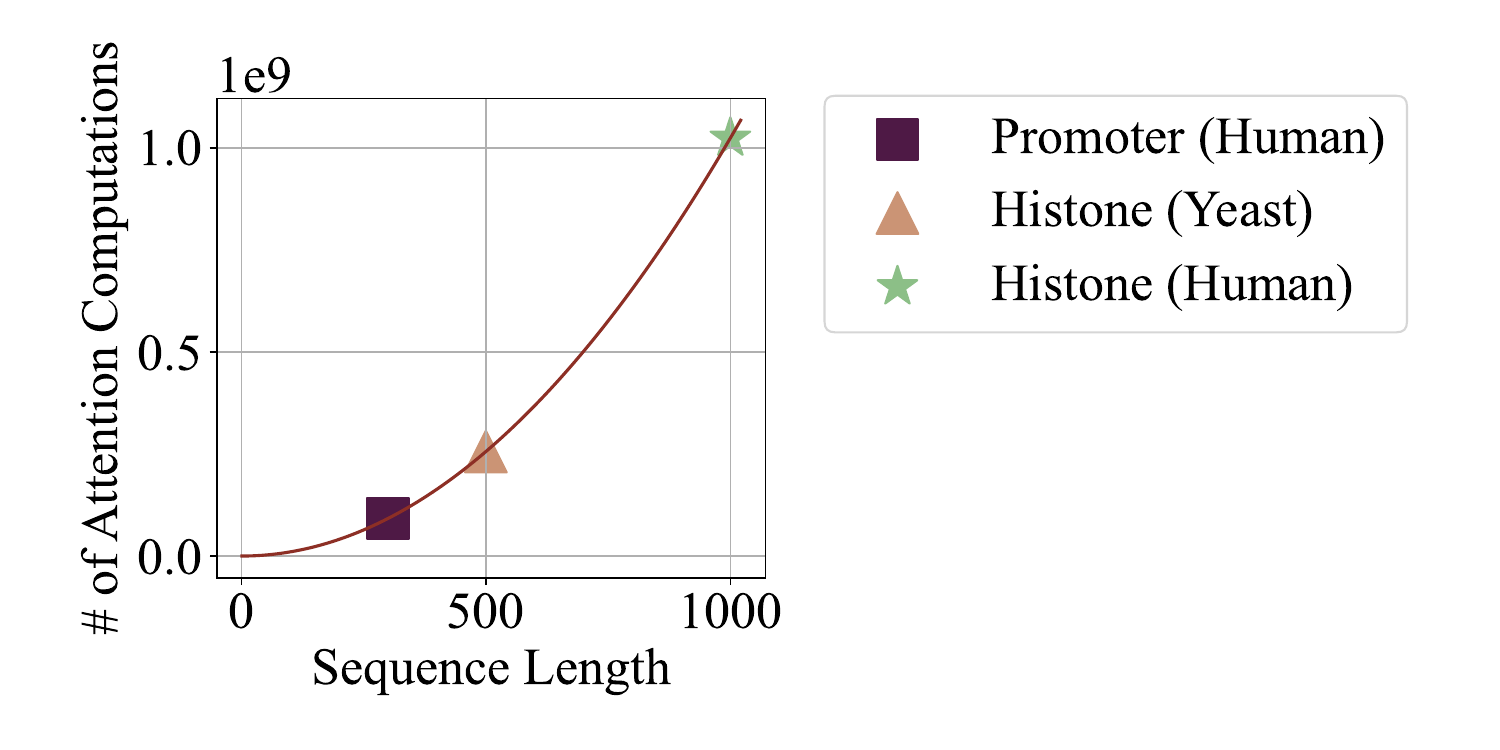}
    \end{subfigure}
             \begin{subfigure}[b]{0.45\linewidth}\captionsetup{justification=raggedright,singlelinecheck=false}
        \caption{}\label{fig:flops}\raisebox{0.0mm}
        {\includegraphics[height=3.8cm,width=5cm]{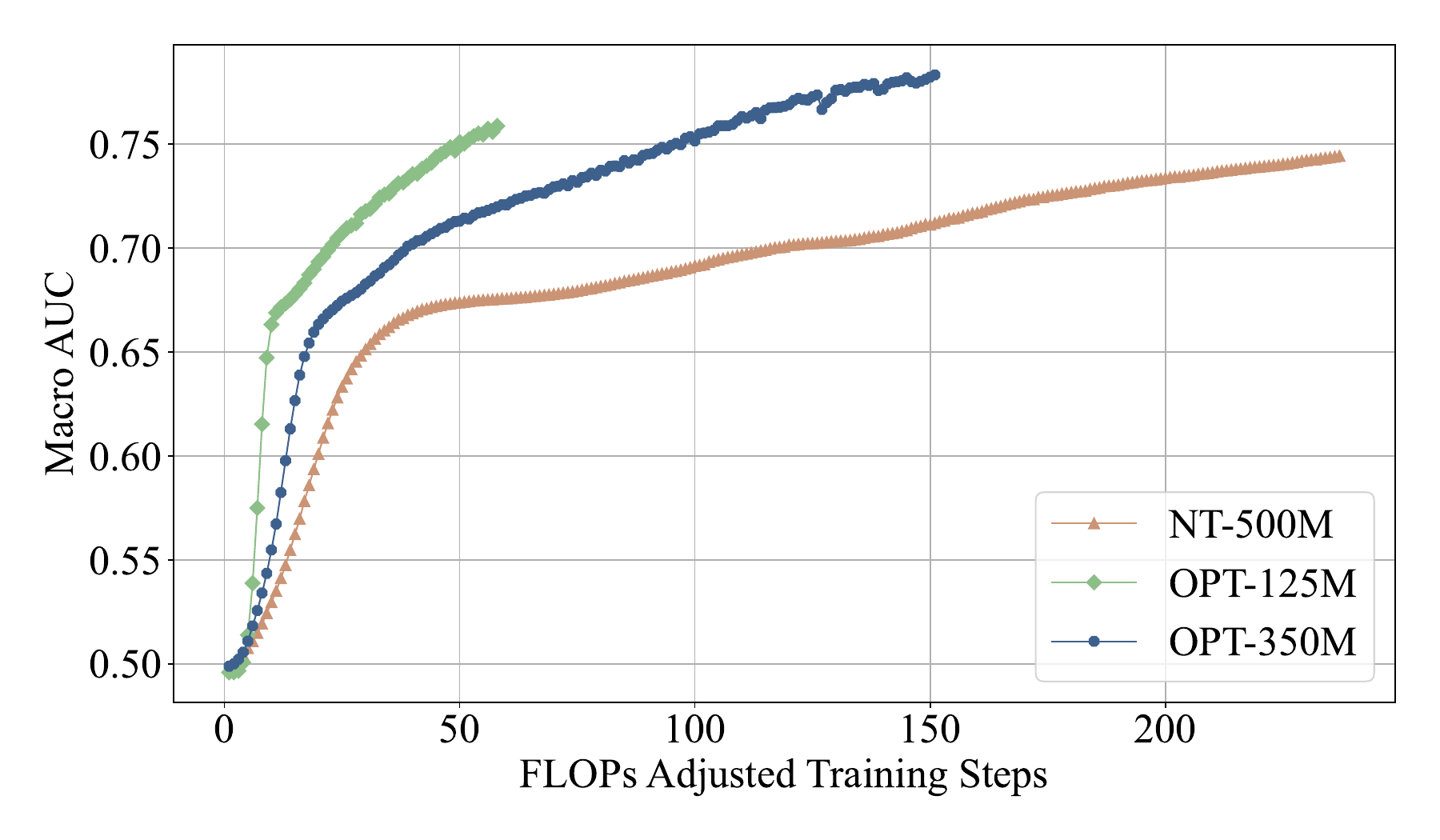}}
    \end{subfigure}

    \begin{subfigure}[b]{0.42\textwidth}
        \centering    \captionsetup{justification=raggedright,singlelinecheck=false}
                \caption{}
        \label{fig:opt-500m}
        \vspace{-3.2mm}
\includegraphics[width=\textwidth]{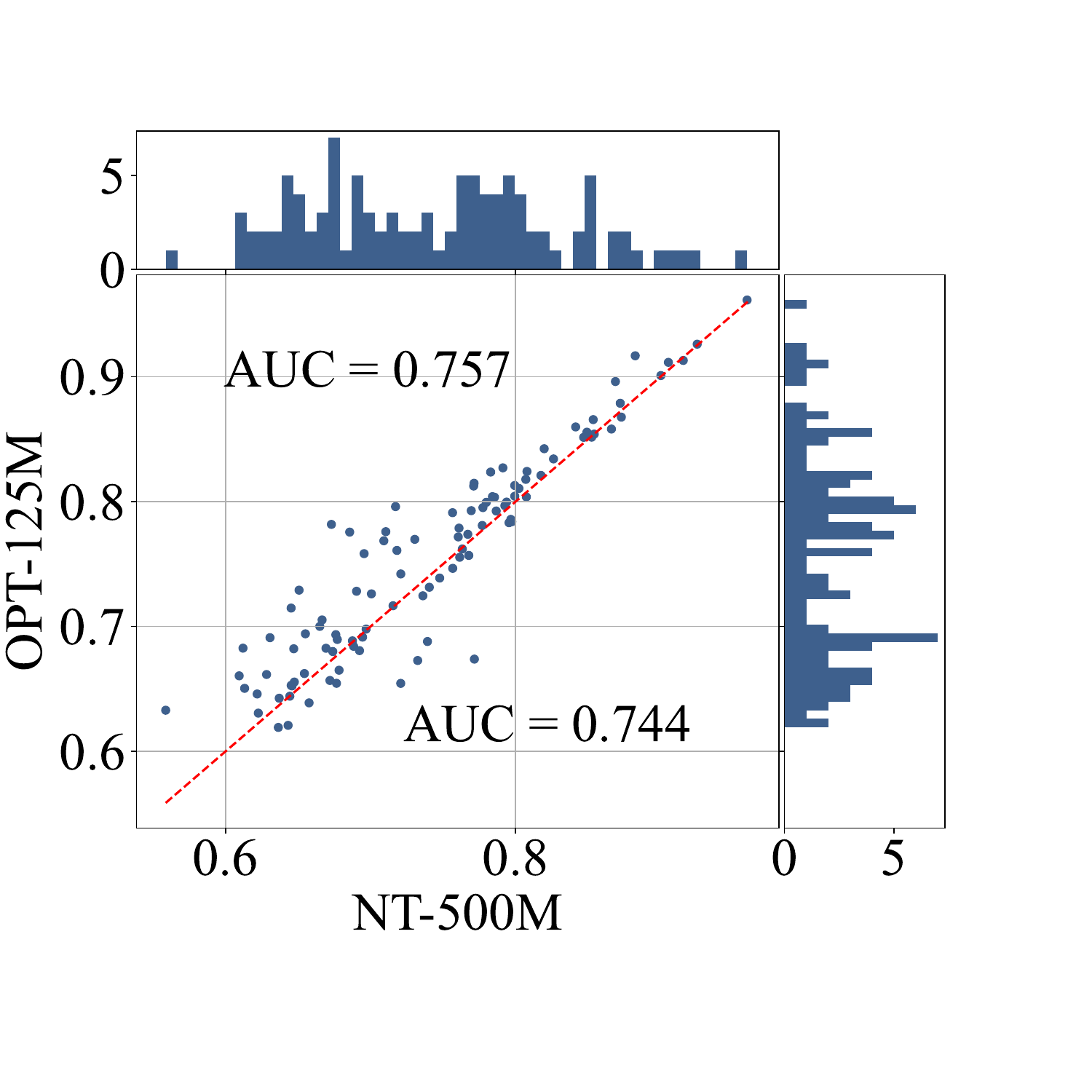}
    \end{subfigure}
    \hspace{1em}
    \begin{subfigure}[b]{0.41\textwidth}
        \centering    \captionsetup{justification=raggedright,singlelinecheck=false}
                \caption{}
        \label{fig:125m-500m}
             \vspace{-3.1mm}\includegraphics[width=\textwidth]{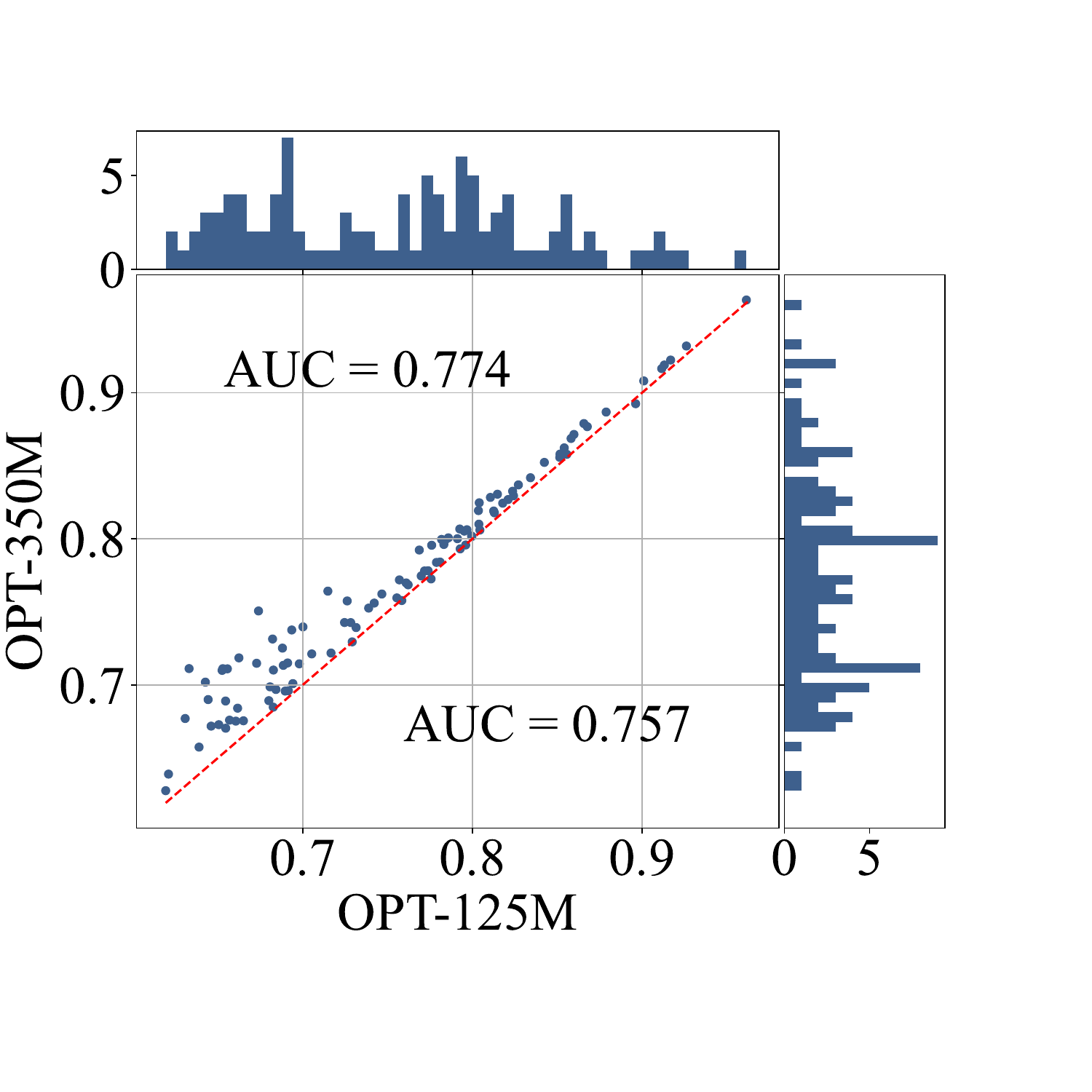}
    \end{subfigure}

        \begin{subfigure}[b]{0.8\textwidth}
        \centering     \captionsetup{justification=raggedright,singlelinecheck=false}
                \caption{}
        \label{fig:pearson}
             \vspace{-3.25mm}\includegraphics[width=6cm]{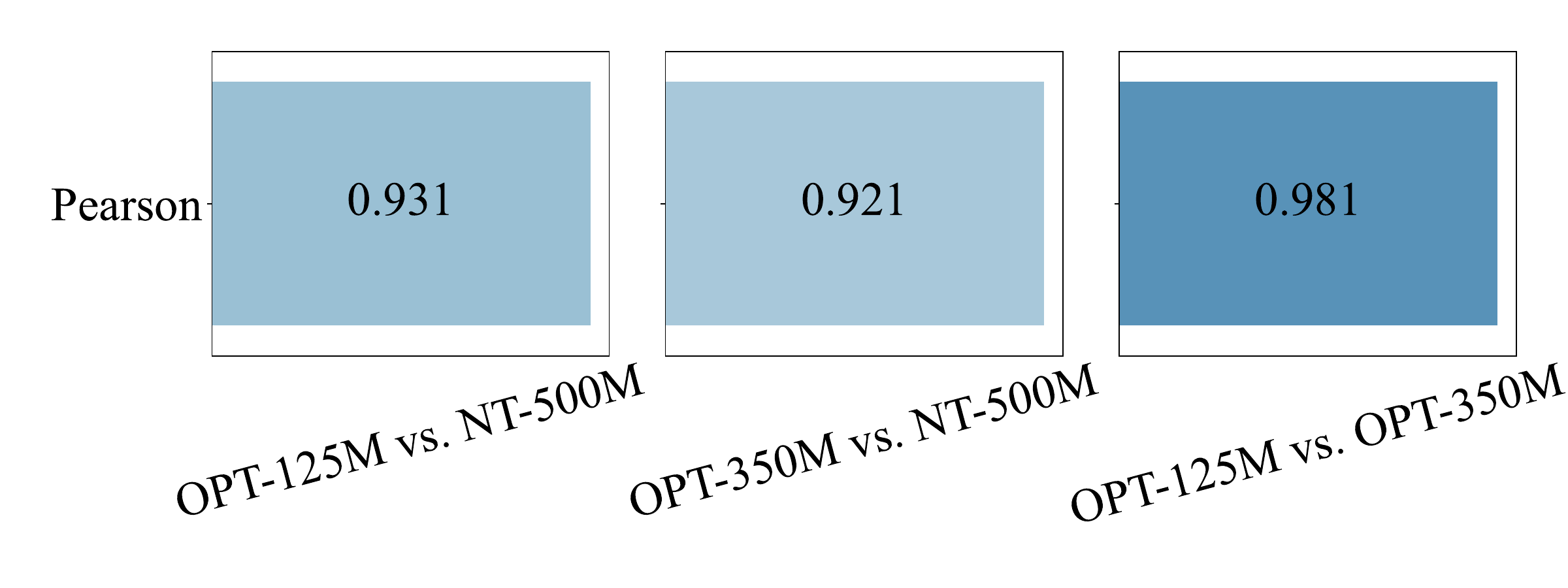}
    \end{subfigure}
    \caption[]{ \textbf{\large{a}} Scaling law of the number of attention computations with increasing sequence length across three tasks: Promoter (Human), Histone (Yeast), and Histone (Human). Notably, the increase in attention computations is quadratic in relation to the sequence length. \textbf{\large{b}} Macro AUC comparison among three models – NT-500M, OPT-125M, and OPT-350M – under FLOPs-adjusted training steps, demonstrating that OPT-350M outperforms OPT-125M, which in turn surpasses NT-500M, within the same training steps. \textbf{\large{c}} Comparison of AUC between OPT-125M and NT-500M for each label in the Histone (Human) task. \textbf{\large{d}} Comparison of AUC between OPT-125M and OPT-350M for each label on Histone (Human). \textbf{\large{e}} Pearson correlation of multitask learning performance for three models: OPT-125M, OPT-350M, and NT-500M. }
\end{figure}

Lastly, we demonstrate \textsc{Lingo} can efficiently scale up to whole-genome scale multitask learning, achieving superior prediction performance within a fixed computational budget compared to Nucleotide Transformer. 
For each 1000-bp DNA sequence in this task, over 100 cell-line specific histone modification markers, annotated based on the ENCODE datasets~\citep{encode2020expanded}, were previously compiled by DeepSEA~\citep{zhou2015predicting}.
As genome-scale high-throughput sequencing assays become a routine to probe biological systems~\citep{przybyla2022new}, this represents a realistic scenario towards deploying DNA foundation models in academic labs, where specific biological questions are of interest, yet huge computational resources to fine-tune DNA foundation models may be restricted.
In Figure~\ref{fig:complexity-hmp}, we initially illustrate that the Histone (Human) task presents significantly greater complexity compared to both the Histone (Yeast) and Promoter (Human) tasks. This figure delves into the computational complexity inherent in the self-attention mechanism of transformer models, a topic frequently noted for its quadratic nature. The complexity is a direct consequence in which self-attention processes the input sequence. As a notable observation from this figure, even a modest increase in the input sequence length from 500 to 1000 leads to a substantial escalation in the number of attention computations, rising to as much as 1e9.

Compared to the NT-500M model, both \textsc{Lingo}-trained OPT-125M and OPT-350M demonstrate superior performance and greater computational efficiency. 
We applied \textsc{Lingo} to OPTs and adaptive rank sampling (as language prefix is not applicable to DNA foundation models) to the NT-500M on the Histone (Human) task, while all models were trained under a fixed budget of 6 ranks per layer on average.
To ensure a fair comparison, we adjusted the computational costs of all three foundation models based on their training FLOPs. This approach allowed for a direct comparison of the models' performance efficacy under a uniform computational constraint, providing an equitable assessment of each model's capabilities when adjusted for FLOPs efficiency. 
The FLOPs efficiency of OPT-125M, denoted as FLOPs$_{\text{OPT}-125\text{M}}$, was designated as the foundational benchmark for computational efficiency. 
Figure~\ref{fig:flops} reveals that although the OPT-125M model's final performance is lower than that of the 350M model, it demonstrates greater computational efficiency. This is evidenced by its higher Macro AUC at FLOPs-adjusted training steps. Similarly, when comparing with the NT-500M model, both OPT-125M and OPT-350M exhibit superior performance, indicating heightened computational efficiency, e.g., OPT-125M achieves $4.443 \times$ efficiency (via $\frac{\text{FLOPs}_{\text{OPT-125M}}}{\text{FLOPs}_{500\text{M}-1000\text{G}}}$) compared to NT-500M. Consequently, our analysis leads to the conclusion that PLMs trained by \textsc{Lingo}, in comparison to the NT-500M model, are more computationally efficient.

Using our \textsc{Lingo} framework, OPT-350M achieved the highest accuracy on this genome-scale understanding task. 
As shown in Figure~\ref{fig:opt-500m}, the \textsc{Lingo}-trained OPT-125M model achieves an AUC of 0.757, already surpassing the 0.744 AUC achieved by the NT-500M model. Furthermore, Figure~\ref{fig:125m-500m} demonstrates that \textsc{Lingo}-trained OPT-350M, with an AUC of 0.774, outperforms OPT-125M. 
These results indicate that \textsc{Lingo}-trained PLMs exhibit more accurate performance over NT-500M model in multi-label classification tasks within the Histone (Human), with the improvement of OPT-350M concentrated in the poorly-performed tasks in OPT-125M (bottomleft corner, Figure~\ref{fig:125m-500m}). 
Finally, we analyzed the Pearson correlation coefficients~\cite{cohen2009pearson} of the AUC across 104 labels in the Histone (Human) for the three different models (Figure~\ref{fig:pearson}). This figure illustrates that the Pearson correlation coefficients for the OPT models (OPT-125M and OPT-350M) are more closely aligned with each other, indicating a higher degree of similarity in their performance patterns. In contrast, the correlation between OPT models and NT-500M is less pronounced. 
This disparity suggests that the \textsc{Lingo}-trained OPT models, despite variations in their sizes, tend to follow a more consistent performance trend compared to the divergence observed when compared with the NT-500M model.

\section{Discussion}
 Recently, there have been significant advancements in the field of genomic domain, attributed to the development of DNA foundation models such as DNABERT-2~\citep{zhou2023dnabert} and Nucleotide Transformer (NT)~\citep{dalla2023nucleotide}. However, the development of these models in the domain of genomics has been hindered by challenges in data scalability. This contrasts with the significant advancements made by pre-trained foundation models in fields such as NLP and CV, where they have demonstrated remarkable progress. To effectively adapt PLMs for applications in the genome domain, we introduce a novel \textsc{Lingo} method designed to prime PLMs, specifically OPT models, for genome understanding tasks. This approach diverges from the straightforward input of DNA sequences. Instead, \textsc{Lingo} harnesses the inherent contextual learning capabilities of PLMs. It guides their transition from processing natural language to interpreting genomic sequences. By inserting text prefix and suffix, this strategy enables PLMs to recalibrate their extensive linguistic knowledge to the unique complexities of genomic sequences, leveraging their existing expertise in a novel context.
 
 In the realm of PEFT, additive fine-tuning, such as prefix tuning~\citep{li2021prefix,liu2021p} and prompt tuning methods~\cite{lester2021power}, in PLMs involves adding new parameters to the existing model, enabling it to learn additional information while retaining its original knowledge. Partial fine-tuning, such as SAM~\citep{fu2023effectiveness}, in contrast, adjusts only a subset of the model's existing parameters, aiming for quicker adaptation to specific tasks with less computational cost.
 
 Reparameterization-based fine-tuning offers a distinct advantage over these methods. By reparameterizing the model's weights, it allows for a more flexible and efficient adaptation to new tasks without the need for extensive additional parameters or the risk of overwriting valuable pre-learned representations. Within this reparameterization scheme, low-rank adaptation methods, have garnered considerable attention. While each approach presents unique advantages, a comparative analysis reveals certain limitations inherent to prefix and prompt tuning methods when contrasted with low-rank adaptation strategies. Predominantly, prefix and prompt tuning methods exhibit constraints in their capacity for extensive model reconfiguration. This limitation stems from their focus on input-level modifications, which may not suffice for tasks necessitating deeper model transformations. Additionally, these methods display a pronounced dependency on the nature of pre-training tasks, potentially limiting their effectiveness in scenarios vastly divergent from initial training contexts. While low-rank adaptation is a well-established method in PLMs, there is a notable absence of mature applications of PLMs in DNA foundation models. DNABERT-2 employs low-rank adapters~\citep{hu2021lora} with a fixed number of ranks to curtail the quantity of trainable parameters. However, deterministic PEFTs can yield sub-optimal performance when the initial states are less than ideal. To solve this issue, we propose an adaptive rank sampling to prune and stochastically reintroduce pruned singular vectors. \textsc{Lingo} demonstrates superior potential. Our empirical observations on three genome understanding tasks demonstrate that OPT-350M, when combined with adaptive rank sampling, positions itself on the Pareto front in comparison to its full-model fine-tuning baseline, utilizing merely $0.94\%$ of the trainable parameters. Interestingly, we also observe that PLMs exhibit superior performance and are more computationally efficient in multi-label classification on Histone (Human). For instance, OPT-125M achieves an efficiency that is $4.443 \times$ greater than that of NT-500M.

\section{Methods}\label{sec:methods}

\subsection{Datasets}
 The sequences for Histone (Yeast) and Promoter (Human) are obtained from the GUE framework~\cite{zhou2023dnabert} and the sequences for  Histone (Human) are extracted from the DeepSEA framework~\citep{zhou2015predicting}.  The detailed statistics for the three tasks are shown in Supplementary Table~\ref{tab:dataset}.

\subsection{BBPE tokenization for DNA sequences}

In Figure~\ref{fig:frame-5}, we present three popular tokenizers for DNA sequences, labeled as ($1$), ($2$), and ($3$). The ``words'' tokenizer employs a dictionary derived from the four nucleotides. The tokenized sequence length of an input DNA equates to the number of nucleotides. However, this method lacks contextual information. In contrast, the $6$-mer tokenizer is gaining popularity in DNA sequencing~\citep{dotan2023effect}. The concept of $k$-mer revolves around extracting continuous subsequences of $k$ nucleotides from a DNA sequence. However, one drawback of $k$-mer tokenization is the increased computational complexity, especially with larger $k$ values. To address this, the BBPE tokenizer initializes a dictionary consisting of all individual bytes in UTF-8 encoding. It progressively selects the most frequent pairs of tokens to merge. Each combined pair is then added to the dictionary as a new token (shown in $\textcircled{1}-\textcircled{5}$). OPTs are tailored with GPT-2's BBPE tokenizer~\citep{radford2019language}. This tokenizer is well-suited for DNA sequences because it efficiently captures the recurring patterns of nucleotides. By focusing on the frequency of specific sequences, it offers a nuanced encoding that can illuminate biological motifs. This dictionary only consists of three tokens: ``AAC'', ``TC'', and ``GA''. In our experiments, we conducted \textsc{Lingo} combined with one-hot/BBPE tokenization. This was done to investigate whether one-hot encoding effectively resolves the challenge of semantic disambiguation, particularly concerning potential overlaps between genomic sub-sequence token identifiers and those found in the conventional English lexicon in BBPE.

\subsection{Domain-shift genome understanding}
In this subsection, we introduce \textsc{Lingo}, importance scores computation, and the cubic budget schedule with adaptive rank sampling
\subsubsection{Language prefix fine-tuning}
Next, we introduce a novel \textsc{Lingo} approach. This approach primes PLMs, particularly OPTs, to tackle genome understanding tasks. Through \textsc{Lingo}, the model is adeptly fine-tuned. Central to our methodology is the hypothesis that the integration of domain-specific prompts, such as \texttt{"Domain: DNA Promoter"}, markedly enhances the model's ability to discern the unique rules and patterns inherent in genomic sequences, distinct from those in natural language. To implement this, we append a prefix \texttt{"\textcolor{pink}{Domain:} \textcolor{teal}{DNA Promoter}\textcolor{pink}{\textbackslash{}nSequence:} "} and a suffix \texttt{"\textcolor{pink}{\textbackslash{}nAnnotation:}"} to the DNA sequence inputs, formatted in Markdown to ensure clarity and structure. 

In this example, text highlighted in pink remains constant, while the text in teal, sequences, and labels are variables, tailored to specific genomic instances. This representation not only facilitates a clear demarcation of the input components but also reflects the flexible nature of our \textsc{Lingo} in accommodating diverse genome understanding tasks.

\subsubsection{Importance score}
PLMs contain many weight matrices to perform matrix multiplication. These weight matrices typically have full-rank. However, performing FMFT is not efficient. Thus, our goal is to reduce the number of ranks to project the high dimensional weights matrices to smaller subspaces. Mathematically, for a pre-trained weight matrix $W_0 \in \mathbb{R}^{d_p \times d_q}$, we approximate the gradient updates using SVD for a low-rank representation, i.e., $W_0 + \Delta W = W_0 + P \Lambda Q$, where $P \in \mathbb{R}^{d_p \times r}$, $Q \in \mathbb{R}^{r \times d_q}$, and the rank $r \ll \min(d_p,d_q)$. Thus, for $n$ matrices in the PLM, we need to perform SVD: $\Delta W_k =P_k\Lambda_kQ_k$ for $k=1, ..., n$. Our importance score computation considers the importance from both singular values, which capture the magnitude of changes and indicate dominant variation directions, and singular vectors, which denote their orientations. Thus, each triplet $\Sigma_i =\{\lambda_{k,i}, P_{k,*i}, Q_{k,i*}\}$ is constructed with the $i$-th singular value $\lambda_{k,i}$ and the corresponding singular vectors $P_{k,*i}, Q_{k,i*}$. The importance score for each singular value is then computed as~\citep{zhang2022platon}:
\begin{equation}\label{eq:S}
            S_{k_i} = s(\lambda_{k,i})+\frac{1}{d_p}\sum_{j=1}^{d_p}s(P_{k,ji}) +\frac{1}{d_q}\sum_{j=1}^{d_q}s(Q_{k,ij}).
\end{equation}

At each time step $t$, each entry in the matrix is associated with an importance score, computed as the product of its sensitivity and uncertainty, i.e., $s^t(w_{ij}) = \bar{I}^t(w_{ij})\bar{U}^t(w_{ij}),$ where $\bar{I}^t(w_{ij})$ denotes the stabilized sensitivity and $\bar{U}^t(w_{ij})$ represents the stabilized uncertainty. These stabilized scores refine the original scores through a weighted adjustment. The updating rules for $\bar{I}^t(w_{ij})$ and $\bar{U}^t(w_{ij})$ are: 
\begin{equation}\label{eq:I}
\bar{I}^t(w_{ij}) = \beta_1 \bar{I}^{t-1}(w_{ij}) + (1-\beta_1) I^t(w_{ij}),
\end{equation}
and 
\begin{equation}\label{eq:U}
\bar{U}^t(w_{ij}) = \beta_2 \bar{U}^{t-1}(w_{ij}) + (1-\beta_2) |I^t(w_{ij}) - \bar{I}^{t-1}(w_{ij})|,
\end{equation} where $I^t(w_{ij}) = |w_{ij} \bigtriangledown_{w_{ij}}\mathcal{L}^t|$ and $\mathcal{L}^t$ denotes the binary cross-entropy for a batch of data $\mathcal{D}$. Therefore, the sensitivity captures how much the loss responds to changes in a specific weight within a training batch. In contrast, uncertainty quantifies the fluctuations in the loss, given by $U^t(w_{ij}) = |I^t(w_{ij}) - \bar{I}^{t-1}(w_{ij})|.$ The importance score is computed by balancing the two factors via $\bar{I}^t(w_{ij})\bar{U}^t(w_{ij})$.

\subsubsection{Cubic budget schedule with adaptive rank sampling}

We introduce a global budget, $b_t$, which diminishes following a cubic budget schedule defined as $b^t = b^T +(b^0-b^T)(1-\frac{t}{T})^3$. For the adaptive rank sampling process, we define masks $R_{k,ii}^t$ for pruning $\lambda_k^t$ and re-introducing the pruned $\lambda_k^t$. These masks are random variables derived from a Bernoulli distribution with parameter $p$, $R_{k,ii}^t \sim \text{Bernoulli}(p)$. 
The singular values to be retained are updated based on the following updating rule:
\begin{equation}\label{eq:rs}
\hat{\Lambda}_{k,ii}^{t} = 
\begin{cases}
\Lambda_{k,ii}^t \cdot (1-R_{k,ii}^t) & \text{if } S_{k,i}^t \text{ is in the top } b^t \text{ of } S^t, \\
\Lambda_{k,ii}^t \cdot R_{k,ii}^t & \text{otherwise.}
\end{cases}
\end{equation}
Note that we here denote $\lambda_k^t$ as $\Lambda_{k,ii}^t$ to more conveniently assign masks based on their position index $i$. Only those singular values, $\hat{\Lambda}_{k,ii}^{t}$, that meet both the adaptive rank sampling criteria and have their importance score $S_{k_i}^t$ is in the top $b^t$ of all scores $S^t$ are retained. By introducing randomness rather than strictly adhering to a deterministic cutoff, the process becomes more robust against potential suboptimal initial states and inaccuracies in importance scores.

In Algorithm~\ref{alg:rs}, we summarize the steps for the adaptive rank sampling algorithm. For each timestep $t$, we first compute the binary cross-entropy loss $\mathcal{L}^t$ for a batch of data $\mathcal{D}$. Then, every $\Delta T$ timesteps, we compute the importance score for each $k$ and $i$, and update $P_k^{t}$ and $Q_k^{t}$. Finally, we update the singular values $\hat{\Lambda}_{k,ii}^{t}$ based on the cubic budget schedule with adaptive rank sampling.

\begin{algorithm}
\caption{Adaptive rank sampling}\label{alg:rs}
\begin{algorithmic}[1]
\Require A batch of data $\mathcal{D}$; budget $b_t$; hyper-parameters $\eta, \lambda, \beta_1, \beta_2$; final timesteps $T_\text{final}$; timesteps $T$ and $\Delta T$ for low rank approximation
\Ensure $\Delta W_k$ 
\For{$t = 1,...,T_{\text{final}}$}
 \State Compute the binary cross-entropy loss $\mathcal{L}^t$ for a batch of data $\mathcal{D}$
        \If{$t \, \% \, \Delta T = 0$ and $t<T$}
            \State Compute the stabilized sensitivity $\bar{I}^t(w_{ij})$ via Equation~\eqref{eq:I} and uncertainty $\bar{U}^t(w_{ij})$ via Equation~\eqref{eq:U}
            \State Compute $S_{k_i}$ for all $k$ and $i$ via Equation~\eqref{eq:S}
            \State Update $ P_k^{t} = P_k^{t-1} - \eta \nabla_{P_k}\mathcal{L}^t - \lambda P_k^{t-1} $ 
            \State Update $ Q_k^{t} = Q_k^{t-1} - \eta \nabla_{Q_k}\mathcal{L}^t - \lambda Q_k^{t-1} $ 
            \State Update $\Lambda_{k,ii}^t = \Lambda_k^{t-1} - \eta \nabla_{\Lambda_k}\mathcal{L}^t - \lambda \Lambda_k^{t-1}$
            \State Update $\hat{\Lambda}_{k,ii}^{t}$ via Equation~\eqref{eq:rs}
        \EndIf
\EndFor
\end{algorithmic}
\end{algorithm}

\subsection{Analysis setup}
For FMFT, LoRA, AdaLoRA, and adaptive rank sampling, we use a batch size of $8$, while the evaluation batch size is set to $16$. The model is trained over $5$ epochs on all datasets for both tasks. We employ the Adam optimizer~\cite{kingma2014adam}, with a learning rate of $\eta = 3e-5$. Additionally, we implement a warmup ratio of $0.1$ followed by linear decay. The L2 regularization weight decay is set at $\lambda = 5e-3$. For LoRA, we set the rank =$8$. For AdaLoRA and adaptive rank sampling, the additional hyper-parameters are shown in Supplementary Table~\ref{tab:hyperparameters}. In this Table, Avg. $b_0$ and Avg. $b_T$ denote the average number of singular values in each matrix. $(T_{\text{total}}- T)/T_{\text{total}}$ denotes the final fine-tune ratio after pruning and reintroducing, and $\Delta T$ indicates the intervals at which pruning and reintroducing are performed. The additoinal hyper-parameters for AdaLoRA are shown in gray columns and the additoinal hyper-parameters for adaptive rank sampling is shown in all columns including the random sampling ratio $p$.

In the Histone (Human) dataset, to facilitate a rigorous comparison, we calibrated the performance metrics of OPT-350M and NT-500M, which require FLOPs$_{\text{OPT}-350\text{M}}$ and FLOPs$_{500\text{M}-1000\text{G}}$ respectively. This calibration involved adjusting their performance steps to the computational effort equivalent to FLOPs$_{\text{OPT-125}\text{M}}$, thereby normalizing their outputs to a unified computational standard. We calibrated the performance metrics of OPT-350M and NT-500M for a rigorous comparison, , which require FLOPs$_{\text{OPT}-350\text{M}}$ and FLOPs$_{500\text{M}-1000\text{G}}$ respectively. This calibration involved adjusting their performance steps to the computational effort equivalent to FLOPs$_{\text{OPT-125}\text{M}}$, thereby normalizing their outputs to a unified computational cost. Specifically, the FLOPs-adjusted training steps ($I$) for OPT-350M and NT-500M, denoted as $n_{\text{OPT-350M}}$ and $n_{500\text{M}-1000\text{G}}$ respectively, were computed as follows:
$
I_{\text{OPT-350M}} = I_{\text{OPT-125M}} \times \frac{\text{FLOPs}_{\text{OPT-125M}}}{\text{FLOPs}_{\text{OPT-350M}}}
$ and
$
I_{500\text{M}-1000\text{G}} = I_{\text{OPT-125M}} \times \frac{\text{FLOPs}_{\text{OPT-125M}}}{\text{FLOPs}_{500\text{M}-1000\text{G}}}.
$
This methodology allowed us to directly compare the performance efficacy of all three models under a common computational constraint, thereby ensuring an equitable assessment of each model’s capabilities within the parameters of FLOPs-adjusted efficiency. In this tasks, we evaluate the validation loss every $1e4$ step, and if the best validation loss has not decreased for $10$ evaluations, we early stop the fine-tuning process.

\section{Data Availability}
 The sequences for Histone (Yeast) and Promoter (Human) are obtained from the GUE framework~\cite{zhou2023dnabert} and the sequences for histone marks prediction in multiple cell types are extracted from the DeepSEA framework~\citep{zhou2015predicting}.

\section{Code Availability}
Our \textsc{Lingo} code is available on Github at \url{https://github.com/zhanglab-aim/LINGO}.
\newpage
\section{Supplementary Information}
\setcounter{table}{-1} 
\counterwithin*{table}{section} 
\renewcommand{\thetable}{S\arabic{table}} 
\setcounter{figure}{-1} 
\counterwithin*{figure}{section} 
\renewcommand{\thefigure}{S\arabic{figure}} 

\subsection{Datasets and setting-up}
\begin{table}[!htbp]
    \caption{Genome understanding tasks}
    \label{tab:dataset}
    \centering
    \begin{tabular}{c|c c c c}
    \hline
    Task & Num. Datasets & Class & Num. Classes & Sequence Length\\
    \hline
     Histone (Yeast)    & 10 & Binary & 2 & 500  \\
      Promoter (Human)   & 3 & Binary & 2 & 300 \\
    Histone (Human)   & 104 &Multi-label & 104 & 1000 \\
      \hline
    \end{tabular}
\end{table}

\begin{sidewaystable}
    \centering
        \caption{Additoinal hyper-parameters for AdaLoRA (in gray) and adaptive rank sampling (all columns).}
    \label{tab:hyperparameters}
    \begin{tabular}{l l| >{\columncolor[gray]{.8}\centering\arraybackslash}p{1.4cm}| >{\columncolor[gray]{.8}\centering\arraybackslash}p{1.4cm} >{\columncolor[gray]{.8}\centering\arraybackslash}p{1.4cm} >{\columncolor[gray]{.8}\centering\arraybackslash}p{1.4cm} >{\columncolor[gray]{.8}\centering\arraybackslash}p{1.4cm} >{\columncolor[gray]{.8}\centering\arraybackslash}p{1.4cm} >{\columncolor[gray]{.8}\centering\arraybackslash}p{1.4cm} >{\centering\arraybackslash}p{1.4cm}}
     \hline
      \textbf{Task}   & \textbf{Dataset}  & \textbf{Avg. $b_0$} & \textbf{Avg. $b_T$} & $(T_{\text{total}}- T)/T_{\text{total}}$ & $\Delta T$ & $\beta_1$ & $\beta_2$ & \textbf{Pruned Matrices} & $p$\\
      \hline
      \multirow{10}{*}{Histone (Yeast)}  & H3 &  $12$ & $6$ & $0.25$ & $100$ & $0.85$ & $0.85$ & $W_q$, $W_k$, $W_v$, $W_{f_1}$, $W_{f_2}$ & $0.05$\\
    & H3K4me1 &  $12$ & $6$ & $0.25$ & $100$ & $0.85$ & $0.85$ & $W_q$, $W_k$, $W_v$, $W_{f_1}$, $W_{f_2}$ & $0.05$\\
    & H3K4me2 &  $12$ & $6$ & $0.25$ & $100$ & $0.85$ & $0.85$ & $W_q$, $W_k$, $W_v$, $W_{f_1}$, $W_{f_2}$ & $0.05$\\
    & H3K4me3 &  $12$ & $6$ & $0.25$ & $100$ & $0.85$ & $0.85$ & $W_q$, $W_k$, $W_v$, $W_{f_1}$, $W_{f_2}$ & $0.05$\\
    & H3K9ac &  $12$ & $6$ & $0.25$ & $100$ & $0.85$ & $0.85$ & $W_q$, $W_k$, $W_v$, $W_{f_1}$, $W_{f_2}$ & $0.05$\\
    & H3K14ac &  $12$ & $6$ & $0.25$ & $100$ & $0.85$ & $0.85$ & $W_q$, $W_k$, $W_v$, $W_{f_1}$, $W_{f_2}$ & $0.05$\\
    & H3K36me3 &  $12$ & $6$ & $0.30$ & $100$ & $0.85$ & $0.85$ & $W_q$, $W_k$, $W_v$, $W_{f_1}$, $W_{f_2}$ & $0.05$\\
    & H3K79me3 &  $12$ & $6$ & $0.30$ & $100$ & $0.85$ & $0.85$ & $W_q$, $W_k$, $W_v$, $W_{f_1}$, $W_{f_2}$ & $0.05$\\
    & H4 &  $12$ & $6$ & $0.30$ & $100$ & $0.85$ & $0.85$ & $W_q$, $W_k$, $W_v$, $W_{f_1}$, $W_{f_2}$ & $0.05$\\
    & H4ac &  $12$ & $6$ & $0.30$ & $100$ & $0.85$ & $0.85$ & $W_q$, $W_k$, $W_v$, $W_{f_1}$, $W_{f_2}$ & $0.05$\\
         \hline
    \multirow{3}{*}{Promoter (Human)} & Prom\_all &  $8$ & $6$ & $0.15$ & $5000$ & $0.85$ & $0.85$ & $W_q$, $W_k$, $W_v$, $W_{f_1}$, $W_{f_2}$, $W_o$ & $0.1$\\
         & Prom\_notata &  $8$ & $6$ & $0.15$ & $5000$ & $0.85$ & $0.85$ & $W_q$, $W_k$, $W_v$, $W_{f_1}$, $W_{f_2}$, $W_o$ & $0.1$\\
        & Prom\_tata &  $8$ & $6$ & $0.15$ & $5000$ & $0.85$ & $0.85$ & $W_q$, $W_k$, $W_v$, $W_{f_1}$, $W_{f_2}$, $W_o$ & $0.1$\\
        \hline
        Histone (Human)  & Histone &  $8$ & $6$ & $0.15$ & $5000$ & $0.99$ & $0.99$ & $W_q$, $W_k$, $W_v$, $W_{f_1}$, $W_{f_2}$, $W_o$ & $0.1$\\
         \hline
    \end{tabular}
\end{sidewaystable}
\subsection{Results}
\begin{figure}[!htbp]
    \centering

    \begin{subfigure}[b]{0.2\linewidth}
        \includegraphics[width=\linewidth]{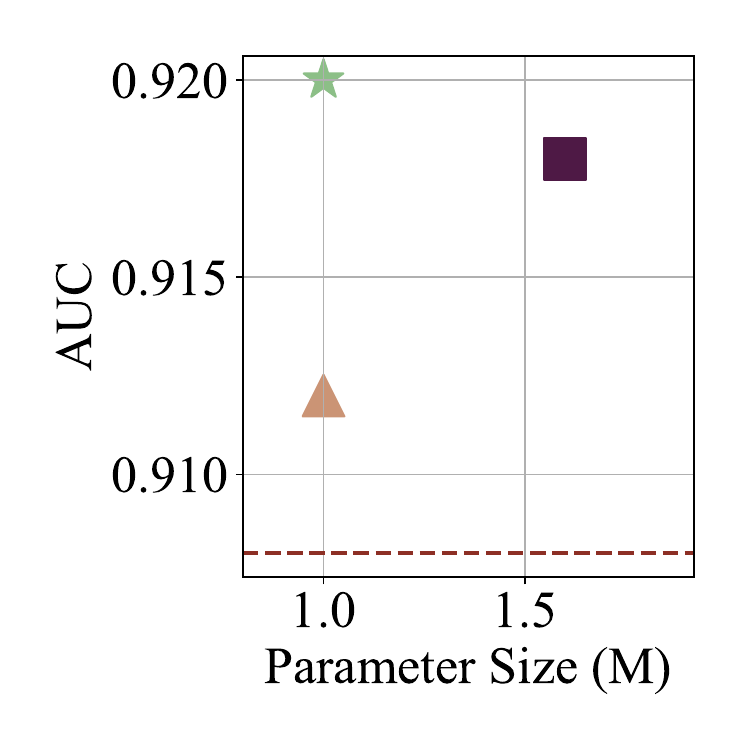}
        \caption{DNABERT-2 on Prom\_all}
    \end{subfigure}
    \begin{subfigure}[b]{0.2\linewidth}
        \includegraphics[width=\linewidth]{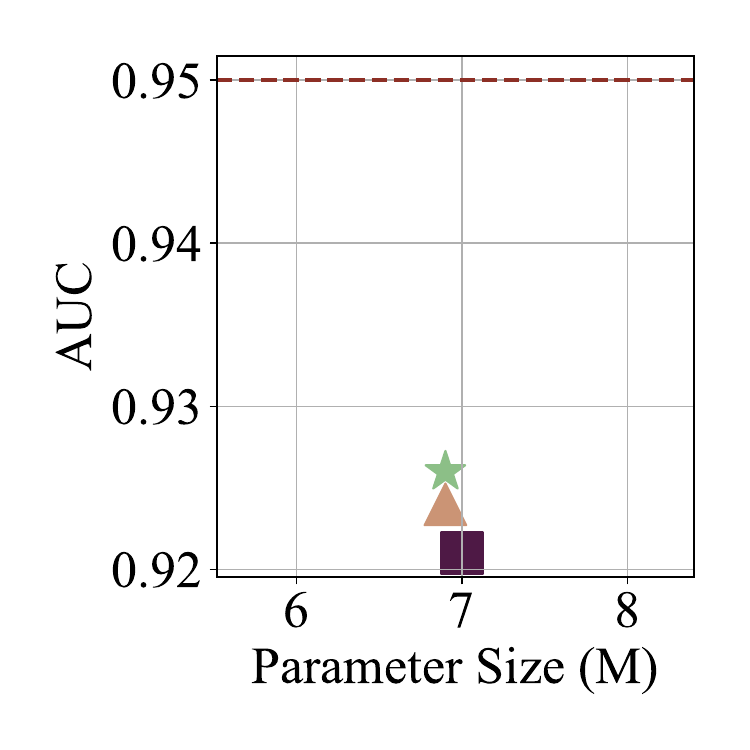}
        \caption{NT-500M on Prom\_all}
    \end{subfigure}
    \begin{subfigure}[b]{0.2\linewidth}
        \includegraphics[width=\linewidth]{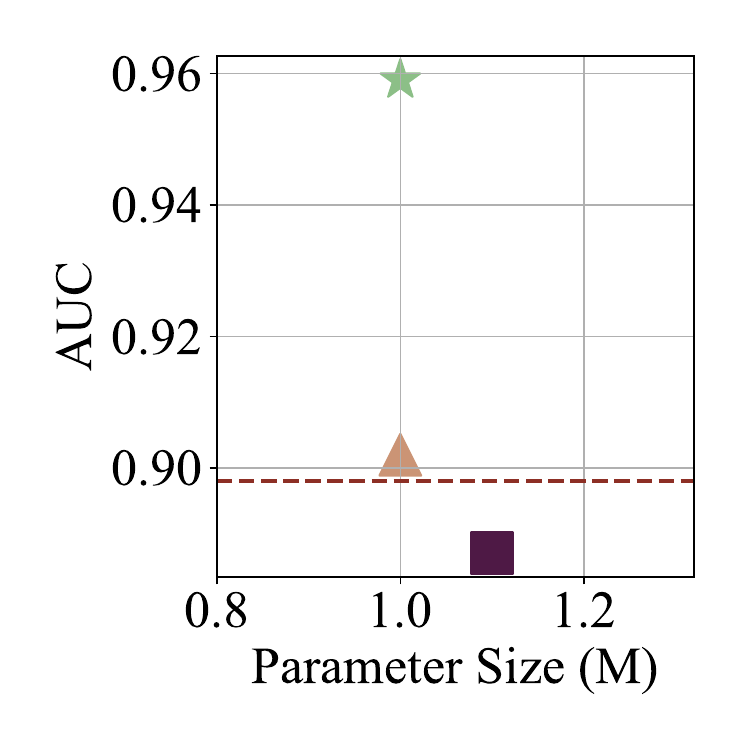}
        \caption{OPT-125M on Prom\_all}
    \end{subfigure}
    \begin{subfigure}[b]{0.2\linewidth}
        \includegraphics[width=\linewidth]{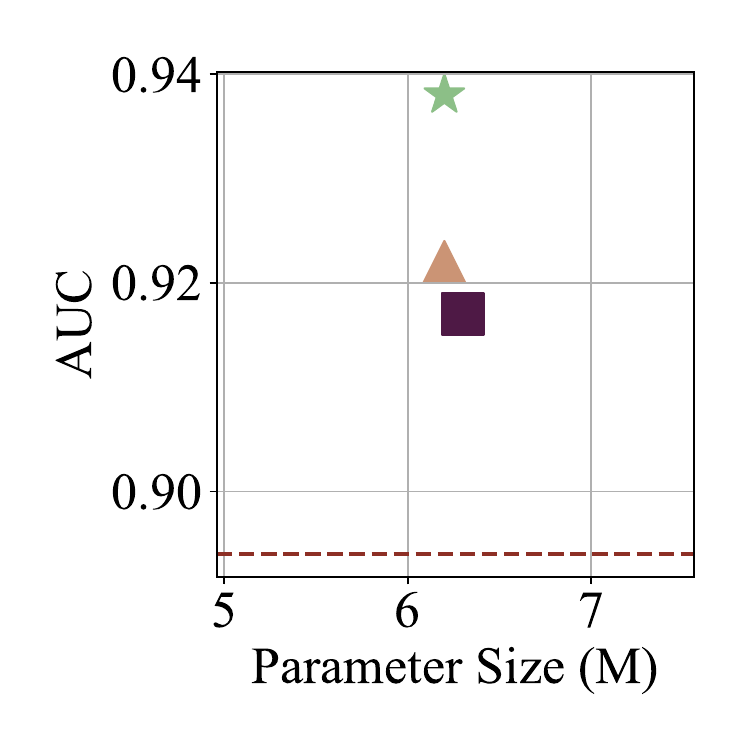}
        \caption{OPT-350M on Prom\_all}
    \end{subfigure}

    \begin{subfigure}[b]{0.2\linewidth}
        \includegraphics[width=\linewidth]{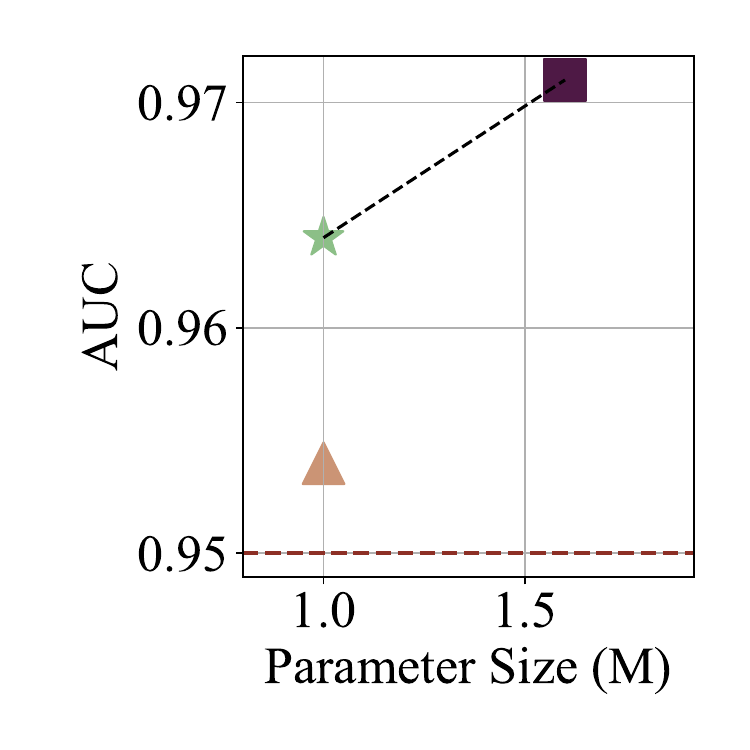}
        \caption{DNABERT-2 on Prom\_notata}
    \end{subfigure}
    \begin{subfigure}[b]{0.2\linewidth}
        \includegraphics[width=\linewidth]{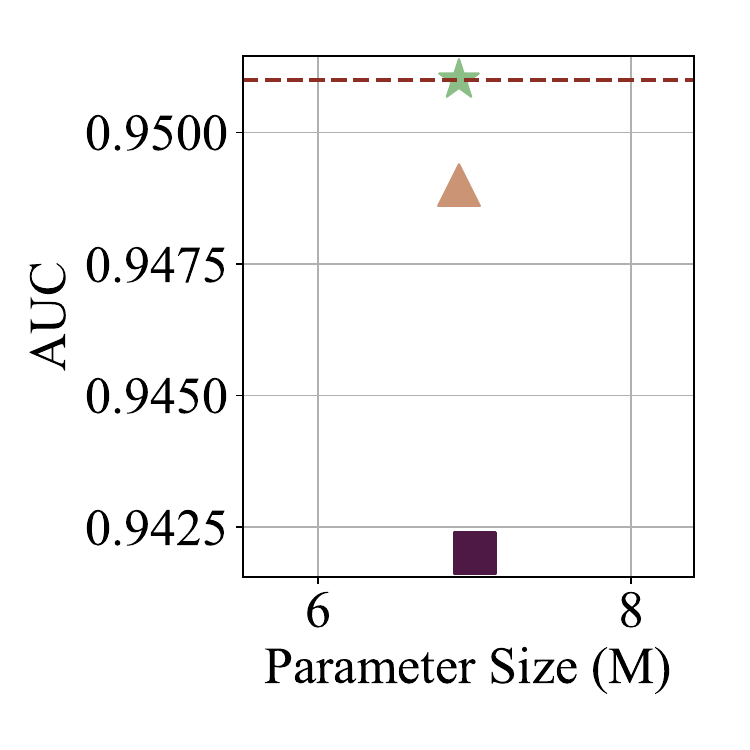}
        \caption{NT-500M on Prom\_notata}
    \end{subfigure}
    \begin{subfigure}[b]{0.2\linewidth}
        \includegraphics[width=\linewidth]{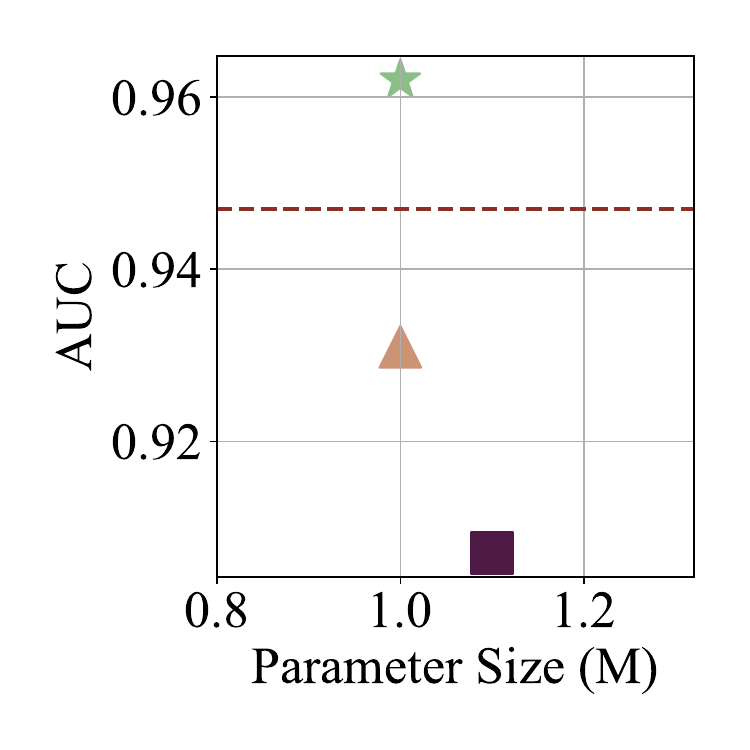}
        \caption{OPT-125M on Prom\_notata}
    \end{subfigure}
    \begin{subfigure}[b]{0.2\linewidth}
        \includegraphics[width=\linewidth]{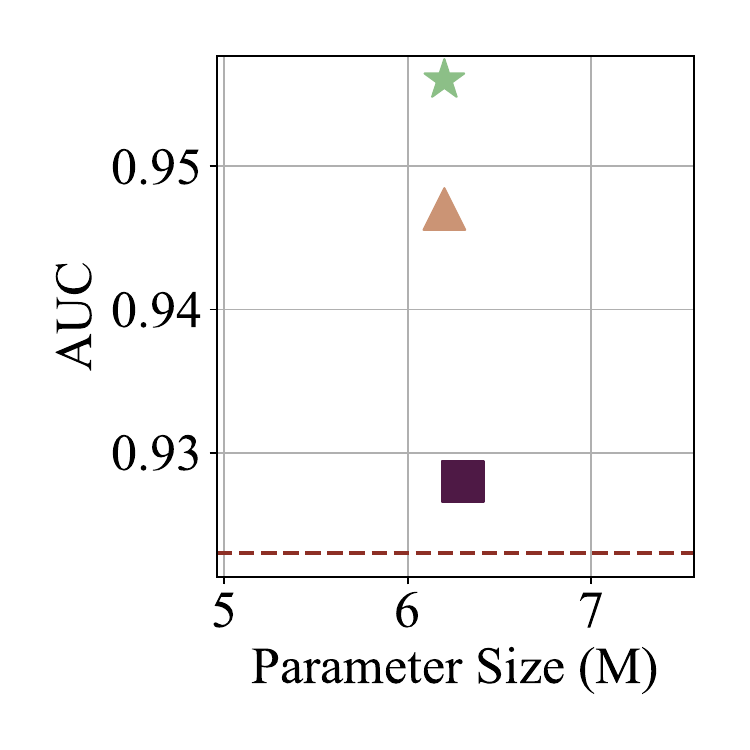}
        \caption{OPT-350M on Prom\_notata}
    \end{subfigure}

    \begin{subfigure}[b]{0.2\linewidth}
        \includegraphics[width=\linewidth]{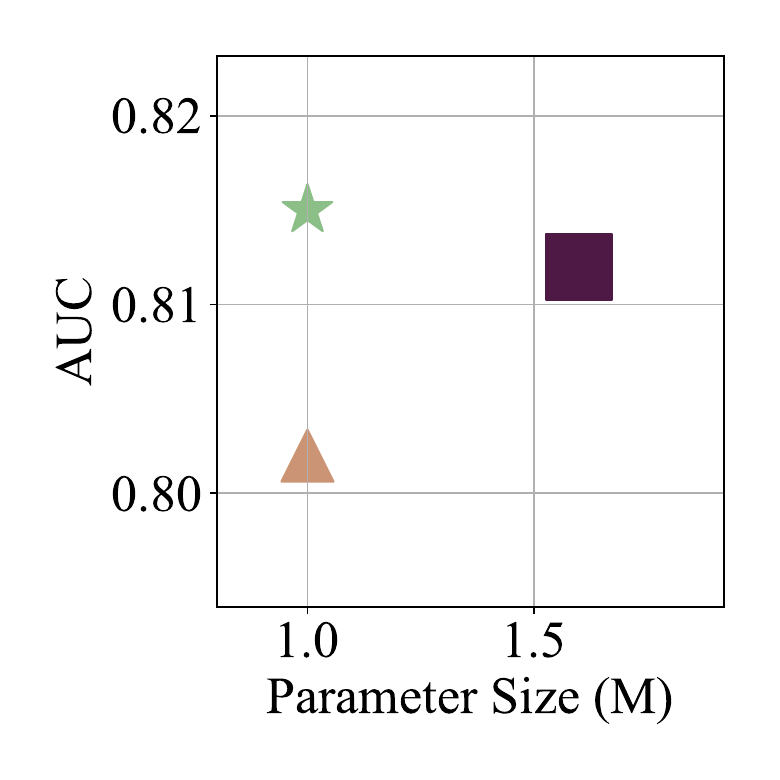}
        \caption{DNABERT-2 on Prom\_tata}
    \end{subfigure}
    \begin{subfigure}[b]{0.2\linewidth}
        \includegraphics[width=\linewidth]{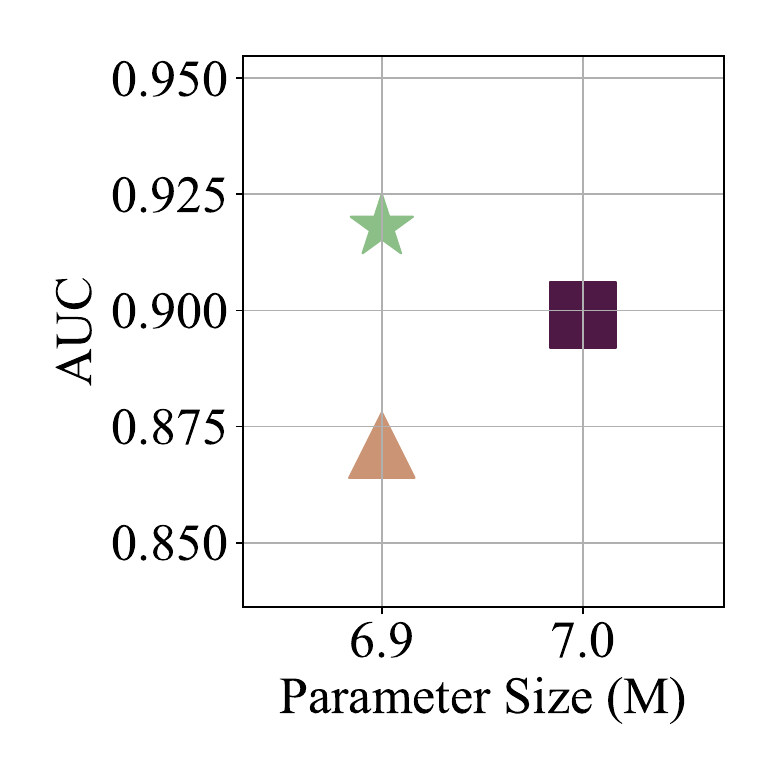}
        \caption{NT-500M on Prom\_tata}
    \end{subfigure}
    \begin{subfigure}[b]{0.2\linewidth}
        \includegraphics[width=\linewidth]{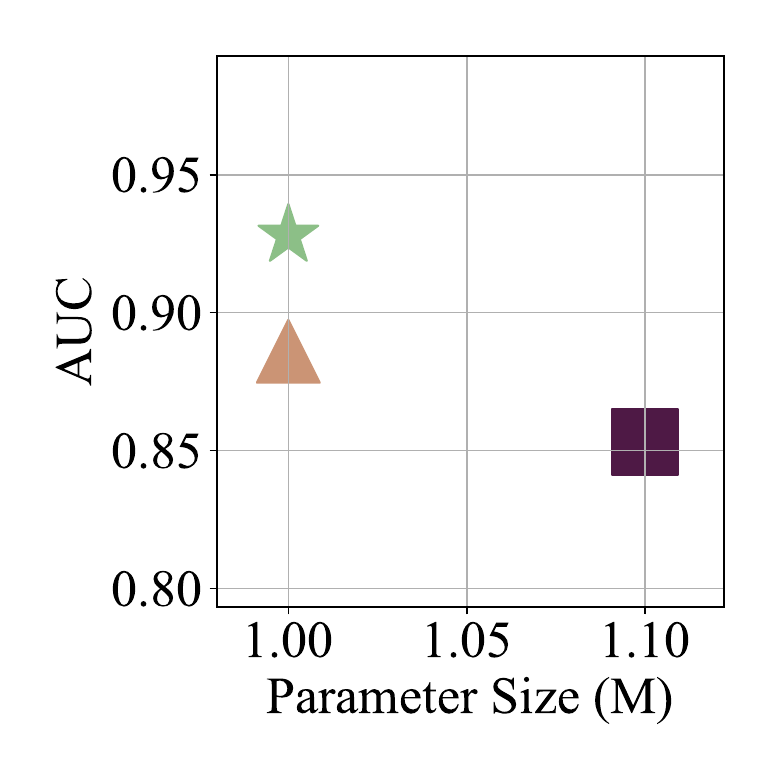}
        \caption{OPT-125M on Prom\_tata}
    \end{subfigure}
    \begin{subfigure}[b]{0.2\linewidth}
        \includegraphics[width=\linewidth]{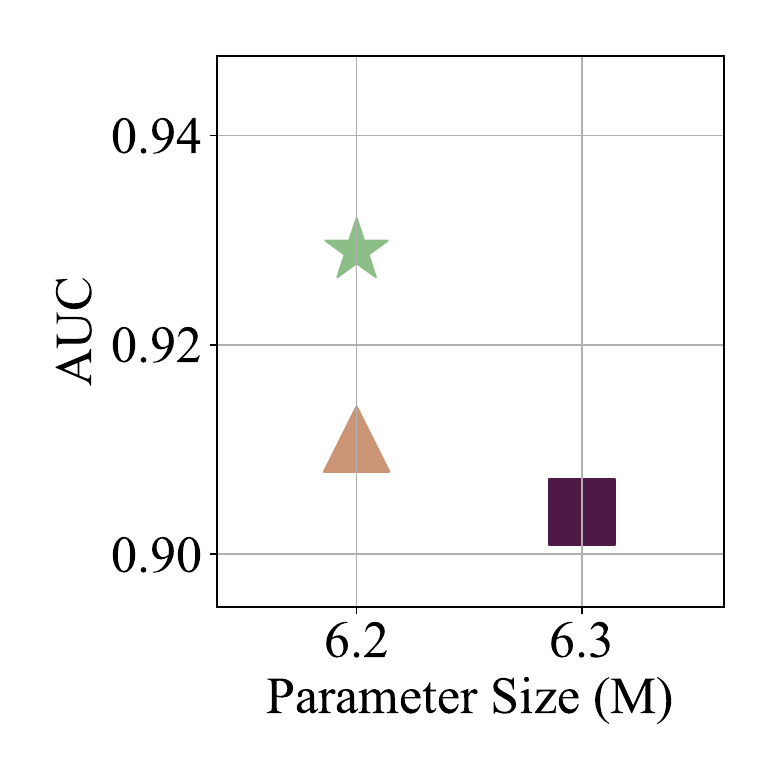}
        \caption{OPT-350M on Prom\_tata}
    \end{subfigure}

    \begin{subfigure}[b]{0.2\linewidth}
        \includegraphics[width=\linewidth]{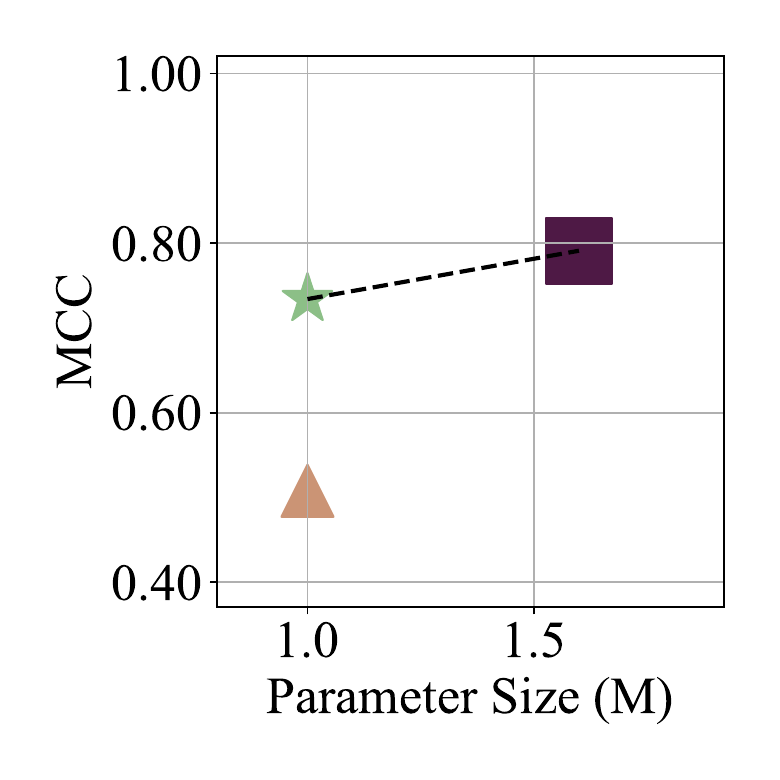}
        \caption{DNABERT-2 on H3}
    \end{subfigure}
    \begin{subfigure}[b]{0.2\linewidth}
        \includegraphics[width=\linewidth]{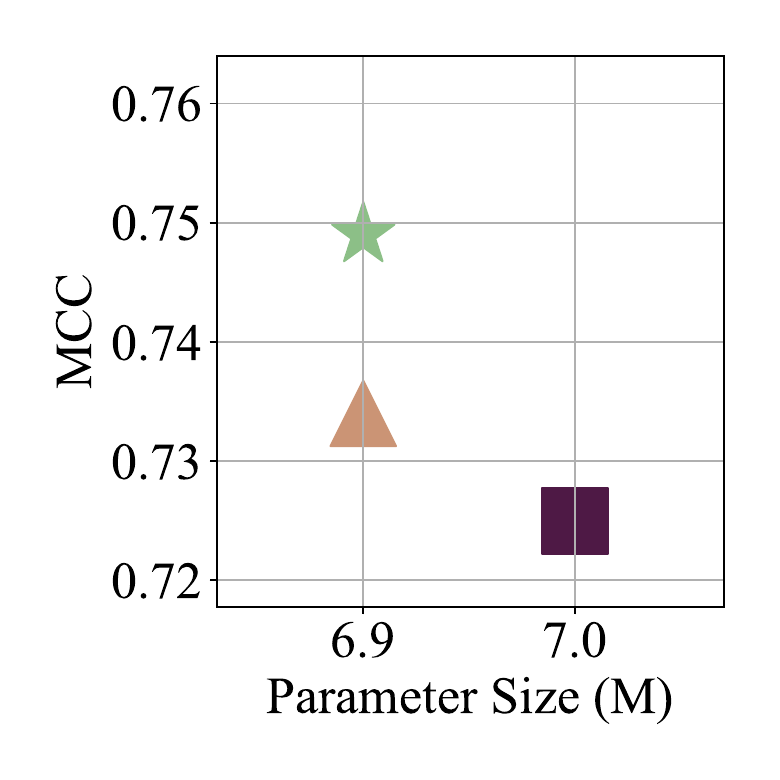}
        \caption{NT-500M on H3 (Yeast)}
    \end{subfigure}
    \begin{subfigure}[b]{0.2\linewidth}
        \includegraphics[width=\linewidth]{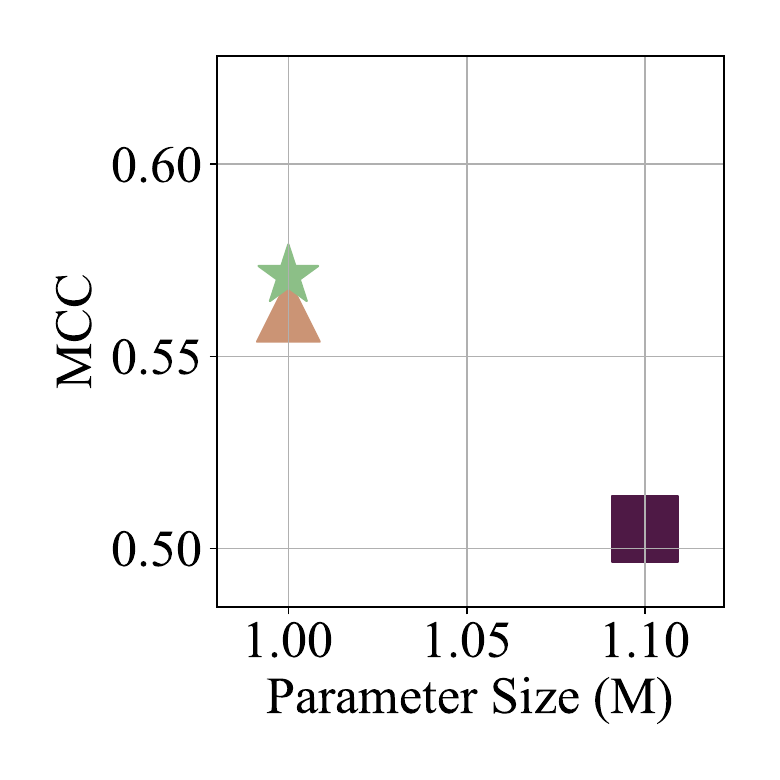}
        \caption{OPT-125M on H3}
    \end{subfigure}
    \begin{subfigure}[b]{0.2\linewidth}
        \includegraphics[width=\linewidth]{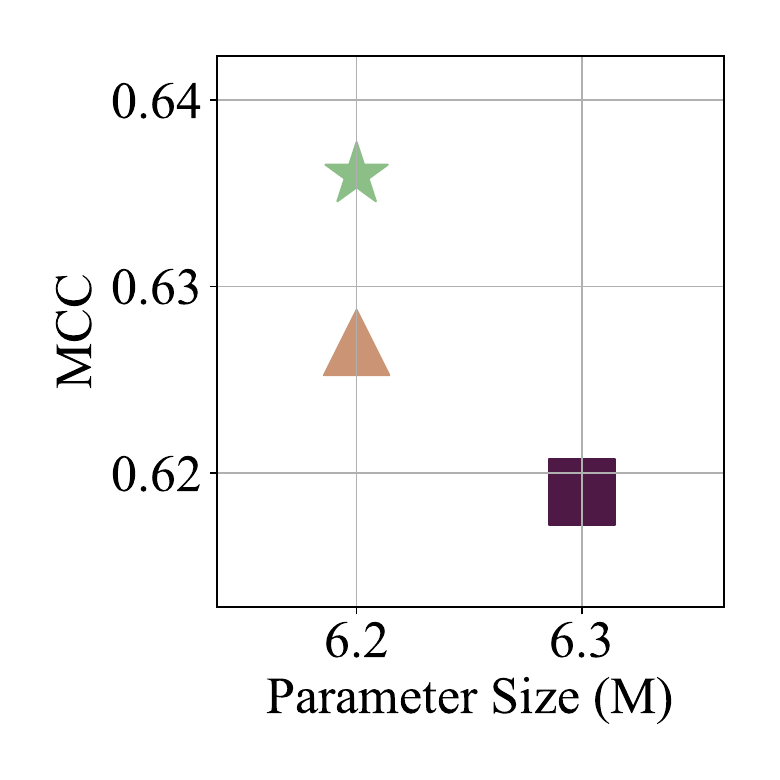}
        \caption{OPT-350M on H3}
    \end{subfigure}

        \begin{subfigure}[b]{0.2\linewidth}
        \includegraphics[width=\linewidth]{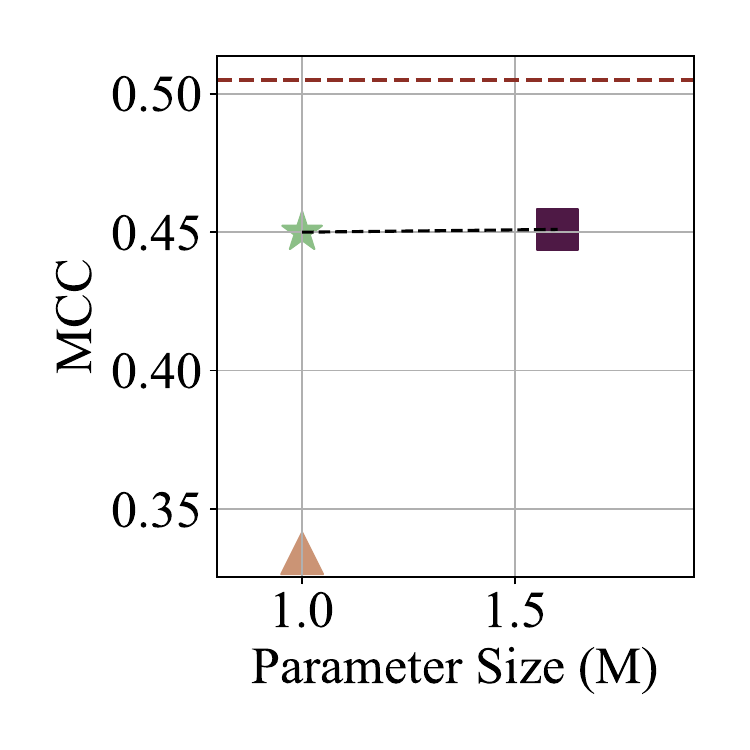}
        \caption{DNABERT-2 on H3K4me1}
    \end{subfigure}
    \begin{subfigure}[b]{0.2\linewidth}
        \includegraphics[width=\linewidth]{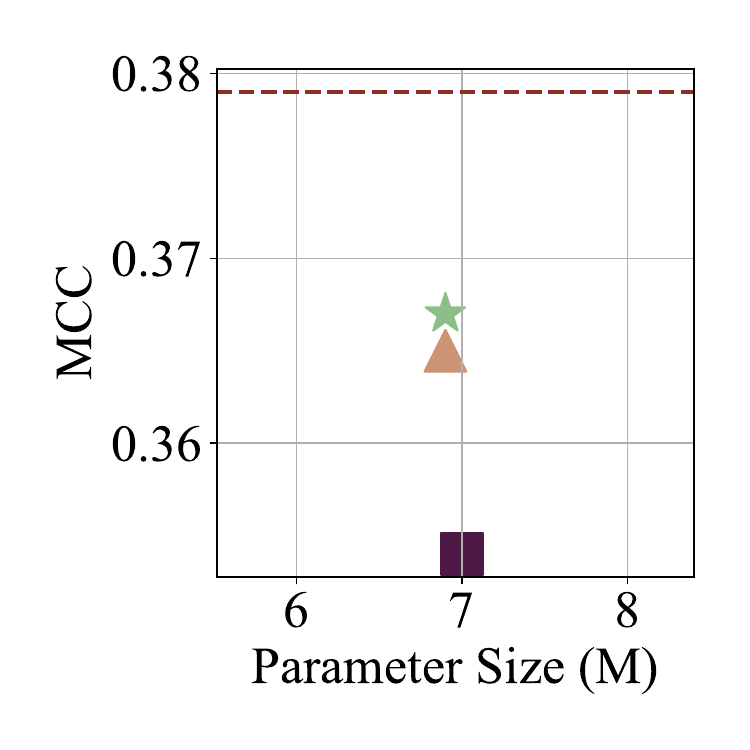}
        \caption{NT-500M on H3K4me1}
    \end{subfigure}
    \begin{subfigure}[b]{0.2\linewidth}
        \includegraphics[width=\linewidth]{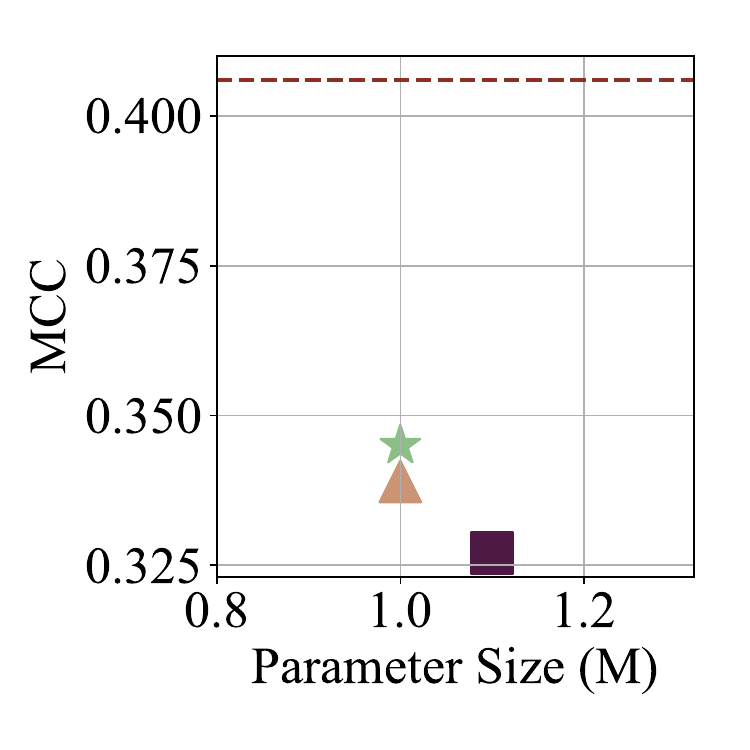}
        \caption{OPT-125M on H3K4me1}
    \end{subfigure}
    \begin{subfigure}[b]{0.2\linewidth}
        \includegraphics[width=\linewidth]{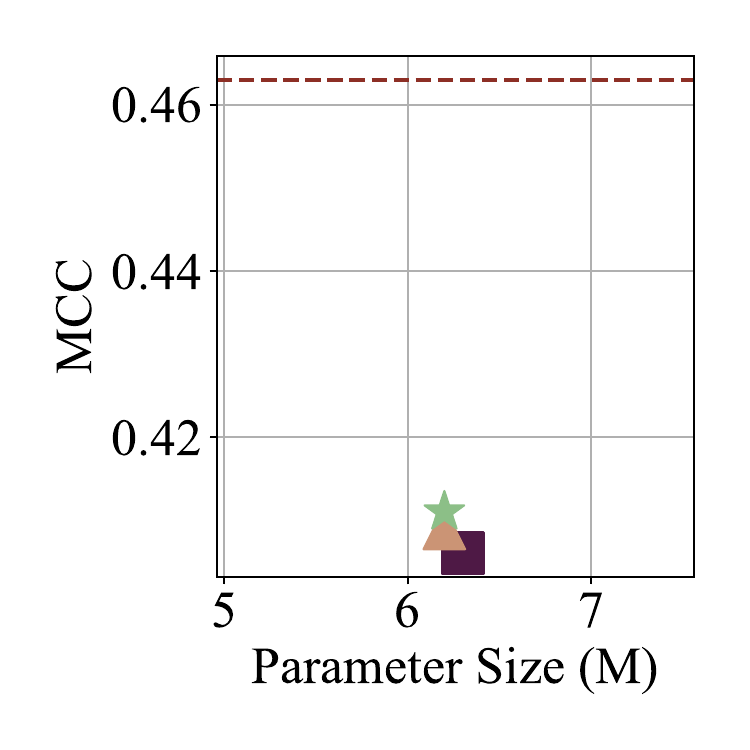}
        \caption{OPT-350M on H3K4me1}
    \end{subfigure}
\end{figure}

\begin{figure}[!htbp]\ContinuedFloat
            \begin{subfigure}[b]{0.2\linewidth}
        \includegraphics[width=\linewidth]{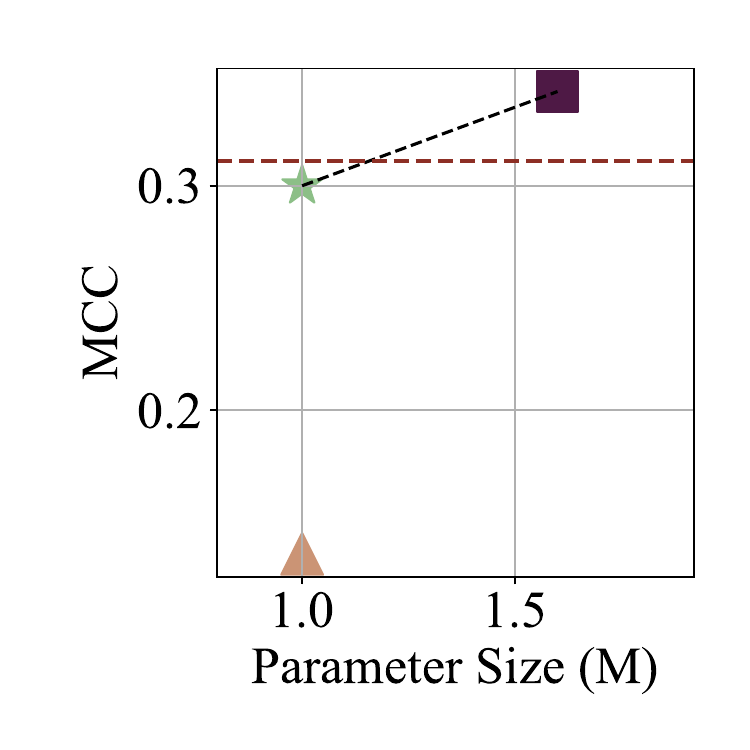}
        \caption{DNABERT-2 on H3K4me2}
    \end{subfigure}
    \begin{subfigure}[b]{0.2\linewidth}
        \includegraphics[width=\linewidth]{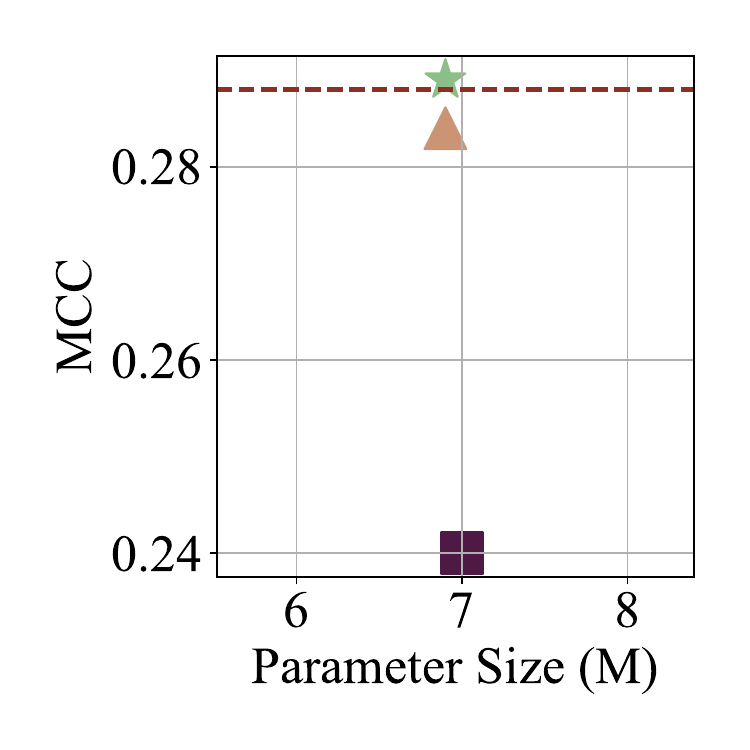}
        \caption{NT-500M on H3K4me2}
    \end{subfigure}
    \begin{subfigure}[b]{0.2\linewidth}
        \includegraphics[width=\linewidth]{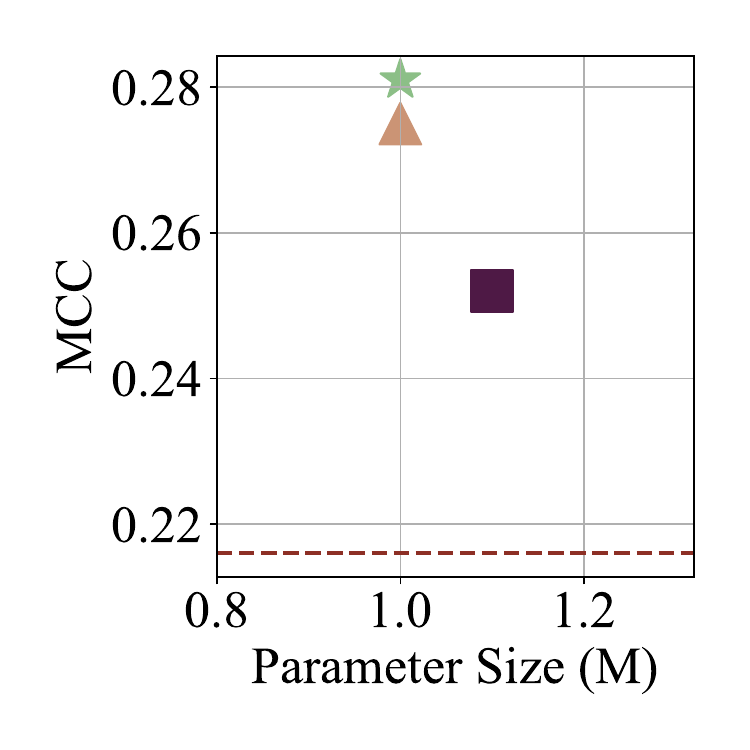}
        \caption{OPT-125M on H3K4me2}
    \end{subfigure}
    \begin{subfigure}[b]{0.2\linewidth}
        \includegraphics[width=\linewidth]{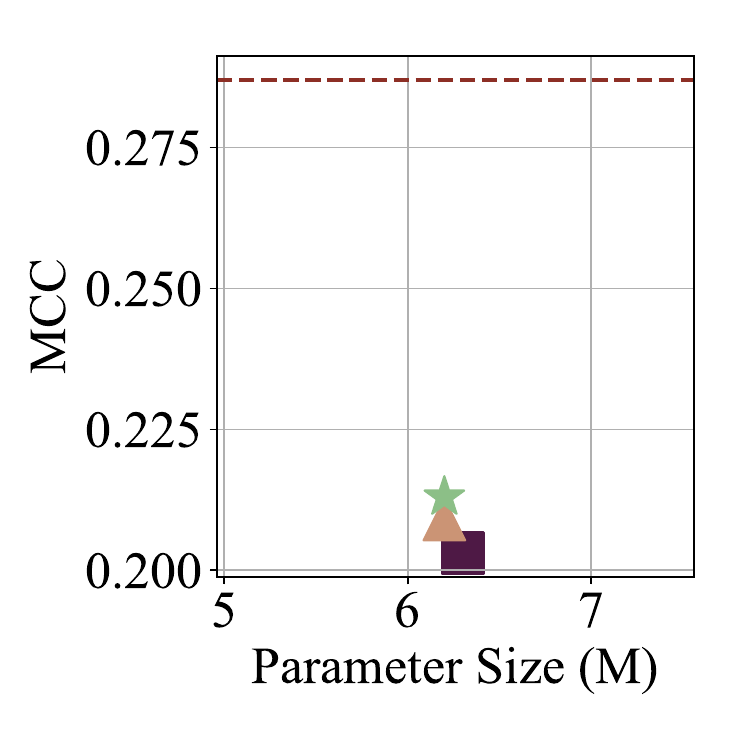}
        \caption{OPT-350M on H3K4me2}
    \end{subfigure}

            \begin{subfigure}[b]{0.2\linewidth}
        \includegraphics[width=\linewidth]{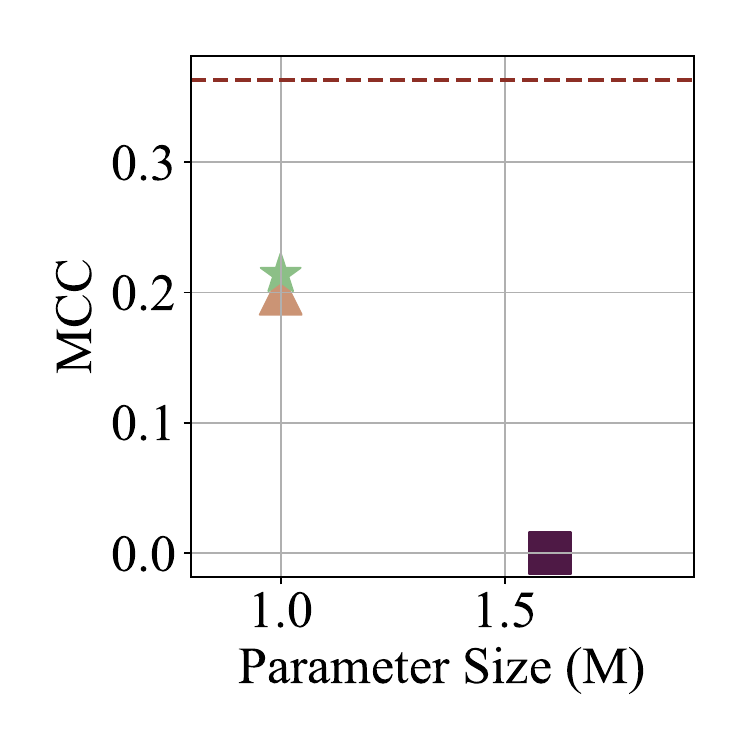}
        \caption{DNABERT-2 on H3K4me3}
    \end{subfigure}
    \begin{subfigure}[b]{0.2\linewidth}
        \includegraphics[width=\linewidth]{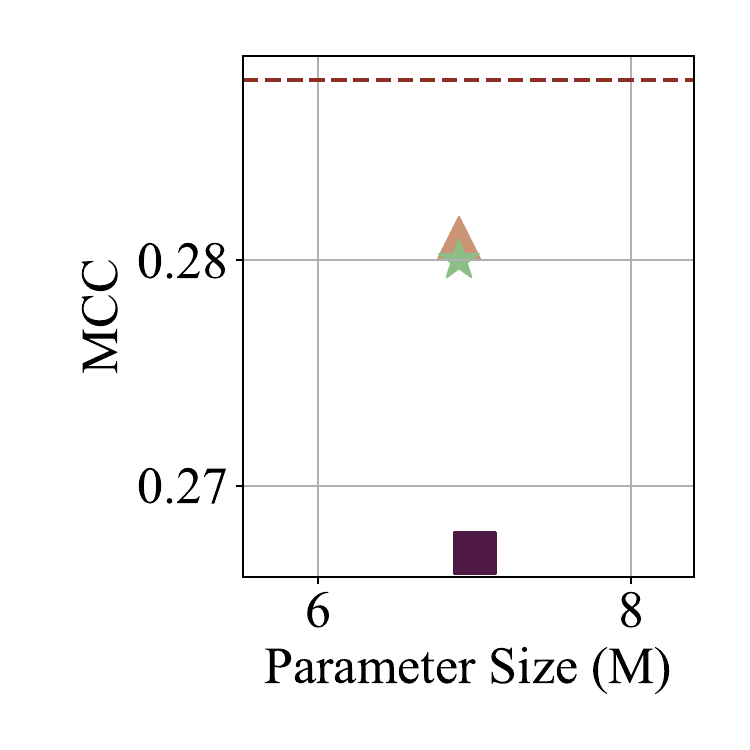}
        \caption{NT-500M on H3K4me3}
    \end{subfigure}
    \begin{subfigure}[b]{0.2\linewidth}
        \includegraphics[width=\linewidth]{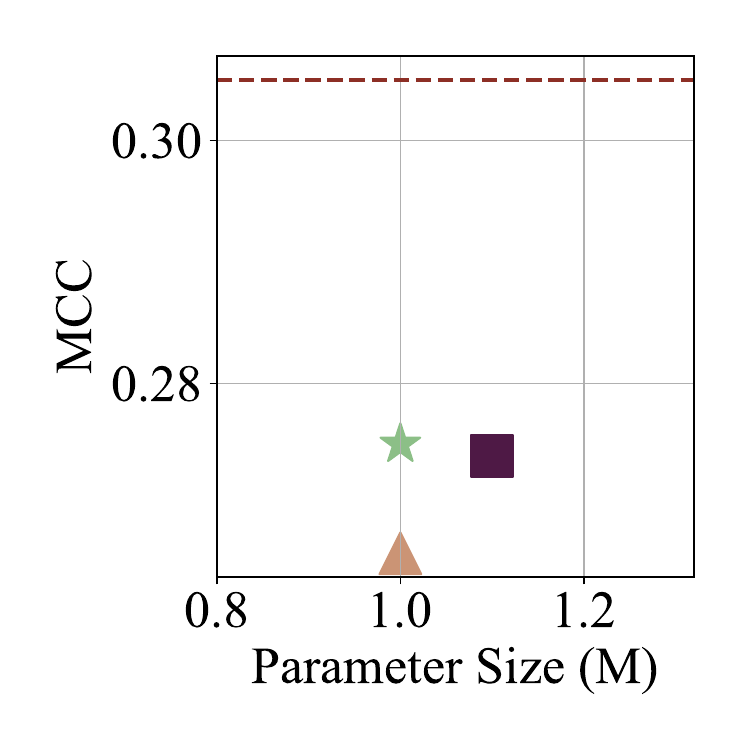}
        \caption{OPT-125M on H3K4me3}
    \end{subfigure}
    \begin{subfigure}[b]{0.2\linewidth}
        \includegraphics[width=\linewidth]{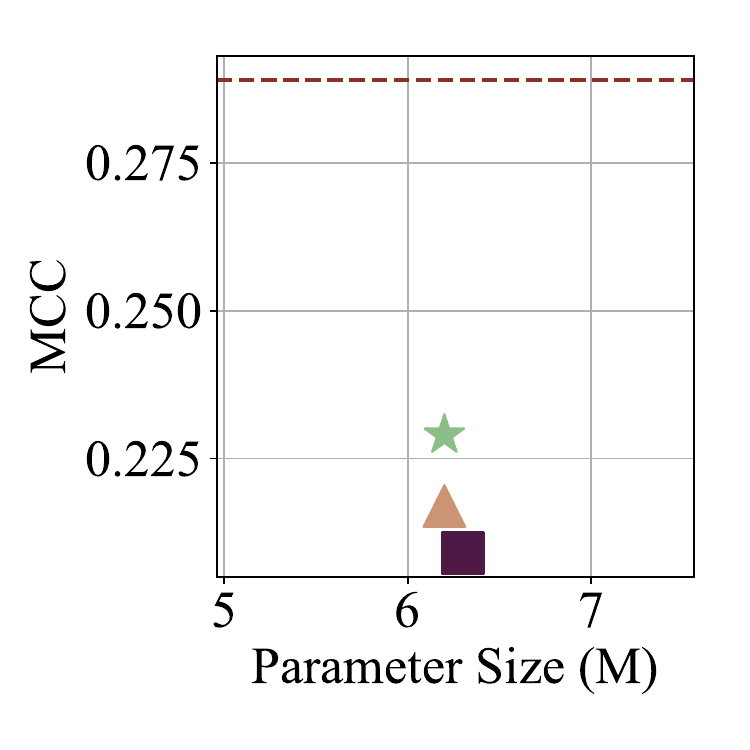}
        \caption{OPT-350M on H3K4me3}
    \end{subfigure}

                \begin{subfigure}[b]{0.2\linewidth}
        \includegraphics[width=\linewidth]{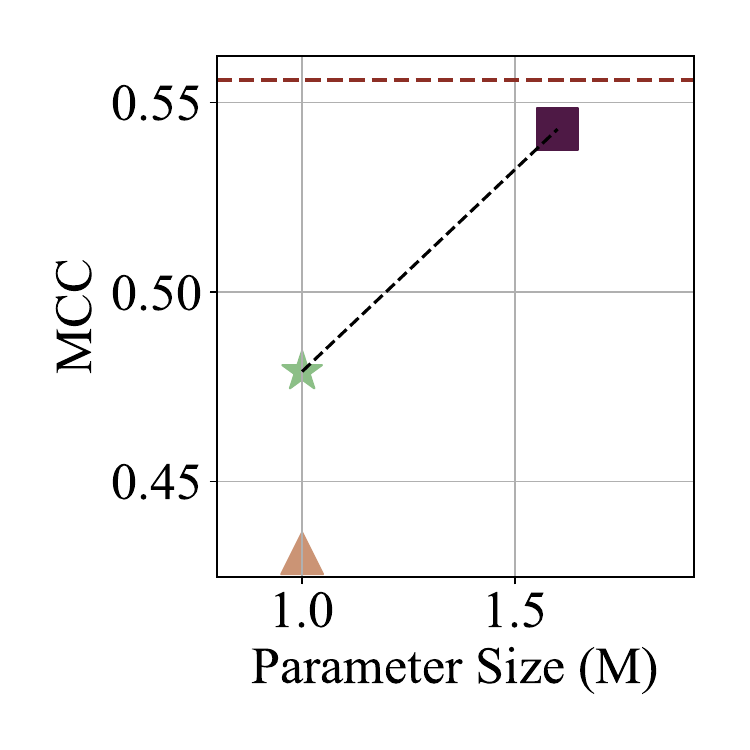}
        \caption{DNABERT-2 on H3K9ac}
    \end{subfigure}
    \begin{subfigure}[b]{0.2\linewidth}
        \includegraphics[width=\linewidth]{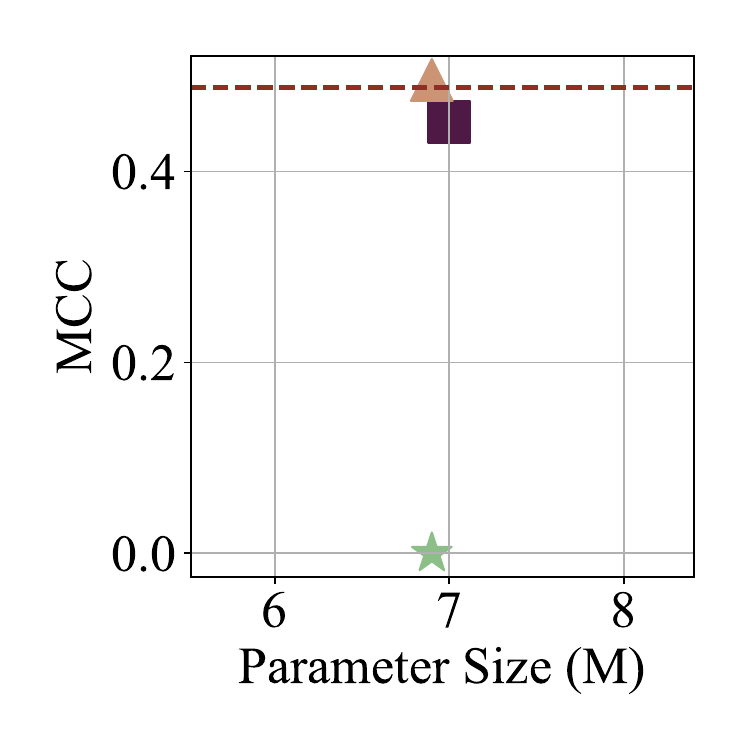}
        \caption{NT-500M on H3K9ac}
    \end{subfigure}
    \begin{subfigure}[b]{0.2\linewidth}
        \includegraphics[width=\linewidth]{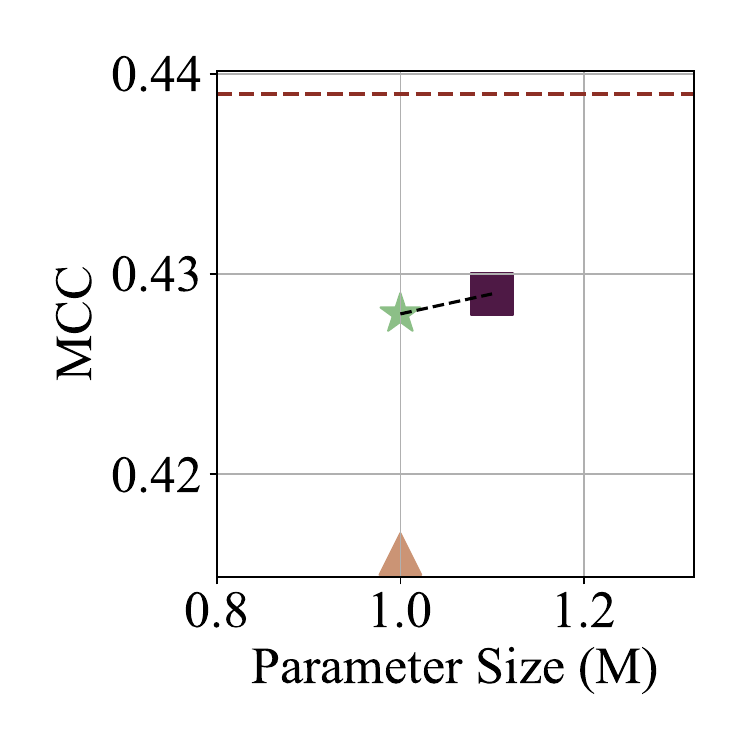}
        \caption{OPT-125M on H3K9ac}
    \end{subfigure}
    \begin{subfigure}[b]{0.2\linewidth}
        \includegraphics[width=\linewidth]{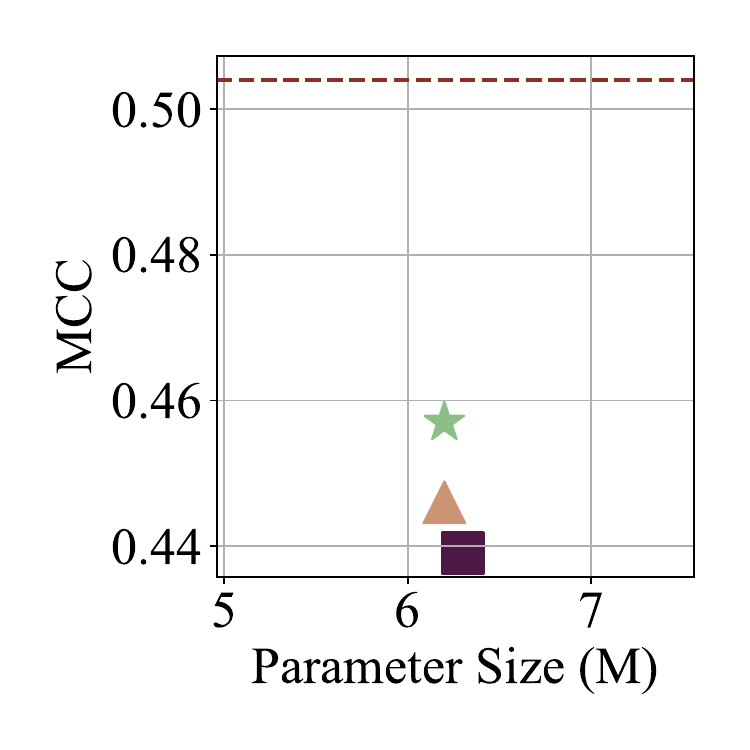}
        \caption{OPT-350M on H3K9ac}
    \end{subfigure}

                    \begin{subfigure}[b]{0.2\linewidth}
        \includegraphics[width=\linewidth]{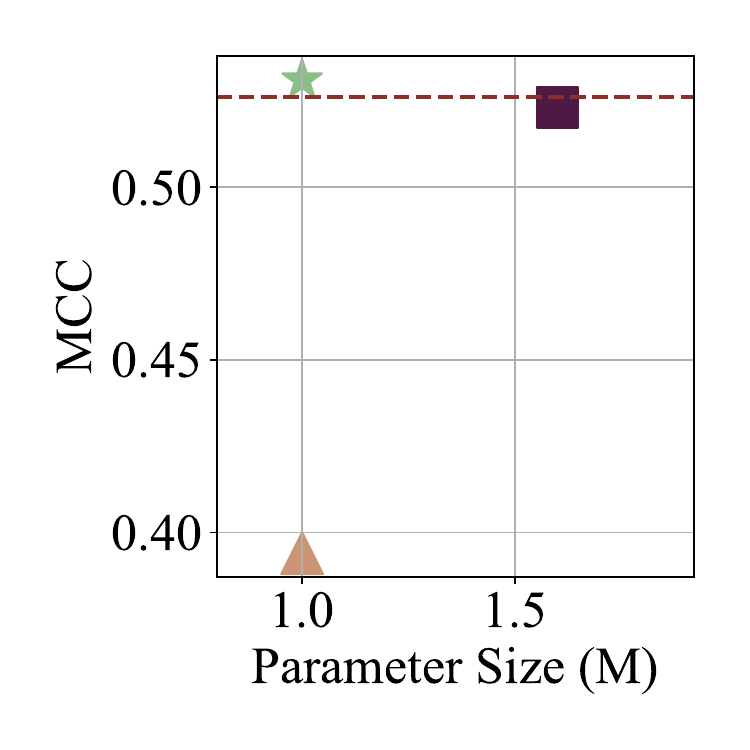}
        \caption{DNABERT-2 on H3K14ac}
    \end{subfigure}
    \begin{subfigure}[b]{0.2\linewidth}
        \includegraphics[width=\linewidth]{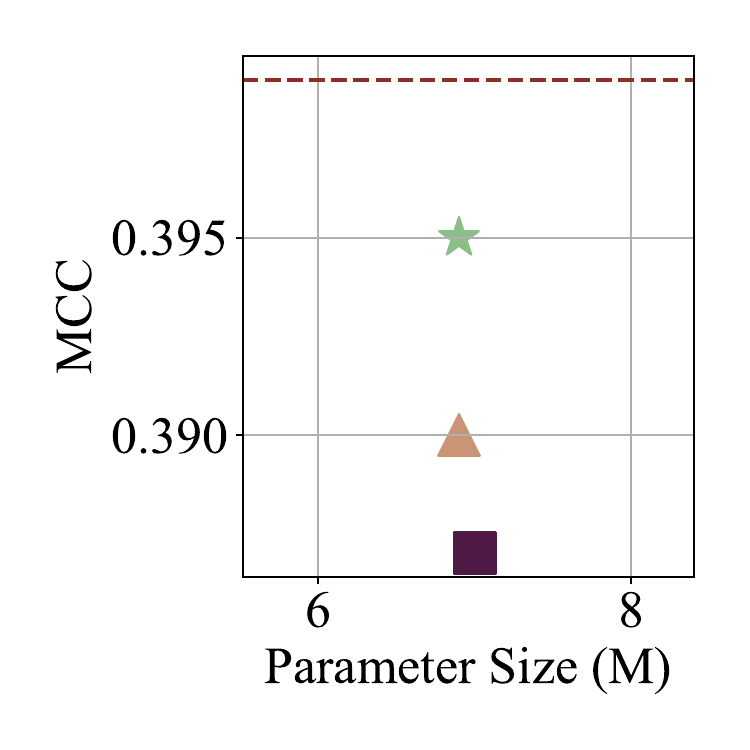}
        \caption{NT-500M on H3K14ac}
    \end{subfigure}
    \begin{subfigure}[b]{0.2\linewidth}
        \includegraphics[width=\linewidth]{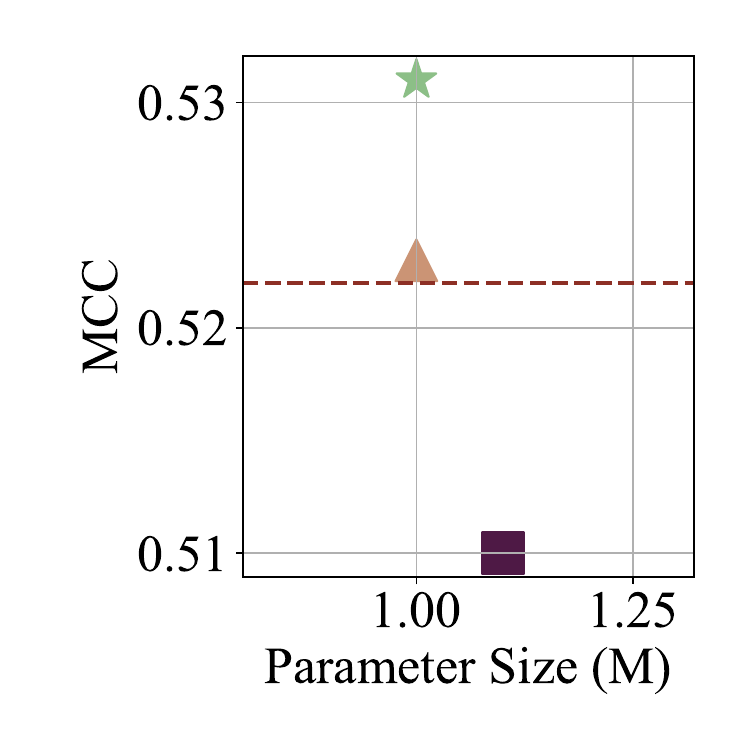}
        \caption{OPT-125M on H3K14ac}
    \end{subfigure}
    \begin{subfigure}[b]{0.2\linewidth}
        \includegraphics[width=\linewidth]{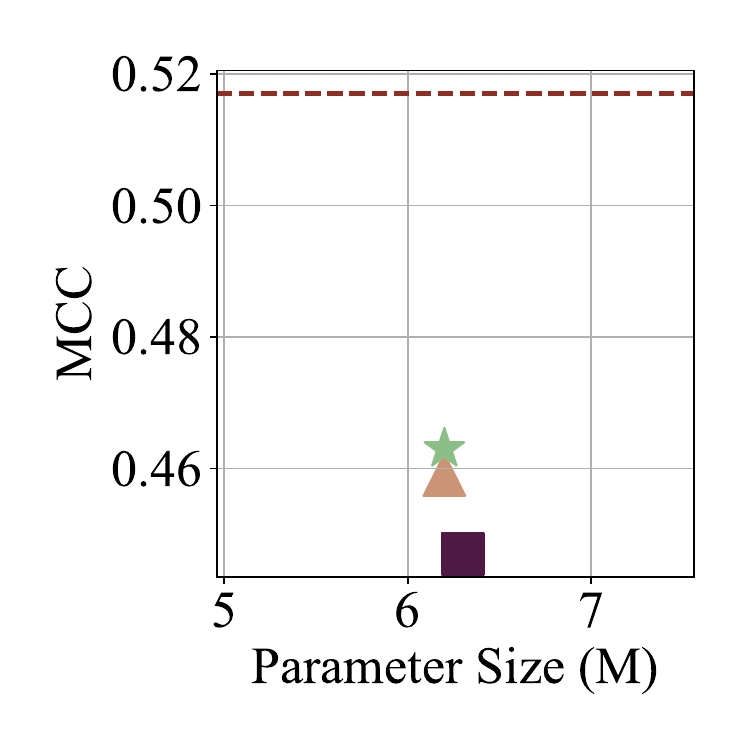}
        \caption{OPT-350M on H3K14ac}
    \end{subfigure}

                        \begin{subfigure}[b]{0.2\linewidth}
        \includegraphics[width=\linewidth]{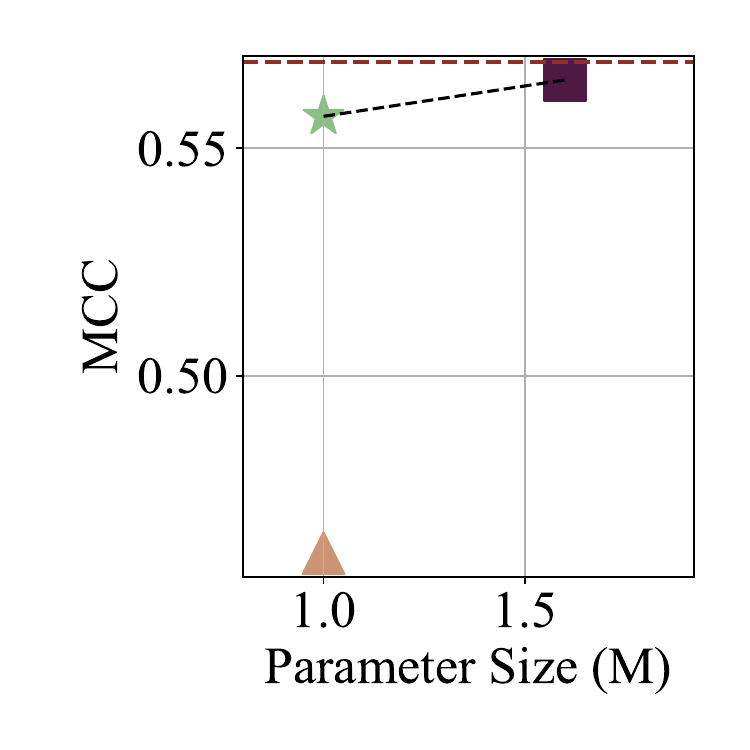}
        \caption{DNABERT-2 on H3K36me3}
    \end{subfigure}
    \begin{subfigure}[b]{0.2\linewidth}
        \includegraphics[width=\linewidth]{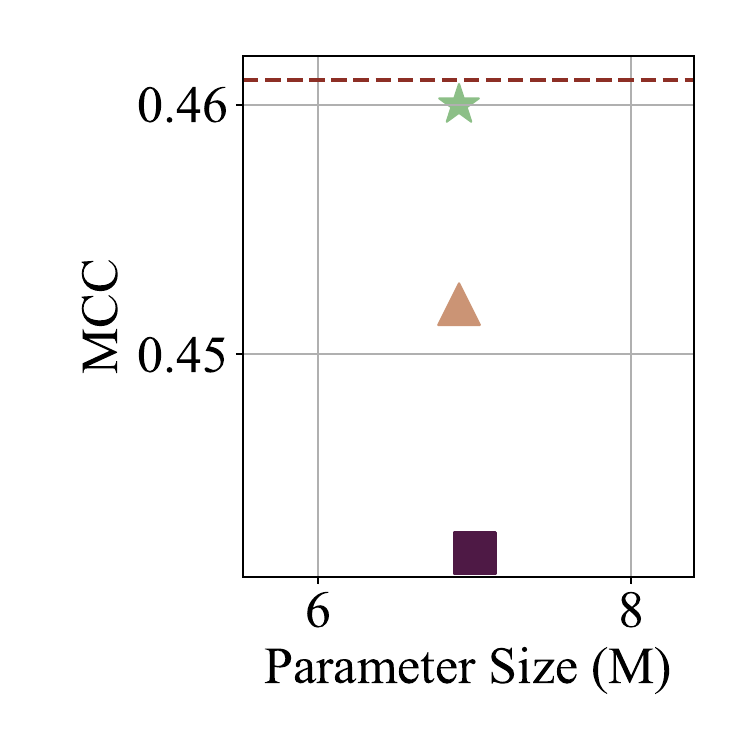}
        \caption{NT-500M on H3K36me3}
    \end{subfigure}
    \begin{subfigure}[b]{0.2\linewidth}
        \includegraphics[width=\linewidth]{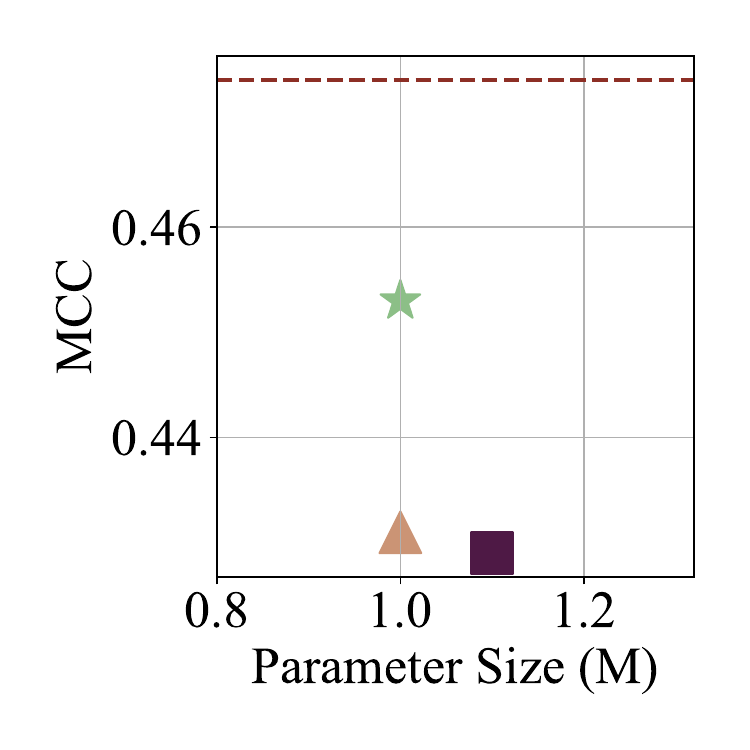}
        \caption{OPT-125M on H3K36me3}
    \end{subfigure}
    \begin{subfigure}[b]{0.2\linewidth}
        \includegraphics[width=\linewidth]{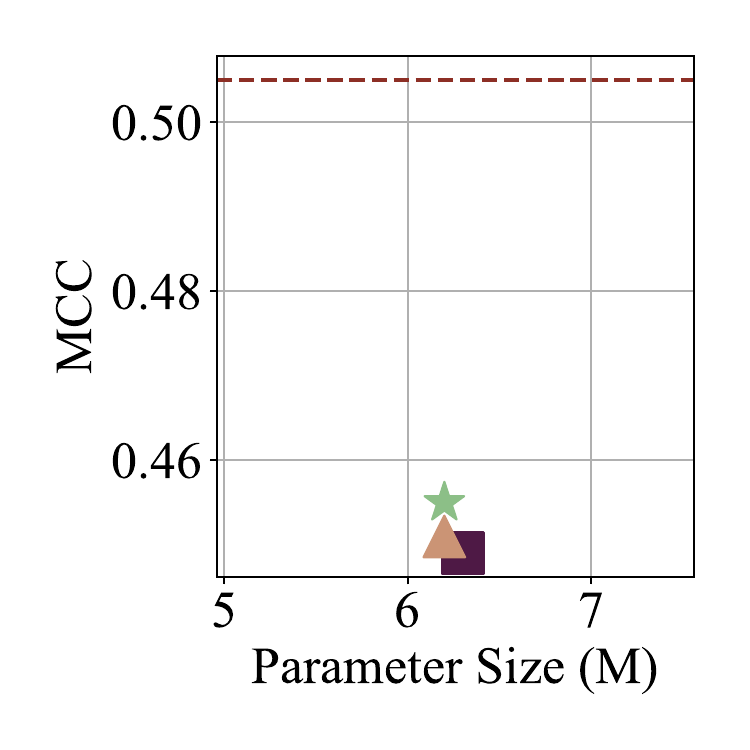}
        \caption{OPT-350M on H3K36me3}
    \end{subfigure}

\end{figure}

\begin{figure}[!htbp]\ContinuedFloat

                            \begin{subfigure}[b]{0.2\linewidth}
        \includegraphics[width=\linewidth]{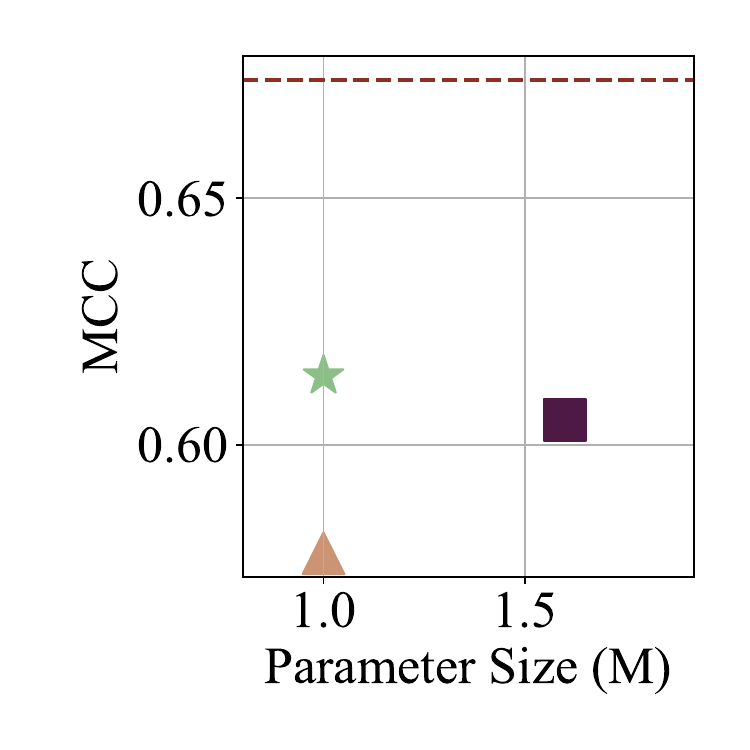}
        \caption{DNABERT-2 on H3K79me3}
    \end{subfigure}
    \begin{subfigure}[b]{0.2\linewidth}
        \includegraphics[width=\linewidth]{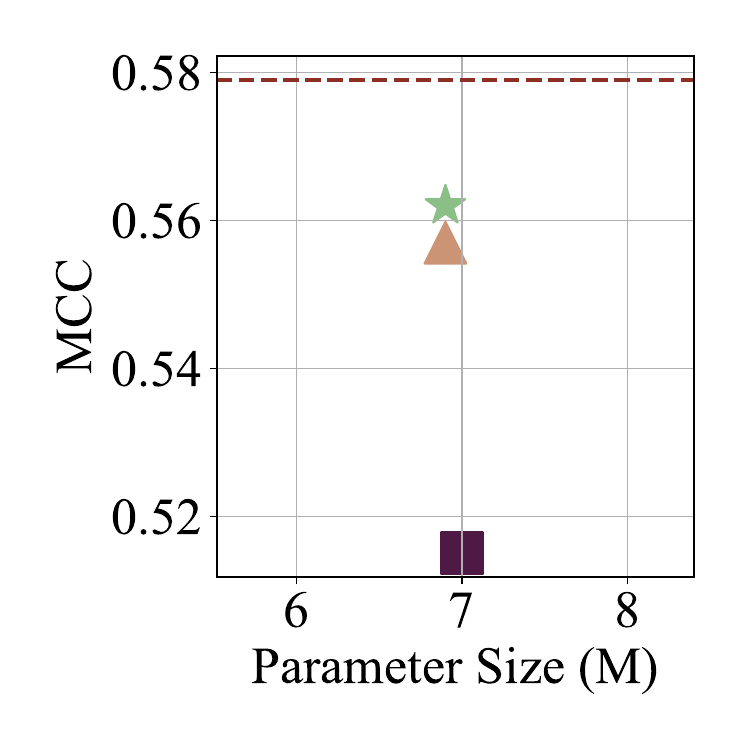}
        \caption{NT-500M on H3K79me3}
    \end{subfigure}
    \begin{subfigure}[b]{0.2\linewidth}
        \includegraphics[width=\linewidth]{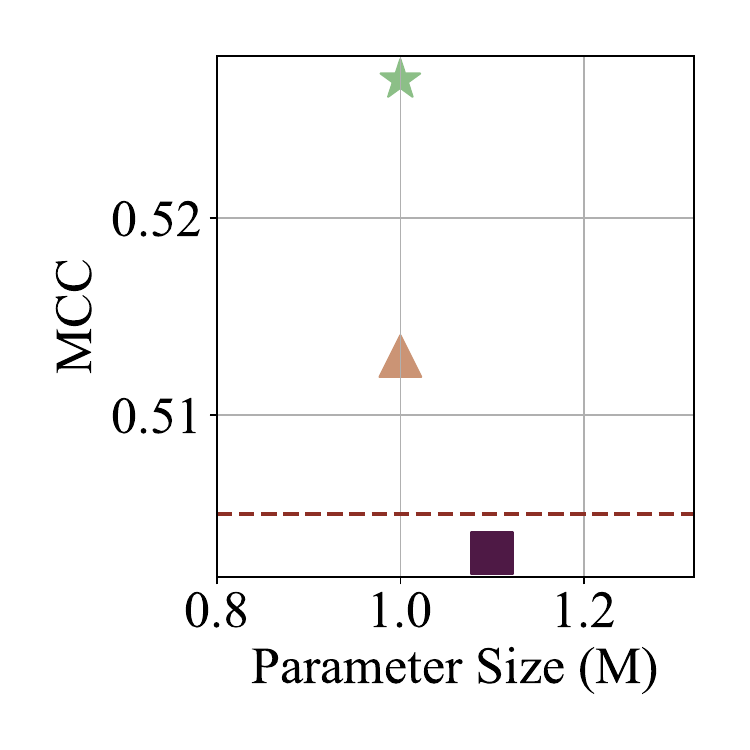}
        \caption{OPT-125M on H3K79me3}
    \end{subfigure}
    \begin{subfigure}[b]{0.2\linewidth}
        \includegraphics[width=\linewidth]{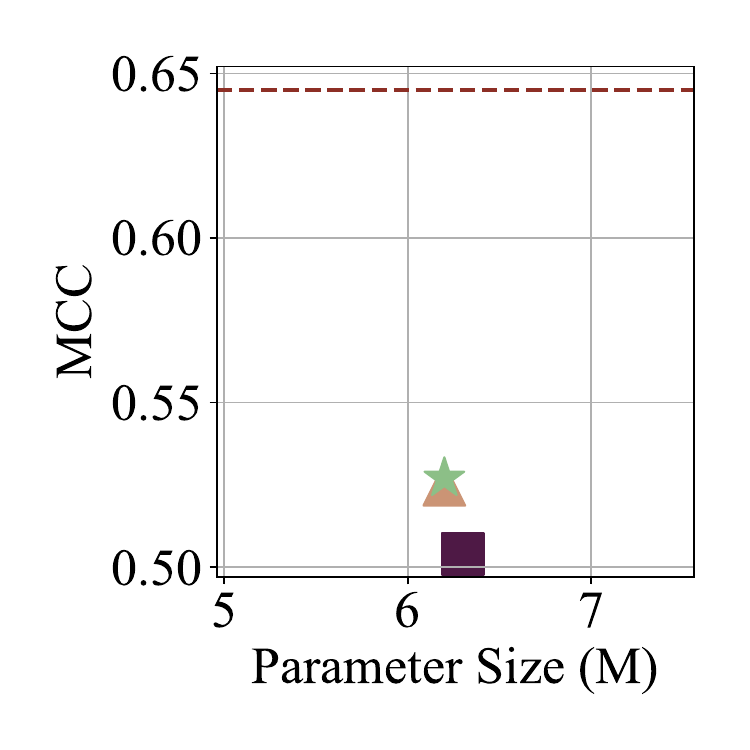}
        \caption{OPT-350M on H3K79me3}
    \end{subfigure}

                                \begin{subfigure}[b]{0.2\linewidth}
        \includegraphics[width=\linewidth]{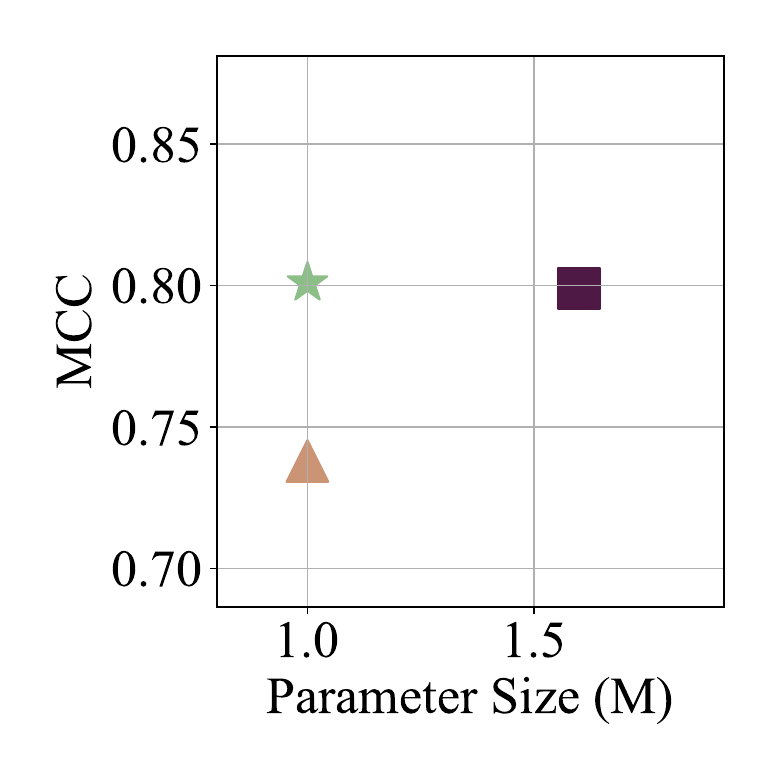}
        \caption{DNABERT-2 on H4}
    \end{subfigure}
    \begin{subfigure}[b]{0.2\linewidth}
        \includegraphics[width=\linewidth]{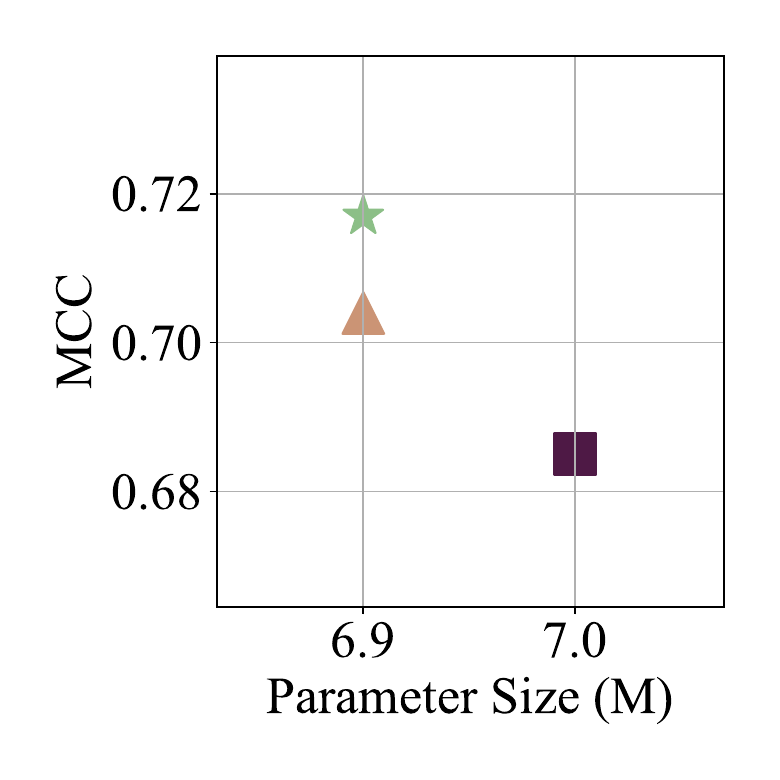}
        \caption{NT-500M on H4}
    \end{subfigure}
    \begin{subfigure}[b]{0.2\linewidth}
        \includegraphics[width=\linewidth]{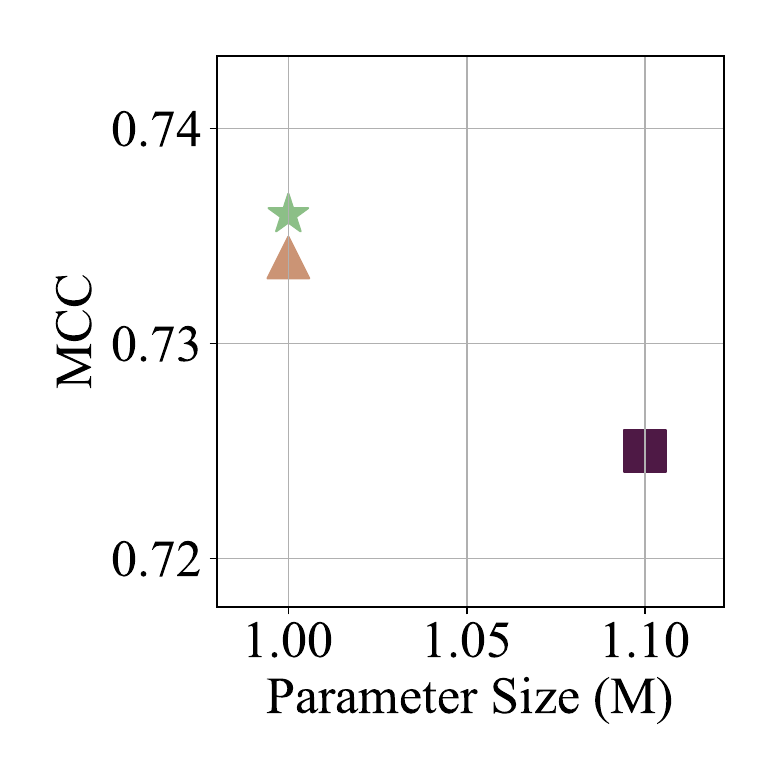}
        \caption{OPT-125M on H4}
    \end{subfigure}
    \begin{subfigure}[b]{0.2\linewidth}
        \includegraphics[width=\linewidth]{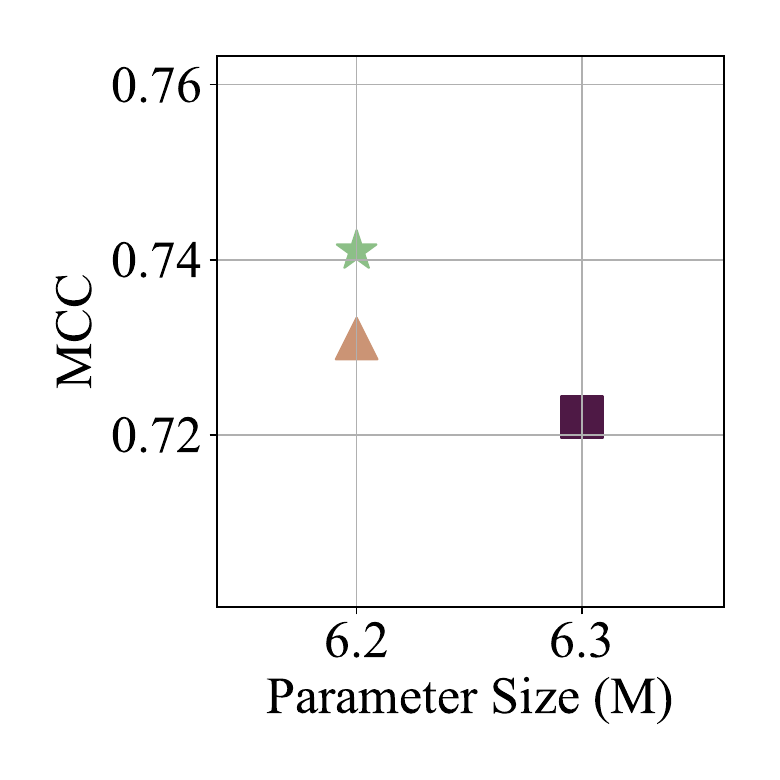}
        \caption{OPT-350M on H4}
    \end{subfigure}

    \begin{subfigure}[b]{0.2\linewidth}
        \includegraphics[width=\linewidth]{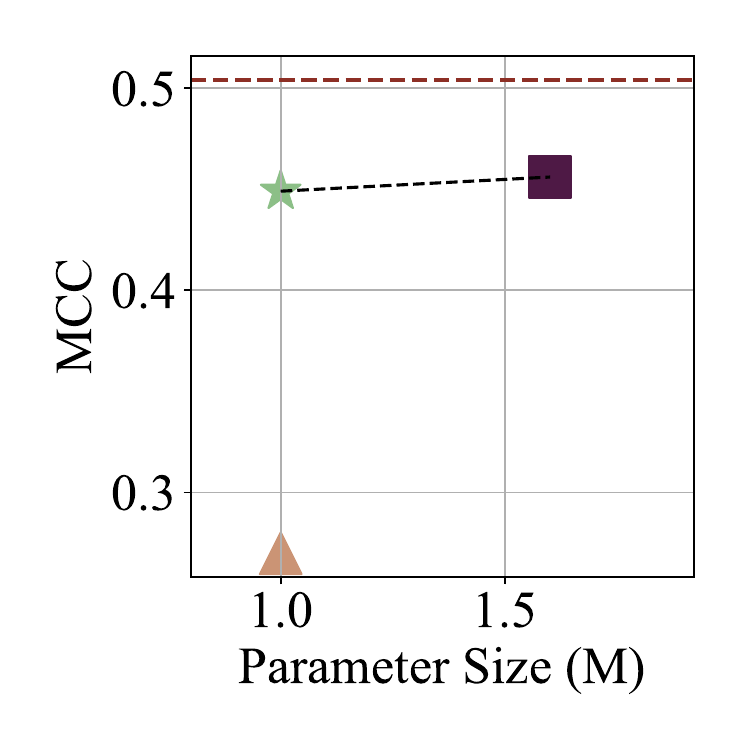}
        \caption{DNABERT-2 on H4ac}
    \end{subfigure}
    \begin{subfigure}[b]{0.2\linewidth}
        \includegraphics[width=\linewidth]{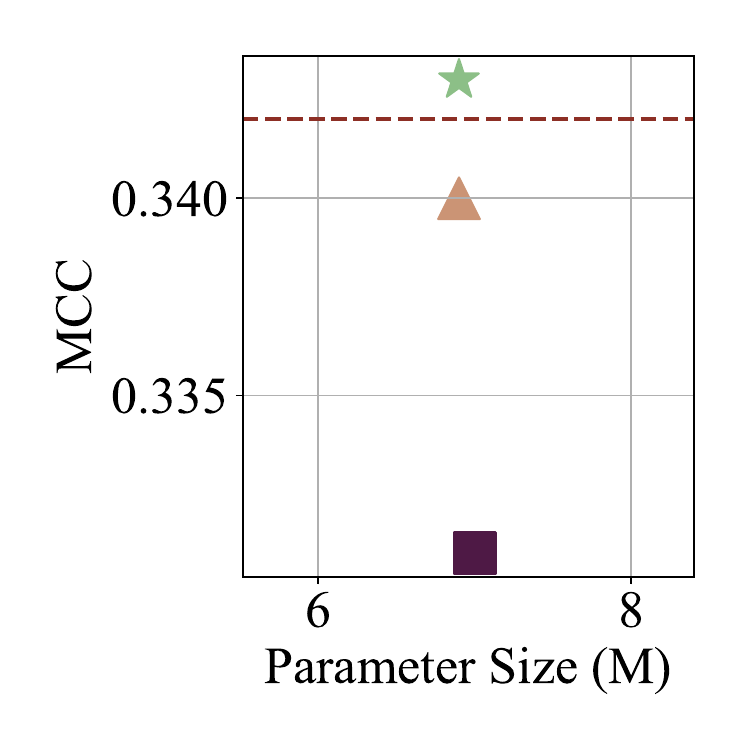}
        \caption{NT-500M on H4ac}
    \end{subfigure}
    \begin{subfigure}[b]{0.2\linewidth}
        \includegraphics[width=\linewidth]{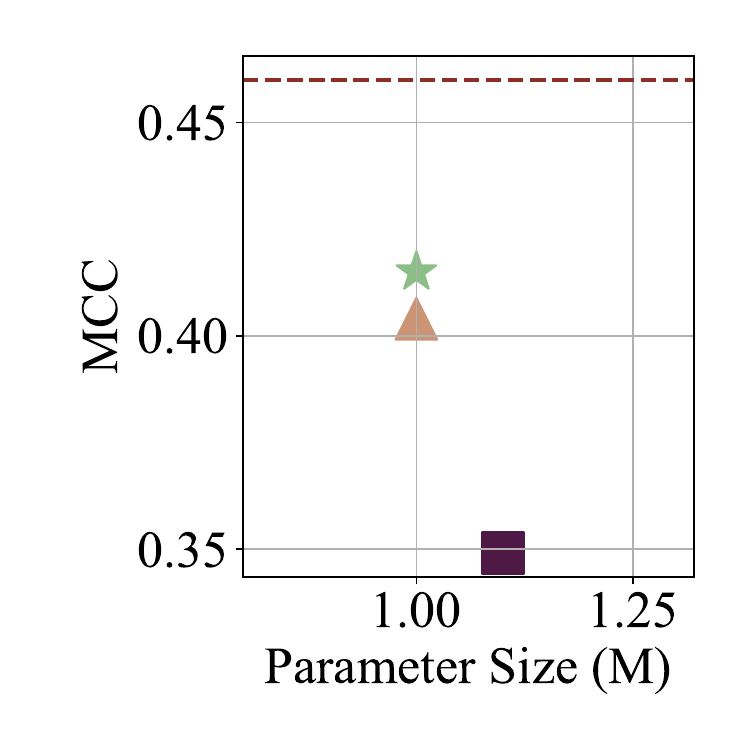}
        \caption{OPT-125M on H4ac}
    \end{subfigure}
    \begin{subfigure}[b]{0.2\linewidth}
        \includegraphics[height=2.7cm]{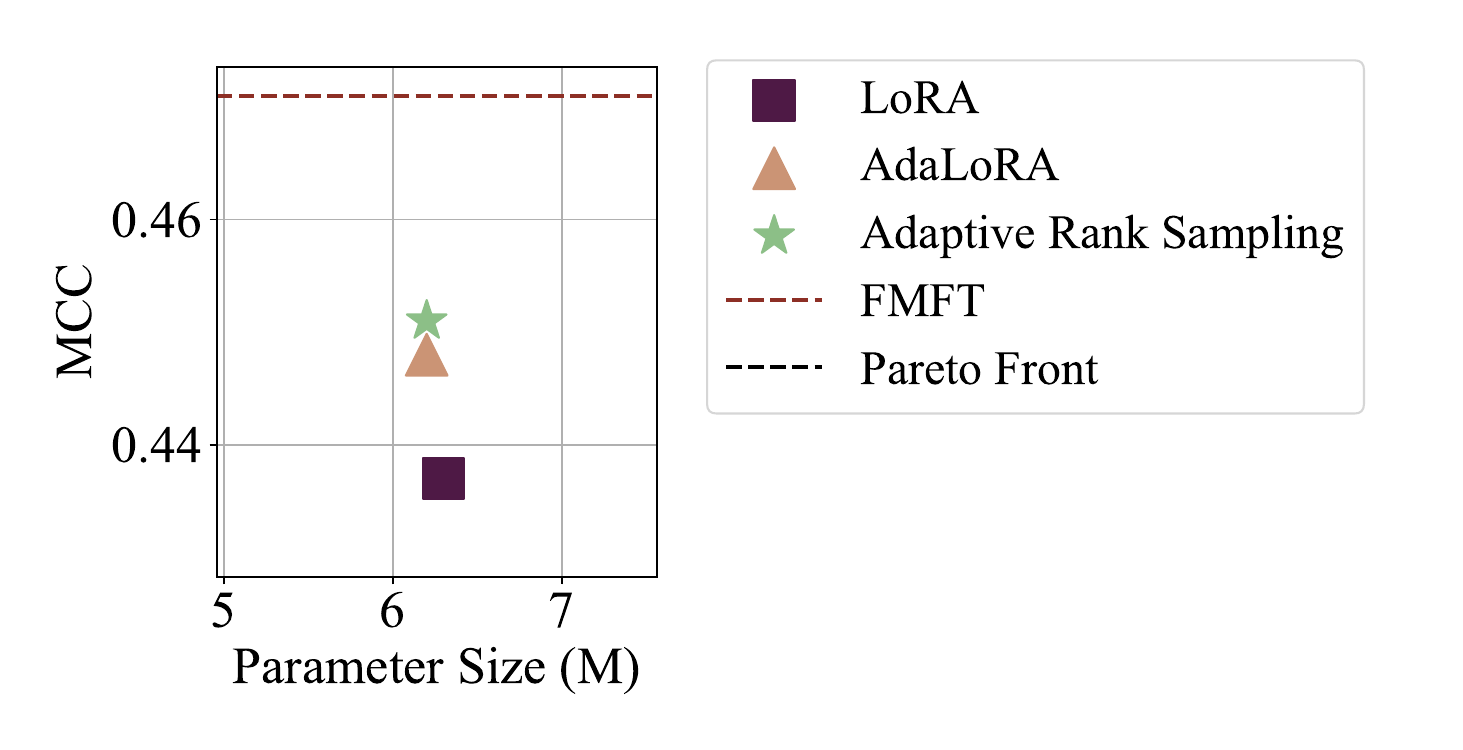}
        \caption{OPT-350M on H4ac}
    \end{subfigure}
    
        \begin{subfigure}[b]{0.49\linewidth}
        \includegraphics[width=6.5cm,height=4cm]{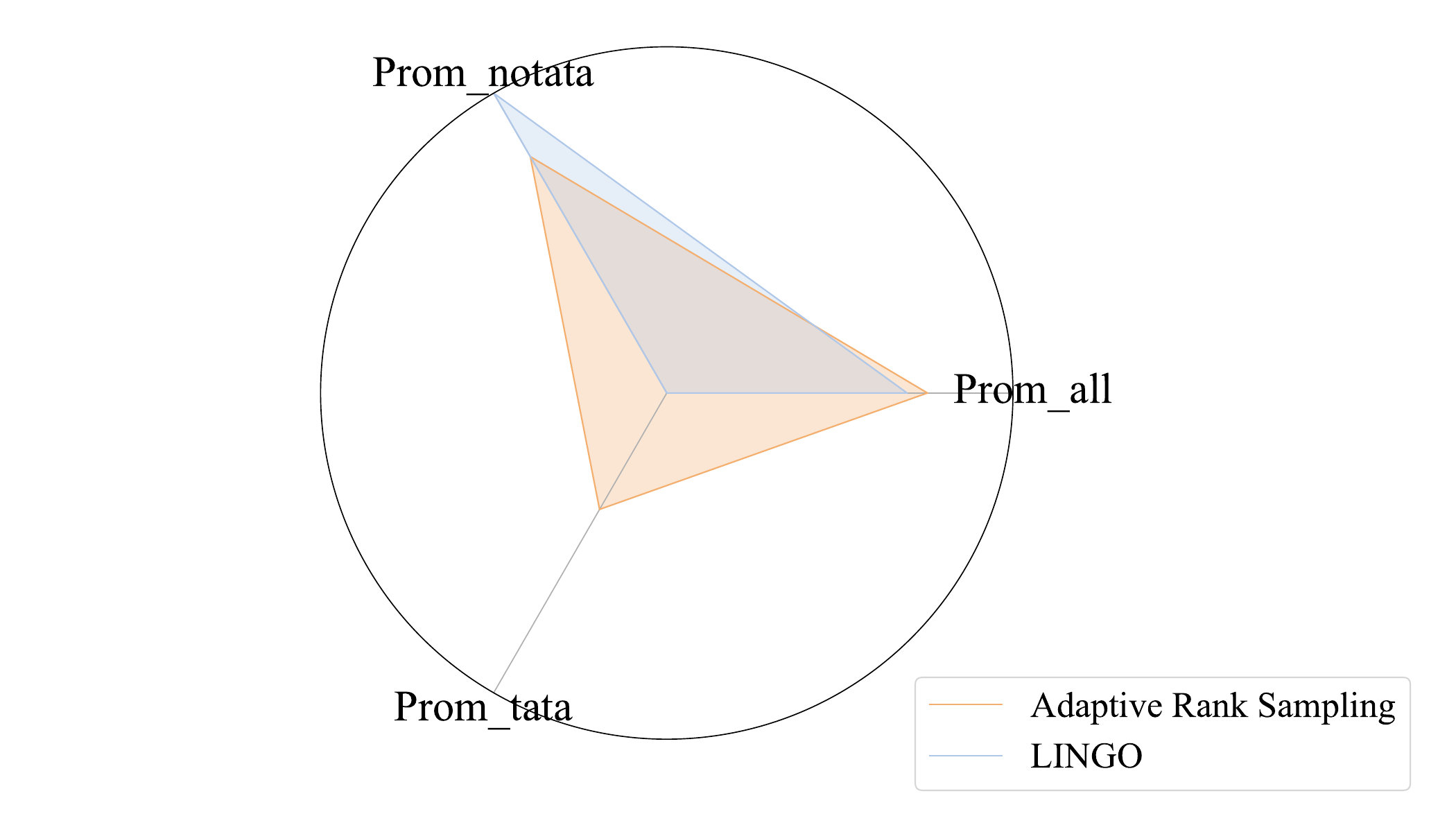}
        \caption{AUC for OPT-125M on the Promoter (Human) task.}
    \end{subfigure}
     \begin{subfigure}[b]{0.4\linewidth}
        \includegraphics[width=6.5cm,height=4cm]{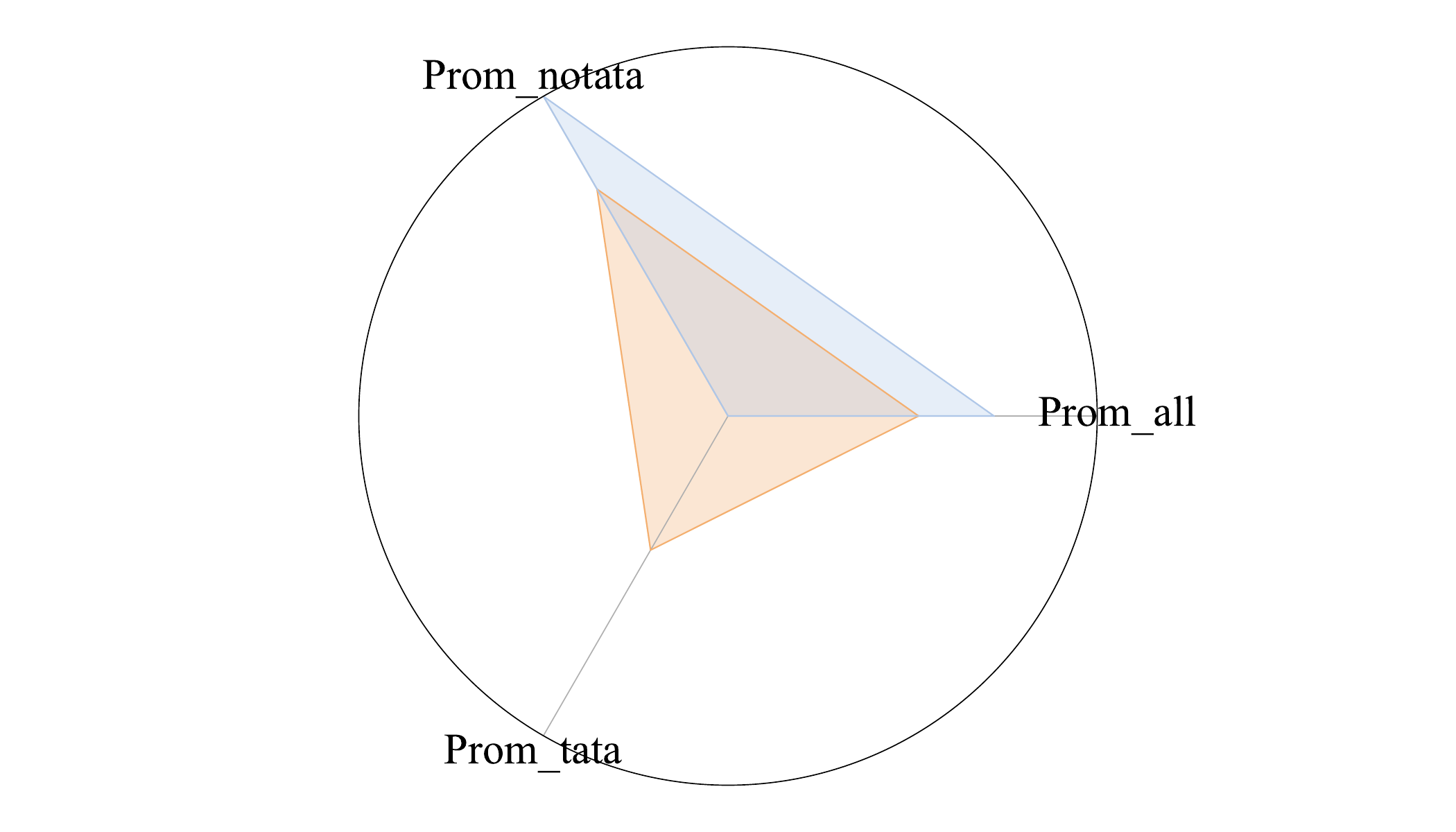}
        \caption{AUC for OPT-350M on the Promoter (Human) task.}
    \end{subfigure}
    \\
            \begin{subfigure}[b]{0.49\linewidth}
        \includegraphics[width=6.5cm,height=4cm]{opt-125m-dsp-emp-radar.pdf}
        \caption{MCC for OPT-125M on the Histone (Yeast) task.}
    \end{subfigure}
     \begin{subfigure}[b]{0.4\linewidth}
        \includegraphics[width=6.5cm,height=4cm]{opt-350m-dsp-emp-radar.pdf}
        \caption{MCC for OPT-350M on the Histone (Yeast) task.}
    \end{subfigure}
    
    \caption{AUCs for various models and methods on the Promoter (Human) task.}\label{fig:peft-more}
\end{figure}

\begin{sidewaystable}[!htbp]
       \caption{MCCs for various models and methods on the Histone (Yeast) task.}
    \label{tab:MCC_EMP}
    \centering
    \begin{tabular}{>{\centering\arraybackslash}p{1.0cm} >{\centering\arraybackslash}p{1.2cm}| >{\centering\arraybackslash}p{1.0cm} >{\centering\arraybackslash}p{1.0cm} >{\centering\arraybackslash}p{1.0cm} >{\centering\arraybackslash}p{1.0cm} >{\centering\arraybackslash}p{1.0cm} >{\centering\arraybackslash}p{1.0cm} >{\centering\arraybackslash}p{1.0cm} 
    >{\centering\arraybackslash}p{1.0cm} >{\centering\arraybackslash}p{1.0cm} >{\centering\arraybackslash}p{1.0cm}
    >{\centering\arraybackslash}p{1.0cm}}
    \hline
      \multirow{4}{*}{\textbf{Model}}   & \multirow{4}{*}{\textbf{Method}}  & \textbf{\# Train. Params.} & \multirow{4}{*}{\textbf{H3}} & \textbf{H3K4me1} & \multirow{4}{*}{\textbf{H3K4me2}}  & \textbf{H3K4me3} & \multirow{4}{*}{\textbf{H3K9ac}} & \textbf{H3K14ac}  & \multirow{4}{*}{\textbf{H3K36me3}} & \textbf{H3K79me3} & \multirow{4}{*}{\textbf{H4}}  & \textbf{H4ac}\\
         &  &  &  &  &   &  &  &  & &  &  & \\
      \hline
      \multirow{4}{*}{DNABERT-$2$}  & FMFT$^1$ &  $117$M & $\mathit{0.783}$ & $\mathbf{0.505}$ & $\mathit{0.311}$ & $\mathbf{0.363}$ & $\mathbf{0.556}$ & $0.526$ & $\mathbf{0.569}$ & $\mathbf{0.674}$ & $\mathbf{0.807}$ & $\mathbf{0.504}$\\
        ~ & LoRA & $1.6$M  & $\mathbf{0.791}$ & $\mathit{0.451}$ & $\mathbf{0.342}$ & $0$ & $\mathit{0.543}$ & $\mathit{0.523}$ & $\mathit{0.565}$ & $0.605$ & $0.799$ & $\mathit{0.456}$\\
         & AdaLoRA & $1.0$M & $0.508$ & $0.334$ & $0.136$ & $0.199$ & $0.431$ & $0.394$ & $0.461$ & $0.578$ & $0.738$ & $0.270$\\
         & Adaptive rank sampling & $1.0$M & $0.734$ & $0.450$ & $0.300$ & $\mathit{0.214}$ & $0.479$ & $\mathbf{0.531}$ & $0.557$ & $\mathit{0.614}$ & $\mathit{0.801}$ & $0.449$\\
         \hline
    \multirow{4}{*}{NT-$500$M}  & FMFT$^2$ & $500$M & $\mathbf{0.756}$ & $\mathbf{0.379}$ & $\mathit{0.288}$ & $\mathbf{0.288}$ & $\mathbf{0.488}$ & $\mathbf{0.399}$ & $\mathbf{0.461}$ & $\mathbf{0.579}$ & $\mathbf{0.752}$ & $\mathit{0.342}$\\
         & LoRA & $7$M & $0.725$ & $0.354$ & $0.240$ & $0.267$ & $0.452$ & $0.387$ & $0.442$ & $0.515$ & $0.685$ & $0.331$\\
                  & AdaLoRA & $6.9$M & $0.734$ & $0.365$ & $0.284$ & $\mathit{0.281}$ & $0.469$ & $0.390$ & $0.452$ & $0.557$ & $0.704$ & $0.340$\\
         & Adaptive rank sampling & $6.9$M & $\mathit{0.749}$ & $\mathit{0.367}$ & $\mathbf{0.289}$ & $0.280$ & $\mathit{0.474}$ & $\mathit{0.395}$ & $\mathit{0.460}$ & $\mathit{0.562}$ & $\mathit{0.717}$ & $\mathbf{0.343}$\\
          \hline
    \multirow{4}{*}{OPT-$125$M}  & FMFT & $125$M & $\mathbf{0.740}$ & $\mathbf{0.406}$ & $0.216$ & $\mathbf{0.305}$ & $\mathit{0.439}$ & $0.522$ & $\mathit{0.474}$ & $0.505$ & $\mathbf{0.750}$ & $\mathbf{0.460}$ \\
         & LoRA & $1.1$M  & $0.505$ & $0.327$ & $0.252$ & $0.274$ & $0.429$ & $0.510$ & $0.429$ & $0.503$ & $0.725$ & $0.349$ \\
         & AdaLoRA & $1.0$M& $0.562$ & $0.339$ & $\mathit{0.275}$ & $0.266$ & $0.416$ & $0.523$ & $0.431$ & $0.513$ & $0.734$ & $0.404$ \\
         & Adaptive rank sampling & $1.0$M & $0.571$ & $\mathit{0.345}$ & $\mathbf{0.281}$ & $0.275$ & $0.428$ & $\mathit{0.531}$ & $0.453$ & $\mathbf{0.527}$ & $\mathit{0.736}$ & $0.415$ \\
         & \textsc{Lingo} & $1.0$M & $\mathit{0.63}$ & $\mathbf{0.406}$ & $\mathit{0.275}$ & $\mathit{0.282}$ & $\mathbf{0.447}$ & $\mathbf{0.554}$ & $\mathbf{0.502}$ & $\mathit{0.523}$ & $0.735$ & $\mathit{0.424}$ \\
          \hline
    \multirow{4}{*}{OPT-$350$M}  & FMFT & $350$M & $\mathbf{0.801}$ & $\mathbf{0.463}$ & $\mathit{0.287}$ & $\mathit{0.289}$ & $\mathbf{0.504}$ & $\mathbf{0.517}$ & $\mathbf{0.505}$ & $\mathbf{0.645}$ & $\mathbf{0.789}$ & $\mathbf{0.471}$ \\
         & LoRA & $6.3$M & $0.619$ & $0.406$ & $0.203$ & $0.209$ & $0.439$ & $0.447$ & $0.449$ & $0.504$ & $0.722$ & $0.437$ \\
         & AdaLoRA & $6.2$M & $0.627$ & $0.409$ & $0.209$ & $0.217$ & $0.446$ & $0.459$ & $0.451$ & $0.525$ & $0.731$ & $0.448$ \\
         & Adaptive rank sampling & $6.2$M & $0.636$ & $0.411$ & $0.213$ & $0.229$ & $0.457$ & $0.463$ & $0.455$ & $0.527$ & $0.741$ & $0.451$ \\
         & \textsc{Lingo} & $6.2$M & $\mathit{0.645}$ & $\mathit{0.42}$ & $\mathbf{0.291}$ & $\mathbf{0.297}$ & $\mathit{0.478}$ & $\mathit{0.479}$ & $\mathit{0.481}$ & $\mathit{0.538}$ & $\mathit{0.753}$ & $\mathit{0.461}$ \\
         \hline
    \end{tabular}
\footnotetext[1]{Model performances (without further pre-training) are taken from the DNABERT-2 paper.}
\footnotetext[2]{Model performances are taken from the NT paper.}
\footnotetext[3]{The best results for each foundation model are highlighted in bold, while the second best are in italic. }
\end{sidewaystable}

\begin{table}[!htbp]
       \caption{\small AUCs for various models and methods on the Promoter (Human) task. The best results for each foundation model are highlighted in bold, while the second best are in italic. }
    \label{tab:PD_AUC}
    \centering
    \begin{tabular}{c c| >{\centering\arraybackslash}p{1.0cm}| >{\centering\arraybackslash}p{1.0cm} >{\centering\arraybackslash}p{1.6cm} >{\centering\arraybackslash}p{1.5cm}}
    \hline
      \multirow{2}{*}{\textbf{Model}}   & \multirow{2}{*}{\textbf{Method}}  & \textbf{\# Train. Params.} & \multirow{2}{*}{\textbf{Prom\_all}} & \multirow{2}{*}{\textbf{Prom\_notata}} & \multirow{2}{*}{\textbf{Prom\_tata}}\\
      \hline
      \multirow{4}{*}{DNABERT-$2$}  & FMFT &  $117$M & $0.908$ & $0.950$ & $0.804$\\
        ~ & LoRA & $1.6$M  &   $\mathit{0.918}$ &  $\mathbf{0.971}$ &  $\mathit{0.812}$\\
         & AdaLoRA & $1.0$M & $0.912$ & $0.954$ & $0.802$\\
         & Adaptive rank sampling & $1.0$M & $\mathbf{0.920}$ & $\mathit{0.964}$ & $\mathbf{0.815}$\\
         \hline
    \multirow{4}{*}{NT-$500$M}  & FMFT & $500$M & $\mathbf{0.950}$ & $\mathbf{0.951}$ & $\mathbf{0.939}$\\
         & LoRA & $7$M& $0.921$ & $0.942$ & $0.899$\\
                  & AdaLoRA & $6.9$M& $0.924$ & $0.949$ & $0.871$\\
         & Adaptive rank sampling & $6.9$M& $\mathit{0.926}$ & $\mathbf{0.951}$ & $\mathit{0.918}$\\
          \hline
    \multirow{5}{*}{OPT-$125$M}  & FMFT & $125$M & $0.898$ & $0.947$ & $0.864$ \\
         & LoRA & $1.1$M& $0.887$ & $0.907$ & $0.853$ \\
         & AdaLoRA & $1.0$M& $0.902$ & $0.931$ & $0.886$ \\
         & Adaptive rank sampling & $1.0$M& $\mathbf{0.959}$ & $\mathit{0.962}$ & $\mathbf{0.928}$ \\
         & \textsc{Lingo} & $1.0$M& $\mathit{0.954}$ & $\mathbf{0.98}$ & $\mathit{0.895}$ \\
          \hline
    \multirow{5}{*}{OPT-$350$M}  & FMFT & $350$M & $0.894$ & $0.923$ & $0.866$ \\
         & LoRA & $6.3$M & $0.917$ & $0.928$ & $0.904$ \\
         & AdaLoRA & $6.2$M & $0.922$ & $0.947$ & $\mathit{0.911}$ \\
         & Adaptive rank sampling & $6.2$M & $\mathit{0.938}$ & $\mathit{0.956}$ & $\mathbf{0.929}$ \\
         & \textsc{Lingo} & $6.2$M& $\mathbf{0.957}$ & $\mathbf{0.983}$ & $0.890$ \\
         \hline
    \end{tabular}
\end{table}

\newpage
\newpage
\bibliography{sn-bibliography}

\end{document}